# Fuzzy Opinion Networks: A Mathematical Framework for the Evolution of Opinions and Their Uncertainties Across Social Networks

Li-Xin Wang*, and Jerry M. Mendel, *Life Fellow, IEEE*

***Abstract*—We propose a new mathematical framework for the evolution and propagation of opinions, called Fuzzy Opinion Network, which is the connection of a number of Gaussian Nodes, possibly through some weighted average, time-delay or logic operators, where a Gaussian Node is a Gaussian fuzzy set with the center and the standard deviation being the node inputs and the fuzzy set itself being the node output. In this framework an opinion is modeled as a Gaussian fuzzy set with the center representing the opinion itself and the standard deviation characterizing the uncertainty about the opinion. We study the basic connections of Fuzzy Opinion Networks, including basic center, basic standard deviation (sdv), basic center-sdv, chain-in-center and chain-in-sdv connections, and we analyze a number of dynamic connections to show how opinions and their uncertainties propagate and evolve across different network structures and scenarios. We explain what insights we might gain from these mathematical results about the formation and evolution of human opinions.**

## I. Introduction

With the irresistible invasion of intelligent mobile phones in our daily life, people are now interconnected more than ever through social networks [32], [54], [60], [72]. Although 'big data' are available from these techno-social networks [56], [67], we must have some good mathematical models for human interactions in order to fully make use of these big data because the origin of these big data is human interactions [12], [25], [34], [63]. Much research has been undertaken to study the formation and evolution of *opinions* by researchers from a variety of disciplines such as physics, mathematics, economics, computer science, electrical engineering, management, social psychology, sociology, philosophy, law and political science, e.g.: [2] (a survey from Bayesian/non-Bayesian angle), [15] (a review from signal processing perspective), [18] (the classic DeGroot model), [22] (a perspective from mathematical sociology), [24] (extensions of the DeGroot model), [26] (the popular Hegselmann-Krause (HK) model), [27] (the latest development of the HK model), [39] (a review of group decision making), [43] (stability of the HK model), [44] (a survey from physics perspective), [47] (a survey from social psychology), [55] (a survey from control theory perspective), and [64] (a perspective from law and political science). In this paper we propose to use fuzzy sets with parametric continuous membership functions to model opinions, where the centers of the membership functions represent the opinions themselves [1] and the shape of the membership functions characterizes the uncertainty about the opinions. Our basic argument is that human opinions are inherently fuzzy [2] (uncertain) so that an opinion and its uncertainty are two sides of the same coin and should be considered *simultaneously* when we study the formation and evolution of opinions.

Where do human opinions come from? Common sense[3] tells us that much of our opinions are acquired through social learning[4] [2], [6], [9], i.e., through communications with other people in the society. Therefore, the formation and evolution of opinions should be studied in a network framework [32]. Since we represent opinions as fuzzy sets, our task in this paper is to study what happens when fuzzy sets are connected through various types of network structures; we call the networked connections of fuzzy sets *fuzzy opinion networks*[5].

To get a flavor of what kind of problems our fuzzy opinion network theory tries to address, let us consider the following example. Person A was planning to go to Beijing to attend a conference on fuzzy systems (WCCI-2014), and he has a half

*Corresponding author.

Li-Xin Wang is with the Department of Automation Science and Technology, Xian Jiaotong University, Xian, P.R. China (e-mail: lxwang@mail.xjtu.edu.cn).

Jerry M. Mendel is with the Signal and Image Processing Institute, University of Southern California, Los Angeles, CA, USA.

[1] That is, we model opinions as continuous variables. Of course, opinions may also be modeled as discrete variables such as in [16] and [42], and we agree with Mason, Conrey and Smith that ([47], page 296): "Continuous attitudes are more realistic models of mental representations but discrete behaviors often serve as the visible cues of internal states."

[2] By which we mean human opinions usually do not have clear-cut boundaries; for example, when we say a person is nice (our opinion about the person), we mean this person belongs to the set "nice people", but "nice people" does not have clear-cut boundaries, i.e., "nice people" is a fuzzy set.

[3] Without a deep understanding of individual-level psychological processes however, (as Mason, Conrey and Smith argued in [47]) we may end up making empirically questionable simplifying assumptions such as that people are rational and accuracy-seeking or that the influence is always assimilative.

[4] Of course, as pointed out by Mason, Conrey and Smith in [47], page 296: " … models should attempt to integrate social influence with other effects on individual decisions rather than to be models solely of social influence that assume people have no other nonsocial reasons to hold one opinion or another."

[5] The concept of fuzzy opinion network was first introduced in [71] (which was a preliminary conference version of the current paper) where it was called a fuzzy network.

day off to visit some interesting places in Beijing. He was looking at the map of Beijing in Fig. 1. As a fuzzy researcher, he used a fuzzy set $X$ = "interesting places in Beijing" with Gaussian membership function:

$$\mu_X(x) = e^{-\frac{|x-c_1|^2}{\sigma_1^2}} \quad (1)$$

to model the problem, where $x = (x_1, x_2)$ is a location on the map of Beijing in Fig. 1, the center $c_1 = (c_{11}, c_{12})$ is the location of the "most interesting place" (e.g., $c_1 = (0,0)$ is TianAnMen Square in Fig. 1), and the standard deviation $\sigma_1$ is a factor quantifying the uncertainty about the choice of $c_1$. He chose the uncertainty factor $\sigma_1 = 1$, but he did not know how to determine the center $c_1$ (the most interesting place) in his model. So he asked his former Ph.D. student, Person B, who is a local resident of Beijing, to provide him the $c_1$.

As a fuzzy researcher himself, Person B again used the Gaussian fuzzy set:

$$\mu_{C_1}(c_1) = e^{-\frac{|c_1-c_2|^2}{\sigma_2^2}} \quad (2)$$

to model the assignment from his advisor. Although a local, the fuzzy Person B knows very little about the attractions in Beijing and he did not know how to determine the center $c_2$ and the standard deviation $\sigma_2$ in his model (2), so he asked his wife for help who recommended their neighbor with the nickname "Beijing Yellow Page (BYP)" to make the suggestion. Consequently, the lazy Person B passed his assignment to other people: his neighbor BYP to provide $c_2$ and his wife (who recommends BYP) to provide $\sigma_2$ which measures the reliability of her recommendation. Fig. 2 illustrates the situation[6].

Although very confident in her knowledge about the attractions in Beijing, BYP knew very little about the taste of the original Person A who was going to visit the places (other than the very brief introduction from Person B's wife that Person A is a professor in Electrical Engineering from the US), so she used yet another Gaussian fuzzy set:

$$\mu_{C_2}(c_2) = e^{-\frac{|c_2-c_3|^2}{\sigma_3^2}} \quad (3)$$

to model her suggestion $C_2$ to Person B, where she gave a particular value for $c_3$ (= the location of Lama Temple = point 2 in Fig. 1) and chose $\sigma_3$ to be a medium value $\sigma_3 = 0.5$ due to her limited knowledge about Person A.

Finally, Person B's wife had to provide $\sigma_2$ which she modeled as a Gaussian fuzzy set $\Sigma_2$ with membership function:

$$\mu_{\Sigma_2}(\sigma_2) = e^{-\frac{|\sigma_2-c_4|^2}{\sigma_4^2}} \quad (4)$$

where she chose $c_4 = 0$ (she had full faith in BYP) and $\sigma_4 = 1$ (measuring her uncertainty about her recommendation). Now the question is: What does Person A's final fuzzy set look like when these fuzzy sets are connected together as in Fig. 2? The goal of this paper is to answer this kind of question in a mathematically rigorous fashion.

[6] We will study this fuzzy opinion network in detail in Section III Connection 6 where the meanings of the variables in Fig. 2 will become clear; here in the Introduction our purpose is just to "get a flavor".

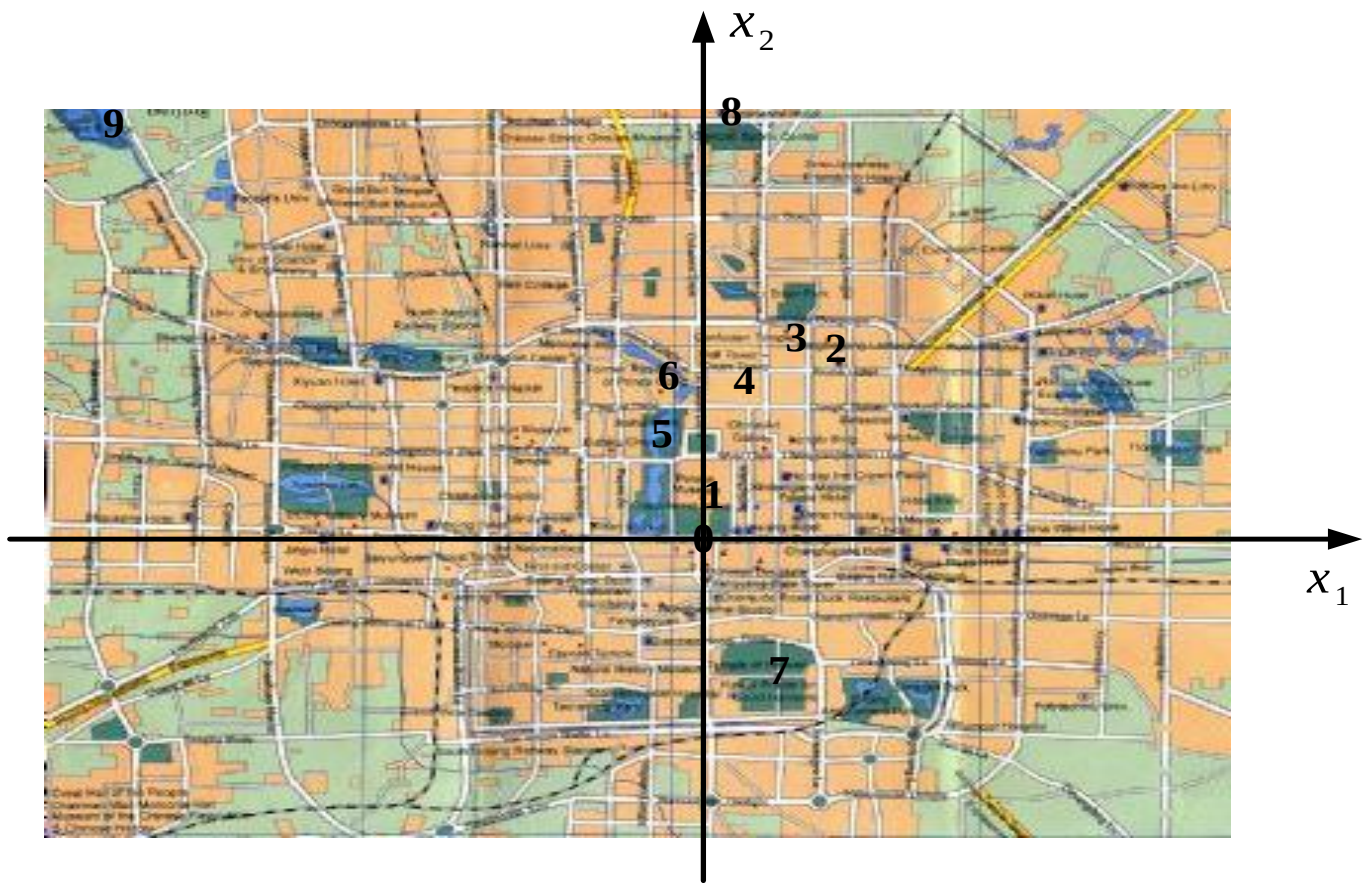

| | |
|---|---|
| 0: TianAnMen Square | 5: North Sea Park |
| 1: Forbidden City | 6: Houhai |
| 2: Lama Temple | 7: Temple of Heaven |
| 3: Imperial College | 8: Olympic Stadium |
| 4: South Luogu Lane | 9: Summer Palace |

Fig. 1: Some interesting places in Beijing.

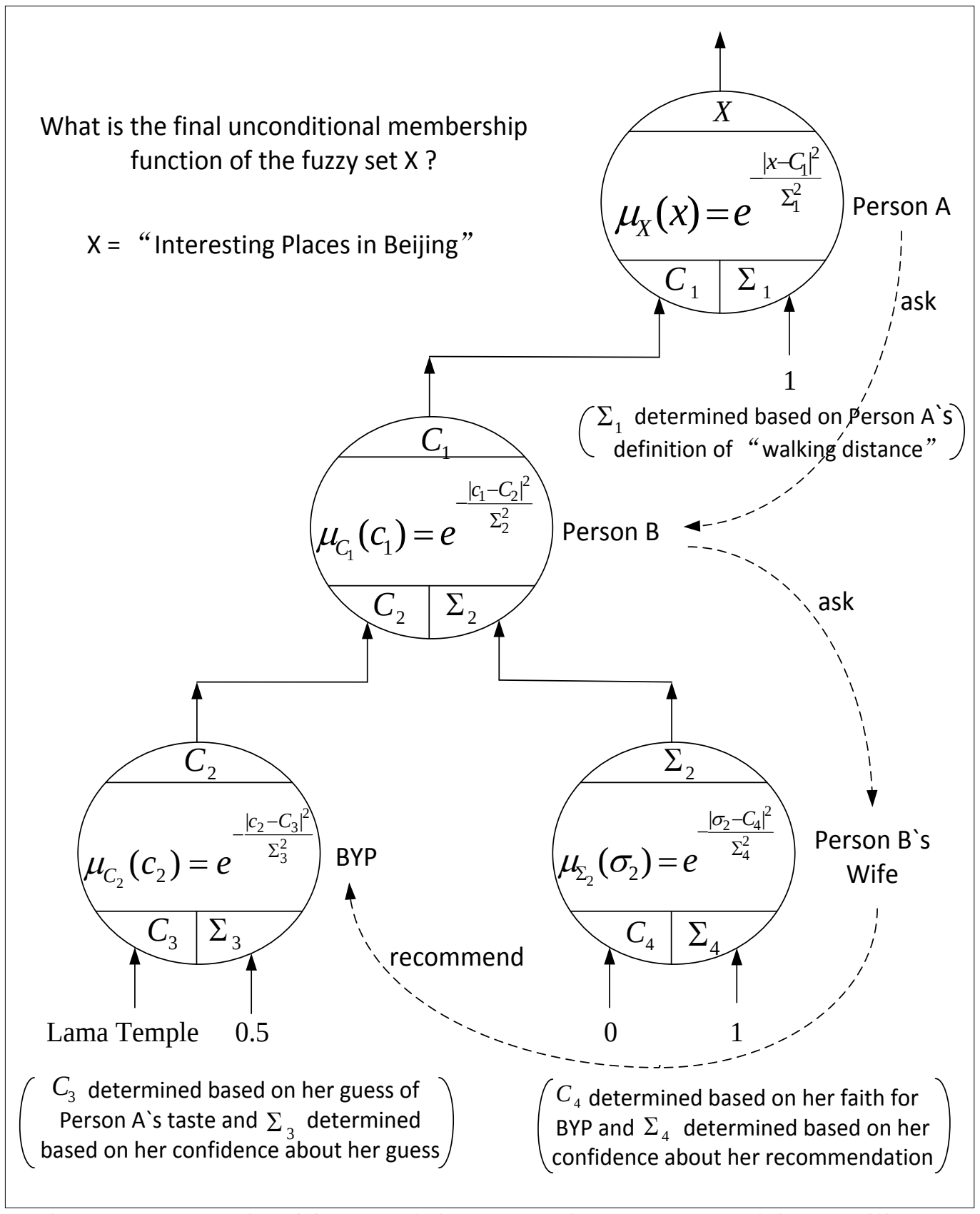

Fig. 2: An example of fuzzy opinion network: Person A's visit to Beijing (Connection 6).

From the example above we can see two major differences between our fuzzy opinion network theory and the traditional network theory [54]:

First, the traditional network theory is based on graph theory [20] where the nodes are simple points without content, whereas the nodes in our fuzzy networks have opinions that are characterized by the central ideas (centers of the Gaussian fuzzy sets) and the uncertainties (standard deviations of the Gaussian fuzzy sets) about these ideas. People communicate with each other using natural languages where fuzzy words and vague expressions are the mainstay rather than exceptions, therefore using a point without content to represent a person oversimplifies the situation and makes it difficult to analyze some key problems in human interactions within the traditional network theory framework. For example, different people have different definitions of "friend" [9], therefore any "friend network" based on survey data or web connection is inherently unreliable. Our fuzzy opinion networks provide a framework to model the contents of human communications and thus are closer to reality than the simple graph representation in traditional network theory.

Second, a large portion of recent network researches [54] study statistical models of graphs which tend to break down when applied to specific, small networks, such as terrorist networks [38]. Statistical models, by their very nature, address the overall questions (such as how many links are needed on average for a person to make connection with any other person in the world – the "small-world problem" [73] or "six degrees of separation") that may be irrelevant to the problems when studying real, specific network structures. In our fuzzy opinion networks, the connections are deterministic and convey specific meanings, making it possible to model content-sensitive connections of human activities.

The fuzzy opinion networks may look similar to Bayesian networks [33], [46], [59], but there are some fundamental differences:

First, the nodes in a Bayesian network are random variables, whereas the nodes in a fuzzy opinion network are fuzzy sets. Random variables can be sampled or be realized to get numerical values, whereas fuzzy sets are membership functions which are the smallest computing elements in the fuzzy opinion network framework and cannot be further reduced. The research on Bayesian networks concentrates on the computation of the conditional probabilities of the nodes when sample values or realizations of some of the random variables (nodes) become available [14], whereas the task of fuzzy opinion network research is to determine the membership functions of the nodes given the structure of the fuzzy opinion network. In short, in Bayesian networks we "compute with numbers", whereas in fuzzy opinion networks we "compute with words" [78], where words are modeled as fuzzy sets characterized by parametric continuous membership functions[7]. That is, in our fuzzy opinion networks "words = parametric continuous membership functions", therefore "computing with words" in our models means "determining the structures and parameters of the membership functions" which is a mathematical and numerical exercise.

Second, the basic computing formula for Bayesian networks is the Bayes' Theorem [57] which uses *integral-product* (see footnote 13) to combine the conditional probability density functions of the connected nodes, whereas the basic computing formula for fuzzy opinion networks is Zadeh's Compositional Rule of Inference [75] which uses *max-min* to get the membership functions of the nodes (see Section II for details). Usually, max-min computation is simpler than integral-product computation, making it possible to obtain simple analytic formulas for the membership functions in many cases where the integral-product operation does not lead to simple solutions (see, e.g., Connections 2, 3, 5 and 6 in Section III).

Third, from a conceptual point of view, Bayesian networks study the propagation of probabilistic uncertainties ("happen or not-happen" , "be or not-to-be") across a structured network, whereas fuzzy opinion networks study the propagation of linguistic vagueness when people communicate with each other through social networks using a natural language. Probabilistic uncertainties can disappear when new information becomes available (the uncertain things have happened or not happened), whereas linguistic vagueness, once created by a person, will be propagated across the network (like a virus) and we cannot expect the original person to provide the "correct information" at a later stage to clarify the vagueness; we have to study what happens to the linguistic vagueness within the structure of the fuzzy opinion network.

The fuzzy opinion networks of this paper may be viewed as *mathematically tractable agent-based models* (ABMs). Each Gaussian node (such as those in Fig. 2) represents an agent, and the agents are connected through different network structures to model different scenarios of opinion diffusion and evolution; then, we let the math lead us into the unknown territories – we push the maths as far as we can[8] and see what messages we can extract from the mathematical results. The conventional ABMs are mostly simulation-based, and their main problem is that there are too many parameters and too many degrees of freedom such that it is often difficult to determine the causes for the observed model outputs [28]. We put mathematical tractability as a constraint to our ABMs (the fuzzy opinion networks), trying to develop some simple and parsimonious heterogeneous agent models to overcome the wilderness of simulation-based ABMs[9].

The rest of the paper is organized as follows. In Section II, we define fuzzy opinion networks and propose to aggregate the conditional fuzzy sets using Zadeh's Compositional Rule of Inference. In Section III, we determine the membership functions of some basic static connections of Gaussian fuzzy nodes, which provide the foundation for more complex fuzzy opinion networks. In Sections IV and V, dynamic Gaussian fuzzy opinion networks with constant and time-varying confidence are studied, respectively, and the implications of the mathematical results in terms of human behavior are discussed. In Section VI, more general time-varying or state-dependent connections are investigated and a bounded confidence co-evolutionary fuzzy opinion network is proposed. Section VII concludes the paper and outlines a research agenda with seven future research directions.

## II. Definition of Gaussian Fuzzy Opinion Networks

First, we define Gaussian Fuzzy Opinion Networks as follows.

[7] As commented in the influential AI textbook [59] on page 551: "there remain many open issues concerning the proper representation of linguistic observations and continuous quantities – issues that have been neglected by most outside the fuzzy community."

[8] We require that analytic formulas must be obtained for the membership functions of the agents' opinions; i.e., we stop when we are unable to obtain such analytic formulas for the membership functions.

[9] Of course, simulations are one of the most important origins of new ideas and new theories and provide a powerful complement to mathematics.

**Definition 1:** A *Gaussian node i* in a Gaussian Fuzzy Opinion Network is a 2-input-1-output node characterized by the Gaussian membership function:

$$\mu_X(x) = e^{-\frac{|x-C_i|^2}{\Sigma_i^2}} \quad (5)$$

where the center $C_i$ and the standard deviation $\Sigma_i$ are two input fuzzy sets to the node defined on the universes of discourse $\Omega_{C_i} = R^n$ and $\Omega_{\Sigma_i} = R^+$, respectively, and the fuzzy set $X$ is the output of the node defined on the universe of discourse $\Omega_X = R^n$. A *Gaussian Fuzzy Opinion Network* (G-FON) is a connection of a number of Gaussian nodes, possibly through some weighted-sum, delay, and/or logic operation elements[10], as shown in Fig. 3[11].

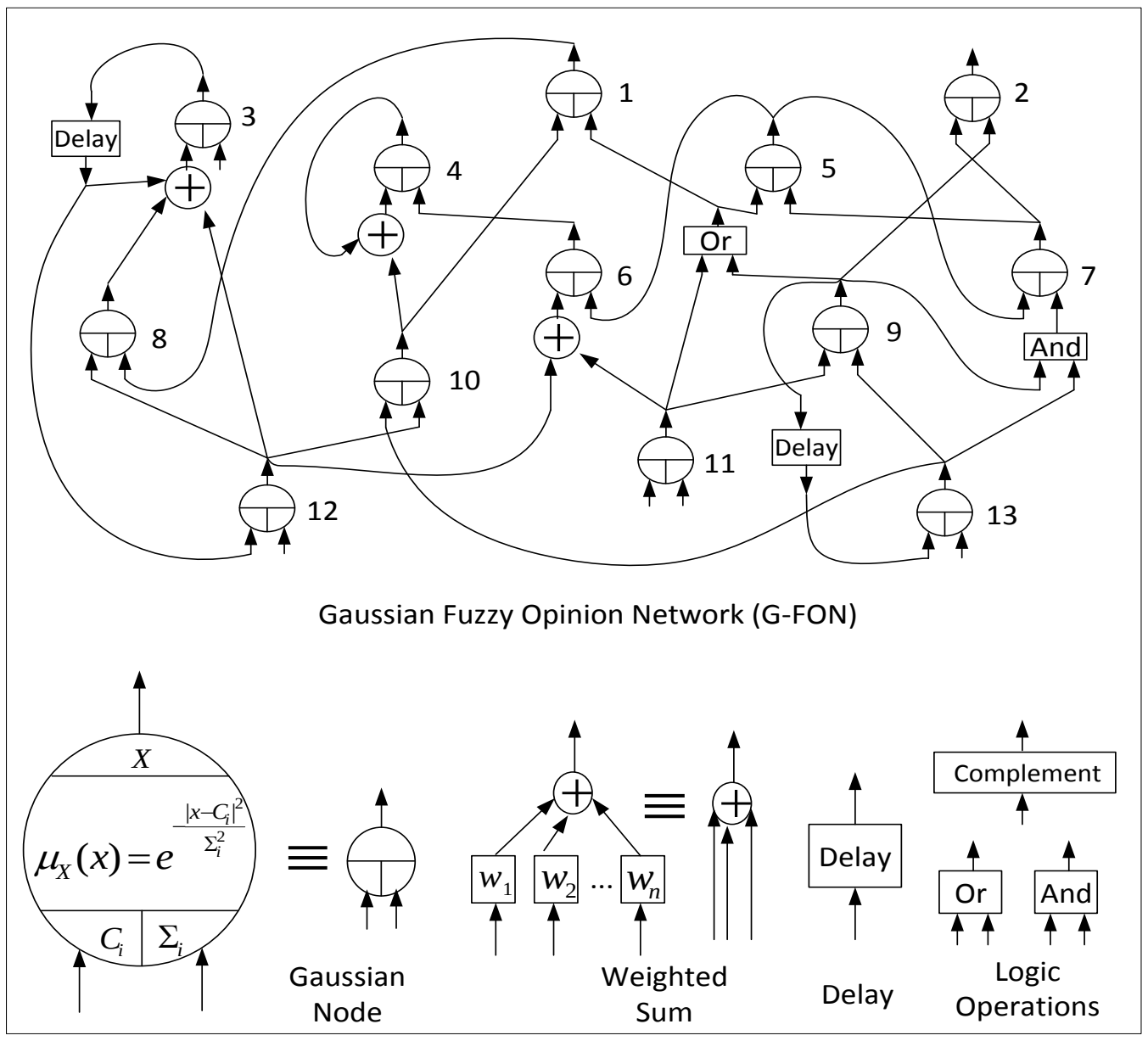

Fig. 3: Gaussian node and Gaussian Fuzzy Opinion Network.

[10] In this paper we consider only weighted-sum and delay connections. The logic operators are useful to model the situations where union, intersection or other logic combinations of the agents' fuzzy outputs are more meaningful; for example, if the Person A in Fig. 2 asked more than one person to recommend "interesting places in Beijing," then the union (max) operator would be most meaningful to connect these recommendations. The weighted-sum combinations usually give assimilative models (as the Connections in later sections of this paper demonstrate), whereas a hierarchy of max/min combinations would lead to contrastive models, especially when more aggressive logic operators such as the Drastic Sum and Drastic Product (for more such operators see e.g. Chapter 3 of [68]) are employed, where group polarization and extremists [65] appear naturally. An anonymous referee suggested to try the fuzzy opinion networks for the group polarization problems and we are very grateful for this important recommendation.

[11] Other membership functions such as triangular, symbolic or trapezoidal membership functions can be used as the basic elements for constructing the fuzzy opinion networks; for example, if we replace the Gaussian membership function (5) by the triangular function $\mu_X(x) = \begin{cases} 1 - \frac{|x-C_i|}{B_i}, & |x - C_i| \le B_i \\ 0 & otherwise \end{cases}$, then we will have a *Triangular Fuzzy Opinion Network* (T-FON). Furthermore, if we use the asymmetric triangular membership function $\mu_X(x) = \begin{cases} 1 + \frac{x-C_i}{B_i^l}, & C_i - B_i^l \le x \le C_i \\ 1 - \frac{x-C_i}{B_i^r}, & C_i \le x \le C_i + B_i^r \\ 0 & otherwise \end{cases}$, then we will be able to handle opinions with *asymmetric uncertainty*. Since the current paper is the first formal paper on fuzzy opinion networks, we will concentrate on the Gaussian membership function and leave the other options to future research.

We see from Figs. 2 and 3 that the key to a fuzzy opinion network is that the membership function of one node depends on the fuzzy sets from other nodes. We define this formally as conditional fuzzy sets in:

**Definition 2:** Let $X, V$ be fuzzy sets defined on the universes of discourse $\Omega_X, \Omega_V$, respectively. A *conditional fuzzy set*[12], denoted as $X|V$, is a fuzzy set defined on $\Omega_X$ with the membership function:

$$\mu_{X|V}(x|V), \ \ x \in \Omega_X \quad (6)$$

depending on the fuzzy set $V$.

For example, $\mu_{X|V}(x|V) = e^{-\frac{|x-V|^2}{\sigma_1^2}}$ defines a conditional fuzzy set whose center $V$ is a Gaussian fuzzy number characterized by $\mu_V(v) = e^{-\frac{|v-c|^2}{\sigma_2^2}}$, where $\sigma_1 > 0, \sigma_2 > 0$ and $c$ are given constants. Other examples of conditional fuzzy sets are the fuzzy sets $\mu_X(x)$ and $\mu_{C_1}(c_1)$ in Fig. 2: $\mu_X(x)$ (Person A's fuzzy set) depends on the fuzzy set $C_1$ obtained from Person B, and $\mu_{C_1}(c_1)$ (Person B's fuzzy set) depends on the fuzzy sets $C_2$ and $\Sigma_2$ obtained from BYP and Person B's wife, respectively.

If the membership function of $V$, $\mu_V(v)$, is given, the two fuzzy sets $X|V$ and $V$ can be combined to get the unconditional fuzzy set $X$ using Zadeh's Compositional Rule of Inference, as follows:

**Combining Conditional Fuzzy Sets Using the Compositional Rule of Inference:** Given $\mu_{X|V}(x|V)$ and $\mu_V(v)$ $(v \in \Omega_V)$, the membership function of the unconditional fuzzy set $X$ can be obtained using Zadeh's Compositional Rule of Inference [75] with minimum t-norm, as follows:

$$\mu_X(x) = \max_{v \in \Omega_V} min\left[\mu_{X|V}(x|v), \mu_V(v)\right] \quad (7)$$

More generally, given two conditional membership functions $\mu_{X|V}(x|V)$ and $\mu_{V|U}(v|U)$ where $U$ is a fuzzy set defined on $\Omega_U$, the intermediate fuzzy set $V$ can be canceled out to get the conditional fuzzy set $X|U$ with membership function:[13]

$$\mu_{X|U}(x|U) = \max_{v \in \Omega_V} min\left[\mu_{X|V}(x|v), \mu_{V|U}(v|U)\right] \quad (8)$$

Using (7) or (8), we now develop the G-FON theory through the detailed analyses of a number of typical Connections in the next four sections, starting from the basic static connections in Section III (Connections 1 to 6), moving to dynamic connections in Sections IV and V (Connections 7 to 11), and finally concluding with the more general time-varying or state-dependent connections in Section VI (Connections 12 and 13).

## III. Basic Connections of Gaussian Nodes

**Connection 1 (Basic Center Connection):** Here, we consider the *basic center connection of Gaussian nodes* in Fig. 4, and determine the unconditional membership function $\mu_X(x)$ of fuzzy set $X$ (notice that the inputs to $C_2, \Sigma_1, \Sigma_2$ in Fig. 4 are crisp numbers (fuzzy singletons): $C_2 = c_2 \in R^n$, $\Sigma_1 = \sigma_1 \in R^+$, $\Sigma_2 = \sigma_2 \in R^+$). Using

[12] Although the conditional possibility distributions were proposed in the possibility theory literature [77], [79] (a very short introduction to possibility theory is Chapter 31 of [68]), up to the authors' knowledge the concept of a conditional fuzzy set as defined here first appeared in [71] which was a preliminary conference version of the current paper.

[13] Recall that in probability theory [17], [57], the conditional densities are combined through $f_{X|U}(x|U) = \int f_{X|V}(x|v) f_{V|U}(v|U) dv$.

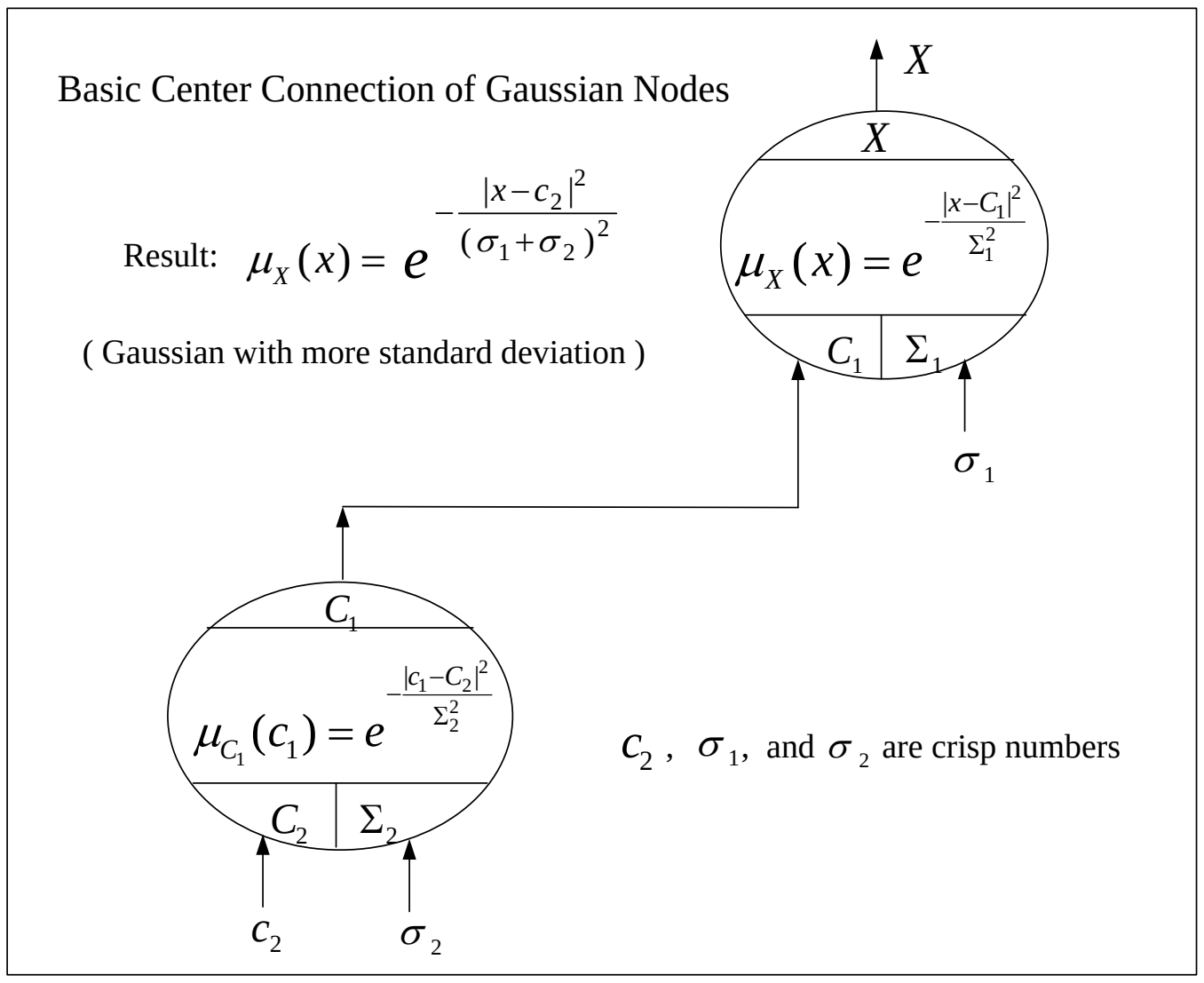

Fig. 4: Basic Center Connection of Gaussian nodes (Connection 1).

$$\mu_{X|C_1}(x|C_1) = e^{-\frac{|x-c_1|^2}{\sigma_1^2}} \tag{9}$$

$$\mu_{C_1}(c_1) = e^{-\frac{|c_1-c_2|^2}{\sigma_2^2}} \tag{10}$$

as the $\mu_{X|V}(x|v)$, $\mu_V(v)$ in the Compositional Rule of Inference (7), respectively, we have from (7) that the

$$\mu_X(x) = \max_{c_1 \in R^n} min\left[e^{-\frac{|x-c_1|^2}{\sigma_1^2}}, e^{-\frac{|c_1-c_2|^2}{\sigma_2^2}}\right] \tag{11}$$

Since $e^{-\frac{|x-c_1|^2}{\sigma_1^2}}$ and $e^{-\frac{|c_1-c_2|^2}{\sigma_2^2}}$ are symmetric functions of $c_1$ around the centers $x$ and $c_2$, respectively, the max in (11) must be achieved at a point on the line which connects the two centers $x$ and $c_2$. Plotting $e^{-\frac{|x-c_1|^2}{\sigma_1^2}}$ and $e^{-\frac{|c_1-c_2|^2}{\sigma_2^2}}$ as functions of $c_1$ over the line connecting $x$ and $c_2$, as shown in Fig. 5, we see that the max in (11) is achieved at the intersection point:

$$\frac{x-c_1}{\sigma_1} = \frac{c_1-c_2}{\sigma_2} \tag{12}$$

which gives

$$c_1 = \frac{c_2\sigma_1 + x\sigma_2}{\sigma_1+\sigma_2} \tag{13}$$

Substituting (13) into (11) yields

$$\mu_X(x) = e^{-\frac{|x-c_2|^2}{(\sigma_1+\sigma_2)^2}} \tag{14}$$

■

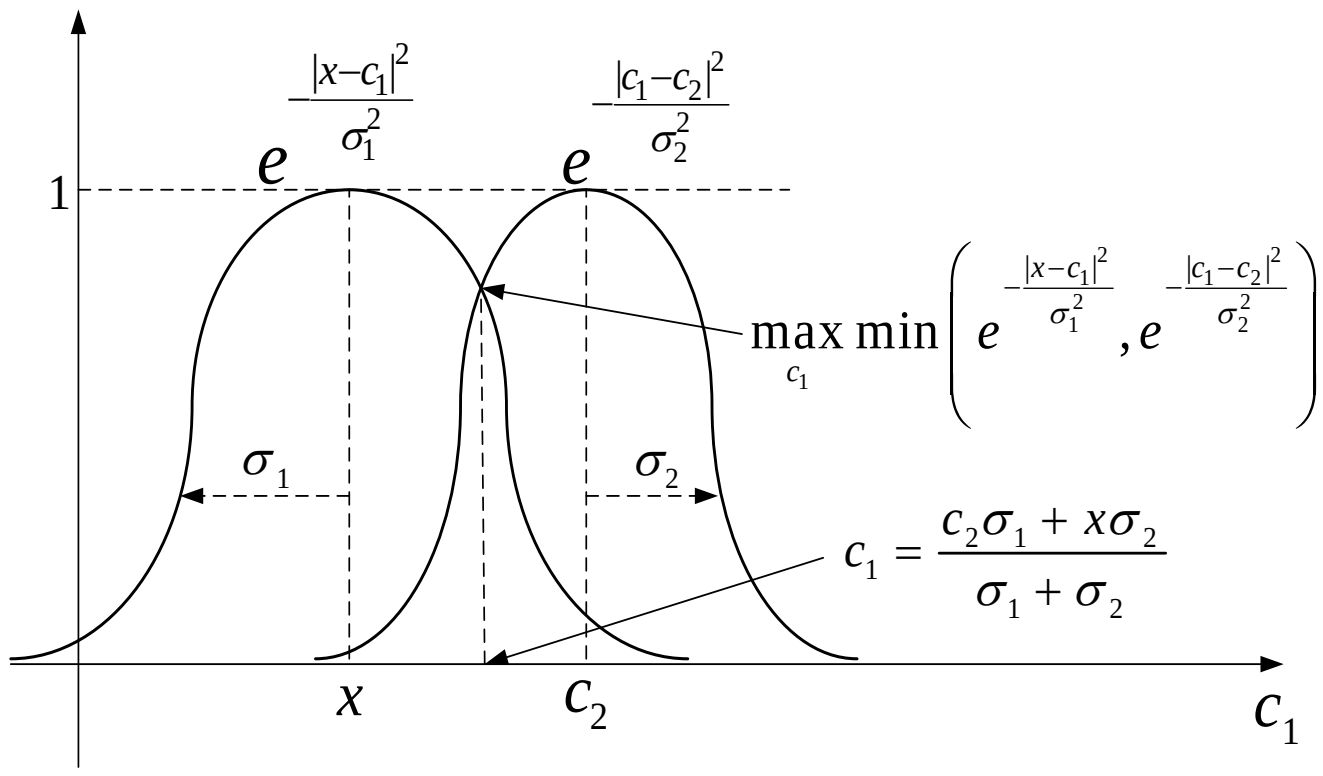

Fig. 5: How the max in (11) is achieved for the Basic Center Connection.

We see from (14) that when two Gaussian fuzzy nodes (9) and (10) are connected in-the-center as in Fig. 4, the Gaussian functional form does not change and the center $c_2$ is transferred from the input node to the output node, but the standard deviation (sdv) increases to the summation of the two individual sdv's; that is, the uncertainty (measured as sdv) increases in a linear fashion when the two fuzzy nodes are connected in-the-center.

**Connection 2 (Basic Sdv Connection):** Here, we consider the *basic sdv (standard deviation) connection of Gaussian nodes* shown in Fig. 6, and determine the unconditional membership function $\mu_X(x)$ of fuzzy set $X$. Using

$$\mu_{X|\Sigma_1}(x|\Sigma_1) = e^{-\frac{|x-c_1|^2}{\Sigma_1^2}} \tag{15}$$

$$\mu_{\Sigma_1}(\sigma_1) = e^{-\frac{|\sigma_1-c_2|^2}{\sigma_2^2}} \tag{16}$$

for $\mu_{X|V}(x|v)$, $\mu_V(v)$ in (7), respectively, we have from (7) that

$$\mu_X(x) = \max_{\sigma_1 \in R^+} min\left[e^{-\frac{|x-c_1|^2}{\sigma_1^2}}, e^{-\frac{|\sigma_1-c_2|^2}{\sigma_2^2}}\right] \tag{17}$$

Plotting $e^{-\frac{|x-c_1|^2}{\sigma_1^2}}$ and $e^{-\frac{|\sigma_1-c_2|^2}{\sigma_2^2}}$ as functions of $\sigma_1$, as shown in Fig. 7, we see that the max in (17) is achieved at the intersection point:

$$\frac{|x-c_1|}{\sigma_1} = \frac{\sigma_1-c_2}{\sigma_2}, \quad \sigma_1 > c_2 \tag{18}$$

which gives

$$\sigma_1 = \frac{c_2 + \sqrt{c_2^2 + 4\sigma_2|x-c_1|}}{2} \tag{19}$$

Substituting (19) into (17) yields

$$\mu_X(x) = e^{-\frac{\left|\sqrt{c_2^2+4\sigma_2|x-c_1|}-c_2\right|^2}{(2\sigma_2)^2}} \tag{20}$$

When $c_2 = 0$, $\mu_X(x)$ of (20) becomes the exponential function $\mu_X(x) = e^{-\frac{|x-c_1|}{\sigma_2}}$. ■

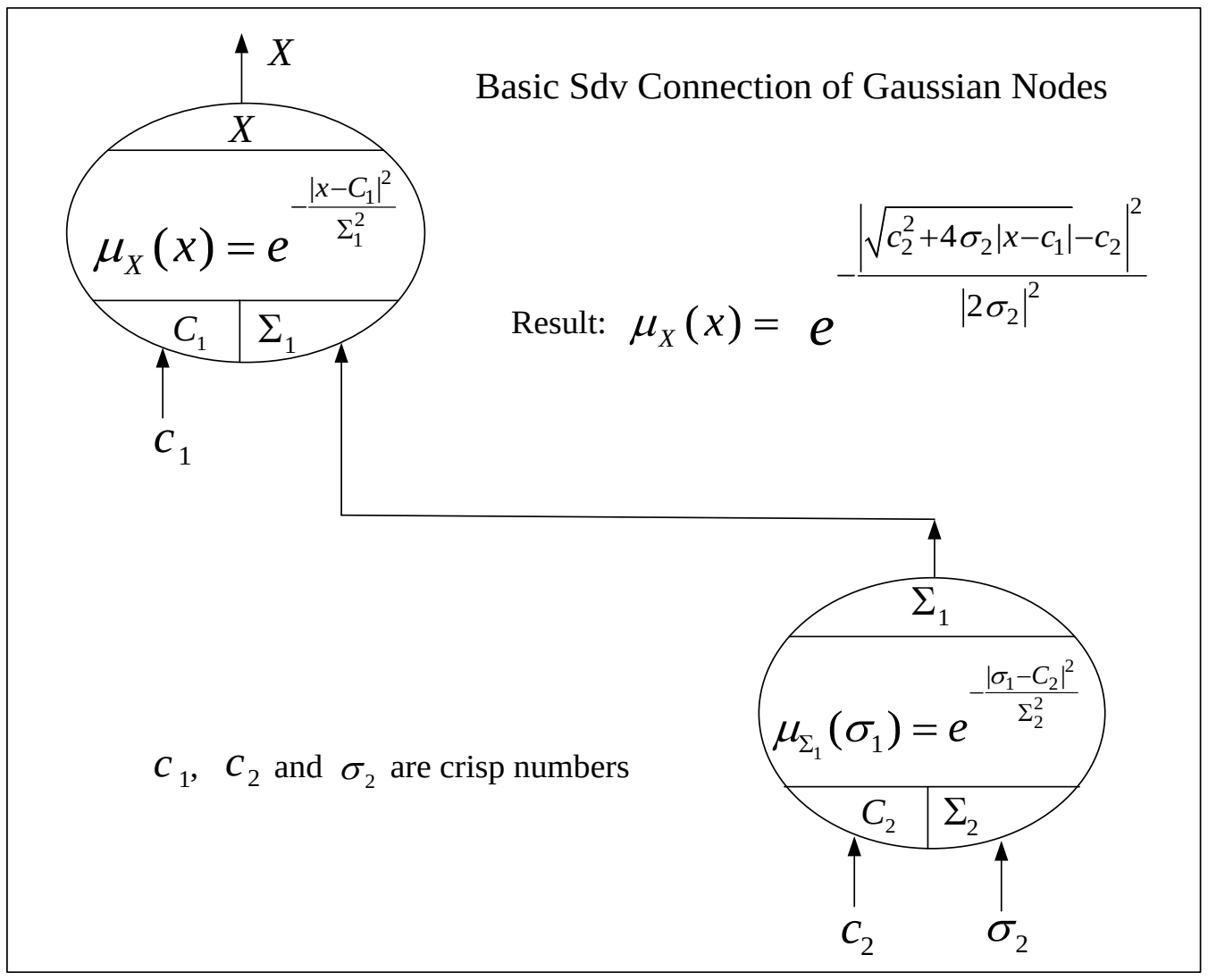

Fig. 6: Basic Sdv Connection of Gaussian nodes (Connection 2).

We see from (20) that its center still equals $c_1$ ($\sqrt{c_2^2+4\sigma_2|x-c_1|} - c_2 = 0$ gives $x = c_1$), which is as expected because the standard deviation connection (Fig. 6) should not alter the central value. Fig. 8 plots the $\mu_X(x)$ of (20) with $c_1 = 4, \sigma_2 = 1$ fixed and $c_2 = 0, 1, 2$, as well as the Gaussian function $e^{-\frac{|x-c_1|^2}{(\sigma_2)^2}}$ for comparison. We see from Fig. 8 that when two Gaussian fuzzy nodes (15) and (16) are connected in-the-sdv as in Fig. 6, the Gaussian function changes to the function (20) which has much heavier tails than the Gaussian function; that is, the Sdv Connection fundamentally changes the fuzzy picture: the membership values of "remote members" in the universe of discourse increase greatly when sdv fuzziness's are connected.

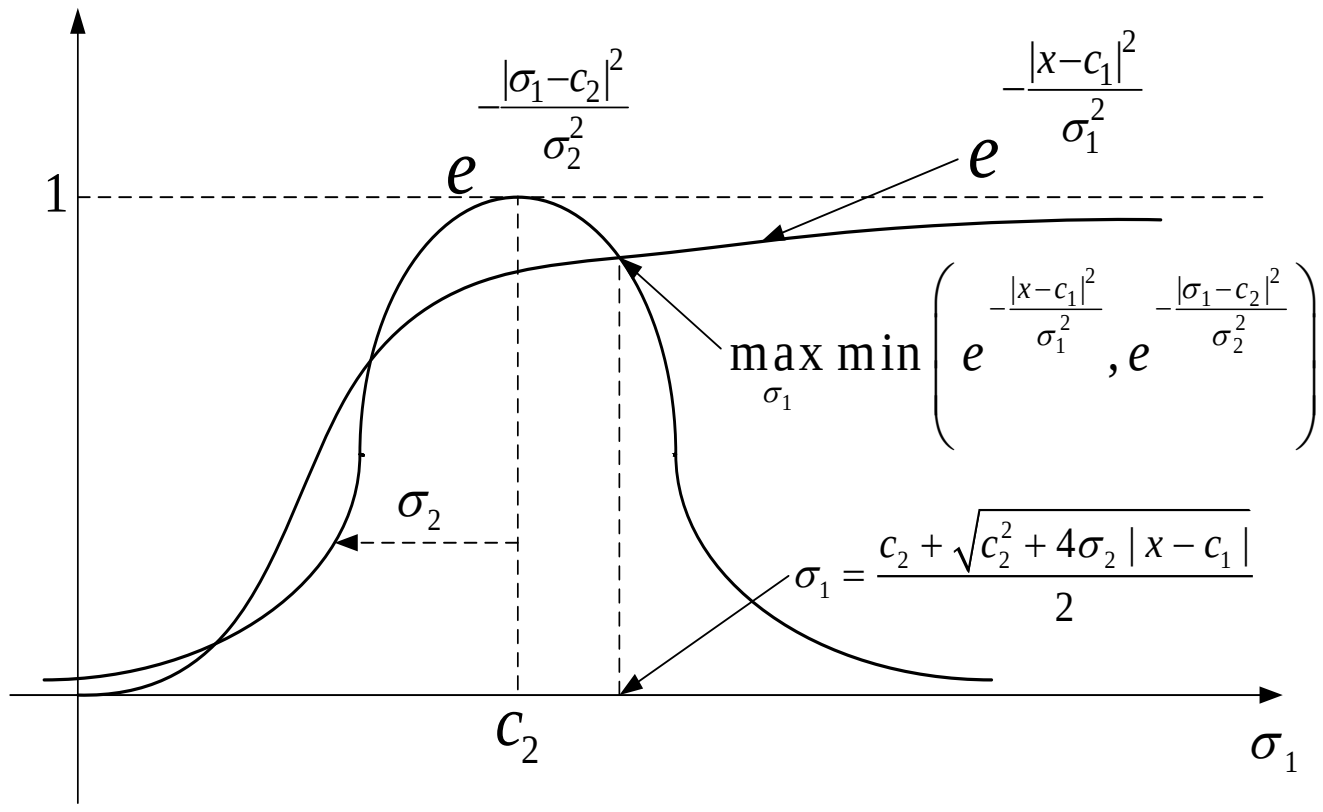

Fig. 7: How the max in (17) is achieved for the Basic Sdv Connection.

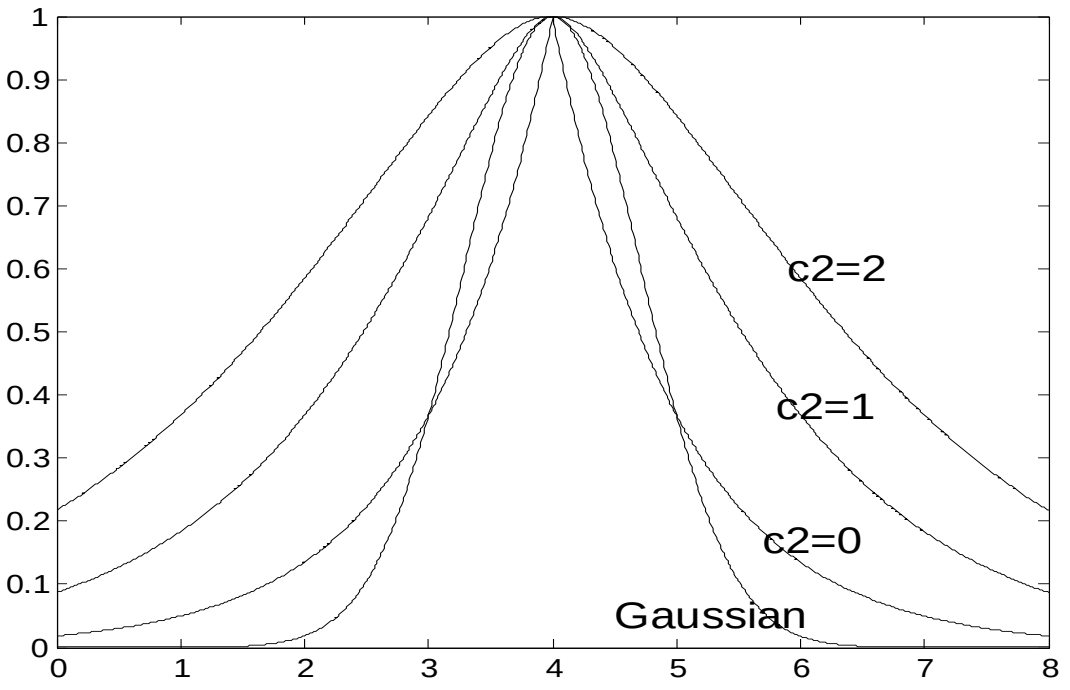

Fig. 8: Plots of $\mu_X(x)$ of (20) (Basic Sdv Connection in Fig. 6) with $c_1 = 4, \sigma_2 = 1$ fixed and $c_2 = 0, 1, 2$; also shown is the Gaussian function $e^{-\frac{|x-c_1|^2}{(\sigma_2)^2}}$ for comparison.

It is clear from Fig. 8 that the fuzzy set $X$ with membership function (20) is "fuzzier" than the Gaussian fuzzy variable, because the $\mu_X(x)$ of (20) is flatter than the Gaussian function in the tail parts and is sharper around the center. Can we define some numbers to characterize the flatness and sharpness of a fuzzy set? In probability theory, the standard deviation and the kurtosis are very useful numbers to measure the uncertainty of a random variable. We now propose similar concepts for a fuzzy set in:

**Definition 3:** Let $X$ be a fuzzy set defined on the universe of discourse $\Omega_X$ with membership function $\mu_X(x)$. Suppose $\mu_X(x)$ is symmetric, normal ( $max_{x\in\Omega_X} \mu_X(x) = 1$ ) and $\arg max_{x\in\Omega_X} \mu_X(x)$ is unique, then the *center*, *standard deviation (sdv)*, *kurtosis* and *sharpness*[14] of $X$, denoted as $c_X$, $\sigma_X$, $k_X$ and $s_X$ respectively, are defined as follows:

$$c_X = \arg \max_{x\in\Omega_X} \mu_X(x) \quad (21)$$

$$\sigma_X = |\arg(\mu_X(x) = e^{-1}) - c_X| \quad (22)$$

$$k_X = \frac{|\arg(\mu_X(x) = e^{-4}) - c_X|}{\sigma_X} \quad (23)$$

$$s_X = \frac{\sigma_X}{|\arg(\mu_X(x) = e^{-1/4}) - c_X|} \quad (24)$$

The center $c_X$ is the point at which the membership function $\mu_X(x)$ achieves its maximum value 1. The meaning of the standard deviation $\sigma_X$ is the distance to the center at which the membership value $\mu_X(x)$ reduces to $e^{-1} \approx 0.36$, or 36% of its maximum value. With this definition, the sdv of the Gaussian membership function $\mu_X(x) = e^{-\frac{|x-c|^2}{\sigma^2}}$ equals $\sigma$, agreeing with our conventional usage of the term. The meaning of the kurtosis $k_x$ is the ratio of the distance to the center at which the membership value $\mu_X(x)$ reduces to $e^{-4} \approx 0.018$ to the distance to the center at which the membership value $\mu_X(x)$ reduces to $e^{-1} \approx 0.36$, which measures how fast the $\mu_X(x)$ declines in the tail part of the membership function. With this definition, the $k_x$ of the Gaussian membership function $\mu_X(x) = e^{-\frac{|x-c|^2}{\sigma^2}}$ equals 2 – a convenient number as a reference point to compare different membership functions. Large $k_x$ implies slow decline of $\mu_X(x)$ in the tail part. Similarly, the sharpness $s_x$ is the ratio of the distance to the center at which the membership value $\mu_X(x)$ reduces to $e^{-1} \approx 0.36$ to the distance to the center at which the membership value $\mu_X(x)$ reduces to $e^{-1/4} \approx 0.77$, which measures how fast the $\mu_X(x)$ declines in the central part of the membership function (i.e., measuring the sharpness of the membership function around the center). With this definition, the $s_x$ of the Gaussian membership function $\mu_X(x) = e^{-\frac{|x-c|^2}{\sigma^2}}$ equals 2. Large $s_x$ implies fast decline of $\mu_X(x)$ around the center (i.e., $\mu_X(x)$ is sharp). Table 1 gives the $c_X$, $\sigma_X$, $k_X$ and $s_X$ of Gaussian, triangular, exponential (Basic Sdv Connection with $c_2 = 0$), Basic Sdv Connection and stretched exponential (Connection 5) membership functions, which are obtained from the definitions (21)-(24) in a straightforward manner.

We now analyze the Basic Sdv Connection result (20) in more details. We see from Table 1 that if $c_2 \gg \sigma_2$ which means the node $\Sigma_1$ is very confident about its center opinion $c_2$, then the kurtosis $k_X = 2 + \frac{2\sigma_2}{\sigma_2+c_2} \approx 2$ and the sharpness $s_X = 2 + \frac{2\sigma_2}{\sigma_2+c_2} \approx 2$, both of which approach the corresponding values of the Gaussian membership function, meaning that in this case the Basic Sdv Connection approaches the Gaussian function. In the other extreme $c_2 \ll \sigma_2$ which means the node $\Sigma_1$ is very uncertain about its center opinion $c_2$, then the kurtosis $k_X = 2 + \frac{2\sigma_2}{\sigma_2+c_2} \approx 4$ and the sharpness $s_X = 2 + \frac{2\sigma_2}{\sigma_2+c_2} \approx 4$, both of which approach the corresponding values of the exponential membership function. These results show that the final membership function $\mu_X(x)=e^{-\frac{\left\lvert\sqrt{c_2^2+4\sigma_2\lvert x-c_1\rvert}-c_2\right\rvert^2}{(2\sigma_2)^2}}$ of the Basic Sdv Connection may be viewed as a function that moves smoothly between the Gaussian function $e^{-\frac{\lvert x-c_1\rvert^2}{\sigma_2^2}}$ and the exponential function $e^{-\frac{\lvert x-c_1\rvert}{\sigma_2}}$.

[14] Although comparing measures were studied in the possibility theory framework [19], [37], specific and easy-to-compute quantities such as the kurtosis and sharpness defined here were not proposed in the literature.

**Table 1: The $c_X$, $\sigma_X$, $k_X$ and $s_X$ of the Gaussian, triangular, exponential, basic sdv connection, and chain-in-sdv with zero centers connection (stretched exponential).**

| | Membership function | Center $c_X$ | Sdv $\sigma_X$ | Kurtosis $k_X$ | Sharpness $s_X$ |
|---|---|---|---|---|---|
| Gaussian | $e^{-\frac{\lvert x-c\rvert^2}{\sigma^2}}$ | $c$ | $\sigma$ | 2 | 2 |
| Triangular | $\begin{cases} 1-\frac{\lvert x-c\rvert}{b}, & x\in[c-b,c+b \\ 0 & \text{otherwise} \end{cases}$ | $c$ | $(1-e^{-1})b$ $\approx 0.63\ b$ | $\frac{1-e^{-4}}{1-e^{-1}}$ $\approx 1.55$ | $\frac{1-e^{-1}}{1-e^{-1/4}}$ $\approx 2.85$ |
| Exponential | $e^{-\frac{\lvert x-c\rvert}{\sigma}}$ | $c$ | $\sigma$ | 4 | 4 |
| Basic Sdv Connection (Fig. 6) | $e^{-\frac{\left\lvert\sqrt{c_2^2+4\sigma_2\lvert x-c_1\rvert}-c_2\right\rvert^2}{(2\sigma_2)^2}}$ | $c_1$ | $\sigma_2+c_2$ | $2+\frac{2\sigma_2}{\sigma_2+c_2}$ | $2+\frac{2\sigma_2}{\sigma_2+2c_2}$ |
| Chain-in-Sdv with Zero Centers Connection (Fig.11) | $e^{-\left\lvert\frac{x-c_1}{\sigma_n}\right\rvert^{\frac{2}{n}}}$ | $c_1$ | $\sigma_n$ | $2^n$ | $2^n$ |

**Connection 3 (Basic Center-Sdv Connection):** Here, we consider the *basic center-sdv connection of Gaussian nodes* shown in Fig. 9, and determine the unconditional membership function $\mu_X(x)$ of fuzzy set $X$. Applying the Basic Center Connection of Fig. 4 to the two nodes with fuzzy sets $X$ and $C_1$ in Fig. 9, we obtain from (14) that

$$\mu_{X|\Sigma_1}(x|\Sigma_1)=e^{-\frac{\lvert x-c_2\rvert^2}{(\Sigma_1+\sigma_2)^2}} \tag{25}$$

Using $\mu_{X|\Sigma_1}(x|\Sigma_1)$ of (25) and

$$\mu_{\Sigma_1}(\sigma_1)=e^{-\frac{\lvert\sigma_1-c_3\rvert^2}{\sigma_3^2}} \tag{26}$$

as the $\mu_{X|V}(x|v)$ and $\mu_V(v)$ in (7), respectively, we have

$$\mu_X(x)=\max_{\sigma_1\in R^+} min\left[e^{-\frac{\lvert x-c_2\rvert^2}{(\sigma_1+\sigma_2)^2}}, e^{-\frac{\lvert\sigma_1-c_3\rvert^2}{\sigma_3^2}}\right] \tag{27}$$

Using the same idea as in Connection 2 (i.e., plotting $e^{-\frac{\lvert x-c_2\rvert^2}{(\sigma_1+\sigma_2)^2}}$ and $e^{-\frac{\lvert\sigma_1-c_3\rvert^2}{\sigma_3^2}}$ as functions of $\sigma_1$), we obtain that the max in (27) is achieved at the intersection point:

$$\frac{\lvert x-c_2\rvert}{\sigma_1+\sigma_2}=\frac{\sigma_1-c_3}{\sigma_3},\qquad \sigma_1>c_3 \tag{28}$$

which gives

$$\sigma_1=\frac{c_3-\sigma_2+\sqrt{(c_3+\sigma_2)^2+4\sigma_3\lvert x-c_2\rvert}}{2} \tag{29}$$

Substituting (29) into (27) yields

$$\mu_X(x)=e^{-\frac{\left\lvert\sqrt{(c_3+\sigma_2)^2+4\sigma_3\lvert x-c_2\rvert}-c_3-\sigma_2\right\rvert^2}{(2\sigma_3)^2}} \tag{30}$$

■

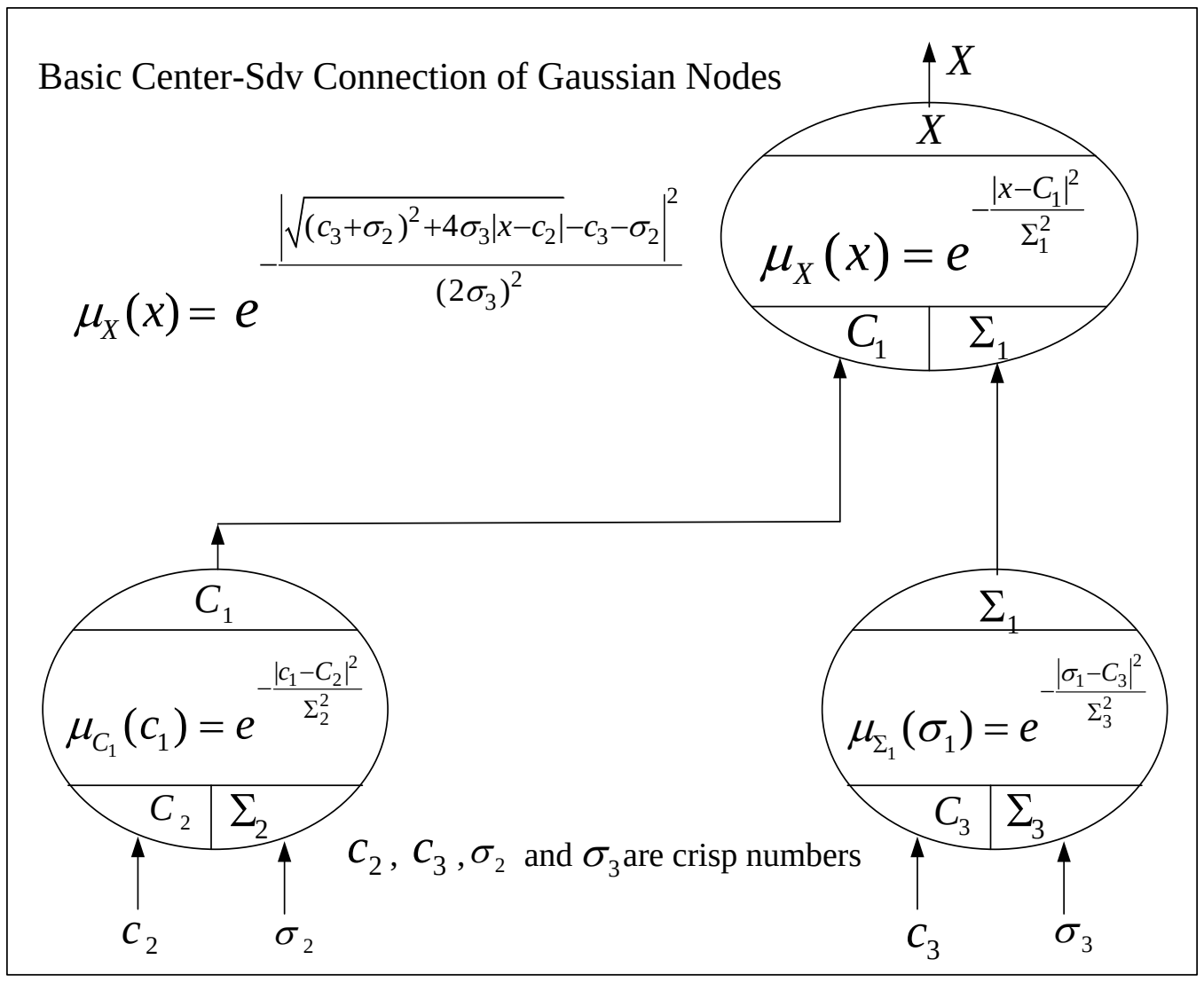

Fig. 9: Basic Center-Sdv Connection of Gaussian nodes (Connection 3).

We see from (30) that (see also Fig. 9): i) the center of the final $\mu_X(x)$ still equals the center $c_2$ of the center-providing node $C_1$ ( $\sqrt{(c_3+\sigma_2)^2+4\sigma_3\lvert x-c_2\rvert}-c_3-\sigma_2=0$ gives $x=c_2$), which is expected because the sdv-providing node $\Sigma_1$ should not modify the central value; and ii) the sdv-input $\sigma_2$ to the center-providing node $C_1$ plays the same role as the center-input $c_3$ to the sdv-providing node $\Sigma_1$. Comparing (30) with the Basic Sdv Connection result (20) (comparing also Figs. 6 and 9), we see that adding a center-input node $C_1$ does not change the functional form of the final membership function, i.e., (30) and (20) are in the same functional form. This shows that, generally speaking, the center connections preserve the functional form of the membership functions, whereas the sdv connections fundamentally change the shape of the membership functions.

**Connection 4 (Chain-in-Center Connection):** Here, we consider the *chain-in-center connection of Gaussian nodes* in Fig. 10, and determine the unconditional membership function $\mu_X(x)$ of fuzzy set $X$. Viewing the bottom two nodes in Fig. 10 as the two nodes in the Basic Center Connection of Fig. 4, then the result (14) shows that this two-node center connection is equivalent to a single Gaussian node with its sdv equal to the summation of the sdv's of the two nodes. Applying this procedure repeatedly bottom-up for the nodes in Fig. 10, we get

$$\mu_X(x) = e^{-\frac{|x-c_n|^2}{(\sigma_1+\sigma_2+\cdots+\sigma_n)^2}} \quad (31)$$

■

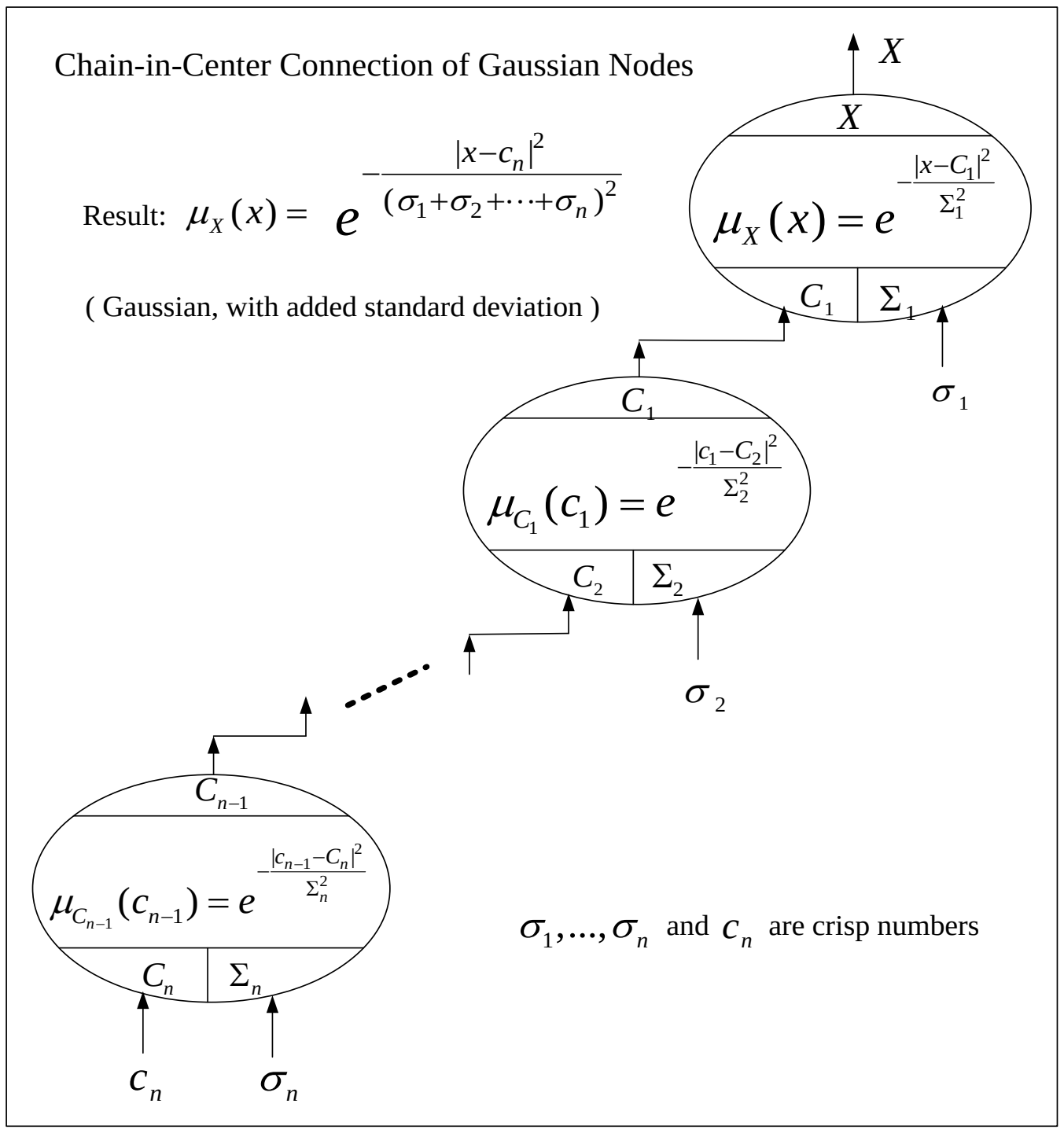

Fig. 10: Chain-in-Center Connection of Gaussian nodes (Connection 4).

The Chain-in-Center Connection in Fig. 10 provides a natural framework to model message transfer among people. Specifically, the original message in Fig. 10 is the non-fuzzy $c_n$, and when $c_n$ is transferred to the next person, it becomes the Gaussian fuzzy set $C_{n-1}$ with center $c_n$ and standard deviation $\sigma_n$. The message transfer continues as in Fig. 10 until the $n$'th person where the message becomes the fuzzy set $X$ with membership function (31). We see from (31) that after being transferred across $n$ people, the original message $c_n$ still occupies the center of the final fuzzy set $X$, but the uncertainty about the message (measured as the standard deviation of the fuzzy set) increases greatly to $\sigma_1+\sigma_2+\cdots+\sigma_n$ : the summation of the standard deviations of the $n$ people who transferred the message. This mathematical result is intuitively appealing.

**Connection 5 (Chain-in-Sdv with Zero Centers Connection):** Here, we consider the *chain-in-sdv with zero centers connection of Gaussian nodes* in Fig. 11 (the centers $c_2, c_3, \ldots, c_n$ are equal to zero), and determine the unconditional membership function $\mu_X(x)$ of fuzzy set $X$. Applying the Basic Sdv Connection formula (20) to the last two nodes in the chain of Fig. 11, we get (notice that $c_{n-1} = c_n = 0$)

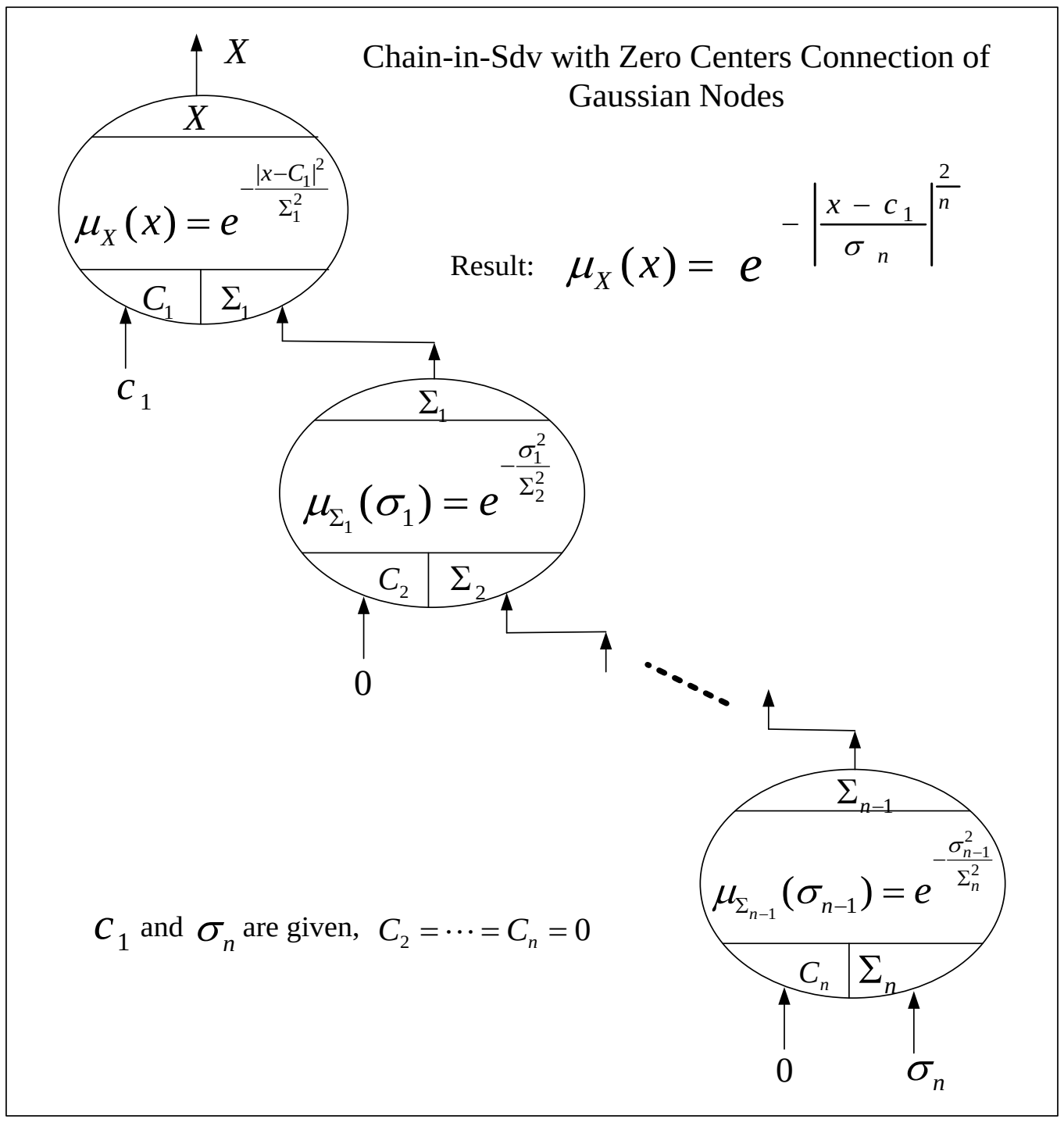

Fig. 11: Chain-in-Sdv with Zero Centers Connection of Gaussian nodes (Connection 5).

$$\mu_{\Sigma_{n-2}}(\sigma_{n-2}) = e^{-\frac{\left|\sqrt{c_n^2+4\sigma_n|\sigma_{n-2}-c_{n-1}|}-c_n\right|^2}{(2\sigma_n)^2}} = e^{-\frac{\sigma_{n-2}}{\sigma_n}} \quad (32)$$

Moving one node up along the chain of Fig. 11 and using the Compositional Rule of Inference (7), we have

$$\mu_{\Sigma_{n-3}}(\sigma_{n-3}) = \max_{\sigma_{n-2}\in R^+} min\left[e^{-\frac{|\sigma_{n-3}-c_{n-2}|^2}{\sigma_{n-2}^2}}, e^{-\frac{\sigma_{n-2}}{\sigma_n}}\right] \quad (33)$$

The max in (33) is achieved at the intersection $\frac{\sigma_{n-3}^2}{\sigma_{n-2}^2} = \frac{\sigma_{n-2}}{\sigma_n}$ (notice that $c_{n-2} = 0$) which gives $\sigma_{n-2} = (\sigma_n\sigma_{n-3}^2)^{1/3}$, and thus we have

$$\mu_{\Sigma_{n-3}}(\sigma_{n-3}) = e^{-\left(\frac{\sigma_{n-3}}{\sigma_n}\right)^{\frac{2}{3}}} \quad (34)$$

Continuing this process up along the chain of Fig. 11, we have $\mu_{\Sigma_{n-4}}(\sigma_{n-4}) = e^{-\left(\frac{\sigma_{n-4}}{\sigma_n}\right)^{\frac{2}{4}}}$, $\mu_{\Sigma_{n-5}}(\sigma_{n-5}) = e^{-\left(\frac{\sigma_{n-5}}{\sigma_n}\right)^{\frac{2}{5}}}$, …, and $\mu_{\Sigma_1}(\sigma_1) = e^{-\left(\frac{\sigma_1}{\sigma_n}\right)^{\frac{2}{n-1}}}$, and the final $\mu_X(x)$ is obtained as

$$\mu_X(x) = \max_{\sigma_1\in R^+} min\left[e^{-\frac{|x-c_1|^2}{\sigma_1^2}}, e^{-\left(\frac{\sigma_1}{\sigma_n}\right)^{\frac{2}{n-1}}}\right] = e^{-\left|\frac{x-c_1}{\sigma_n}\right|^{\frac{2}{n}}} \quad (35)$$

■

The $\mu_X(x)$ of (35) is a stretched exponential function [40] which has much heavier tails and sharper center than the Gaussian and exponential functions for long chains (large $n$). Notice from Table 1 that the standard deviation of the stretched exponential function (35) $\sigma_X = \sigma_n$ which does not change with $n$, but its kurtosis $k_X = 2^n$ and sharpness $s_X = 2^n$ increase exponentially with $n$, revealing the strong heavy tails and very sharp centers of the stretched exponential functions for large $n$. Sharp center means that the center opinion is not robust because the membership value reduces sharply when the opinion moves slightly away from the center. Heavy tail implies that the opinions far away from the center opinion may not be ignored. Putting together (sharp center and heavy tails), the effect of the Chain-in-Sdv Connection is the sharp increase of the overall uncertainty – different opinions look more and more similar to each other. Determining the membership function of the Chain-in-Sdv Connection when the $c_2, c_3, \dots, c_n$ are nonzero is an open problem (exact result requires solving an $n$'th-order equation whose general solution is not available).

To summarize the basic connections of Gaussian nodes (Connections 1 to 5), we have:

i) The Center Connection preserves the Gaussian form for the membership functions and causes the uncertainty (measured by the sdv) to increase in a linear fashion as more uncertain opinions are connected through the center inputs of the Gaussian nodes; and,

ii) The Sdv Connection changes the functional form of the membership functions from Gaussian to stretched exponential type of functions and causes the uncertainty (measured by the kurtosis and the sharpness) to increase exponentially as more uncertain opinions about the uncertainties are connected through the sdv inputs of the Gaussian nodes.

The following connection gives the solution to the problem described in the Introduction.

**Connection 6 (Person A's Visit to Beijing):** Here, we consider the fuzzy network in Fig. 2, and determine the final unconditional membership function of Person A's fuzzy set $X$: "interesting places in Beijing". Applying the Basic Center-Sdv Connection (Fig. 9) formula (30) to the Gaussian nodes Person B, BYP and Person B' Wife in Fig. 2, we get

$$\mu_{c_1}(c_1) = e^{-\frac{\left|\sqrt{(c_4+\sigma_3)^2+4\sigma_4|c_1-c_3|}-c_4-\sigma_3\right|^2}{(2\sigma_4)^2}} \quad (36)$$

where, as Fig. 2 shows, $c_3 =$ the location of Lama Temple = point 2 in Fig. 1, $\sigma_3 = 0.5$, $c_4 = 0$ and $\sigma_4 = 1$. The final unconditional membership function of Person A's fuzzy set $X$ is

$$\mu_X(x) = \max_{c_1\in R^2} min\left[e^{-\frac{|x-c_1|^2}{\sigma_1^2}}, e^{-\frac{\left|\sqrt{(c_4+\sigma_3)^2+4\sigma_4|c_1-c_3|}-c_4-\sigma_3\right|^2}{(2\sigma_4)^2}}\right] \quad (37)$$

Since $e^{-\frac{|x-c_1|^2}{\sigma_1^2}}$ and $e^{-\frac{\left|\sqrt{(c_4+\sigma_3)^2+4\sigma_4|c_1-c_3|}-c_4-\sigma_3\right|^2}{(2\sigma_4)^2}}$ are symmetric functions of $c_1 \in R^2$ in the three-dimensional space around centers at $x$ and $c_3$, respectively, the max in (37) must be achieved at a point on the line connecting the two centers $x$ and $c_3$. With $c_1 \in R^2$ being a point on the line connecting $x$ and $c_3$, let $y = |x - c_1|$, then $|c_1 - c_3|$ in (37) equals $|x - c_3| - y$, we therefore obtain that the max in (37) is achieved at

$$\frac{y}{\sigma_1} = \frac{\sqrt{(c_4+\sigma_3)^2 + 4\sigma_4(|x-c_3| - y)} - c_4 - \sigma_3}{2\sigma_4} \quad (38)$$

whose solution is

$$y = \frac{\sigma_1\left(\sqrt{(c_4+\sigma_3+\sigma_1)^2 + 4\sigma_4|x-c_3|} - (c_4+\sigma_3+\sigma_1)\right)}{2\sigma_4} \quad (39)$$

Substituting (39) into (37) ($|x - c_1| = y$) yields

$$\mu_X(x) = e^{-\frac{\left|\sqrt{(c_4+\sigma_3+\sigma_1)^2+4\sigma_4|x-c_3|}-(c_4+\sigma_3+\sigma_1)\right|^2}{(2\sigma_4)^2}} \quad (40)$$

whose center $c_X = c_3 =$ the location of Lama Temple = point 2 in Fig. 1, standard deviation $\sigma_X = \sigma_4 + c_4 + \sigma_3 + \sigma_1 = 1 + 0 + 0.5 + 1 = 2$, kurtosis $k_X = 2 + \frac{2\sigma_4}{\sigma_4+c_4+\sigma_3+\sigma_1} = 2 + \frac{2}{2.5} = 2.8$, and sharpness $S_X = 2 + \frac{2\sigma_4}{\sigma_4+2(c_4+\sigma_3+\sigma_1)} = 2 + \frac{2}{4} = 2.5$. Fig. 12 plots this $\mu_X(x)$ (also shown in Fig. 12 is the Gaussian function with Person A's own standard deviation $\sigma_1 = 1$ for comparison). ■

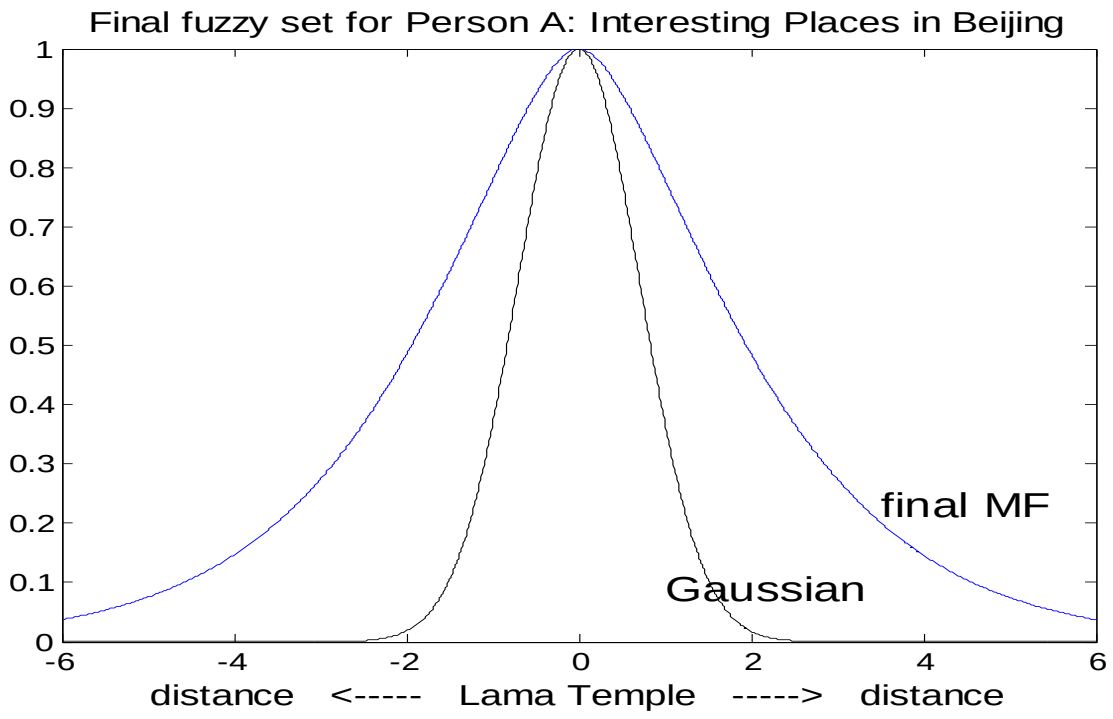

Fig. 12: The final unconditional membership function (40) of the fuzzy set $X$ (interesting places in Beijing) in Fig. 2 (upper curve; the lower curve is the Gaussian function for comparison).

From the final membership function (40) (Fig. 12) we see that after collecting the fuzzy information from Person B, Person B's wife and BYP through the fuzzy network Fig. 2, Person A's fuzzy set $X$ = "interesting places in Beijing" is centered at BYP's suggestion $c_3$ = Lama Temple = point 2 in Fig. 1, but the uncertainty (standard deviation) increases due to the fuzzy nature of the information. Specifically, the standard deviation increases from Person A's own $\sigma_1 = 1$ to the final $\sigma_X = 2.5$. Furthermore, the tail of the membership function becomes fatter with the kurtosis increasing from 2 for Person A's own Gaussian membership function to 2.8 for the final fuzzy set. In conclusion, the most interesting place for Person A to visit is still BYP's suggestion Lama Temple – this is reasonable because Lama Temple is the only concrete place suggested by these fuzzy people, but the uncertainty about this suggestion is high because the uncertainties of all the people involved must be included in the final membership function.

## IV. Dynamic Connections with Constant Confidence

The basic connections studied in the last section are static and the opinions do not change with time. In real life, opinions are changing with time through feedback loops when people communicate with each other. In this section we study a number of typical examples of dynamic connections where feedback loops exist and the opinions are evolving continuously. In particular, we will analyze the basic self-feedback (Connection 7), one feedback loop plus one static node (Connection 8), and two coupling feedback loops (Connection 9).

**Connection 7 (Self-Feedback Connection):** Here, we consider the dynamic fuzzy opinion network in Fig. 13 where the initial $X(0)$ is a fuzzy singleton at $x(0)$ and the $\sigma$ is a constant, and determine the unconditional membership function $\mu_{X(t)}(x(t))$ of fuzzy set $X(t)$. We see from Fig. 13 that

$$\mu_{X(t)|X(t-1)}(x(t)|X(t-1)) = e^{-\frac{|x(t)-X(t-1)|^2}{\sigma^2}} \tag{41}$$

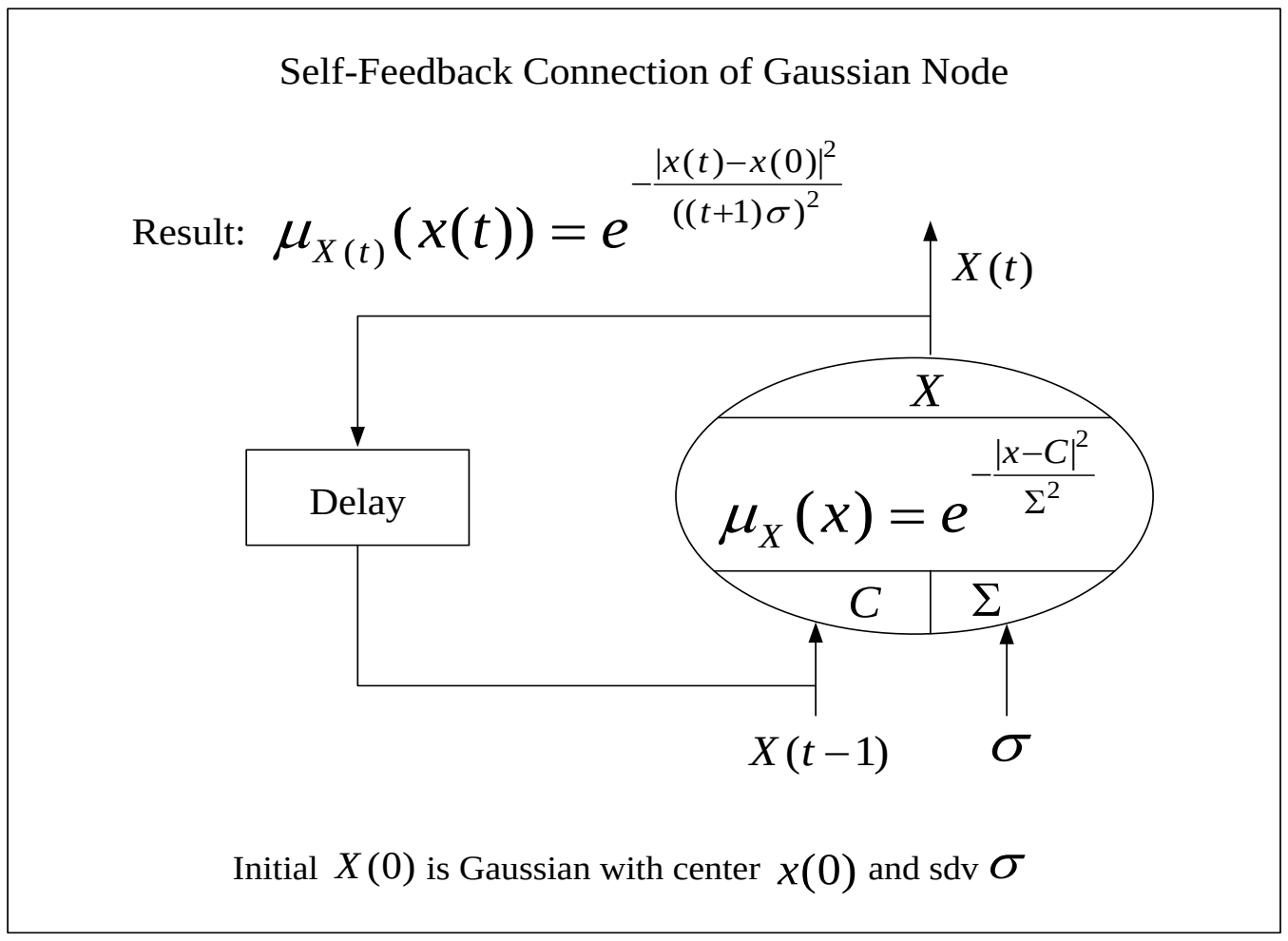

Fig. 13: Self-Feedback Connection of Gaussian node (Connection 7).

Using the compositional rule of inference (8), we get

$$\mu_{X(t)|X(t-2)}(x(t)|X(t-2)) = \max_{x(t-1)\in R^n} min\left[e^{-\frac{|x(t)-x(t-1)|^2}{\sigma^2}}, e^{-\frac{|x(t-1)-X(t-2)|^2}{\sigma^2}}\right] \tag{42}$$

The max in (42) is achieved when $\frac{x(t)-x(t-1)}{\sigma} = \frac{x(t-1)-X(t-2)}{\sigma}$ which gives

$$\mu_{X(t)|X(t-2)}(x(t)|X(t-2)) = e^{-\frac{|x(t)-X(t-2)|^2}{(2\sigma)^2}} \tag{43}$$

Continuing this process we obtain

$$\mu_{X(t)}(x(t)) = e^{-\frac{|x(t)-x(0)|^2}{((t+1)\sigma)^2}} \tag{44}$$

■

We see from (44) that the standard deviation of $X(t)$ equals $(t+1)\sigma$ which goes to infinity as $t$ increases. This means that without other people providing information, self-feedback will make the fuzzy information fuzzier and fuzzier. Specifically, the initial condition $x(0)$ may be viewed as a person's original opinion about something and $\sigma$ represents his/her uncertainty about the judgment. If the person thinks about the problem again and again just by himself /herself, as modeled by the feedback loop in Fig. 13, then one more uncertainty $\sigma$ will be added to the fuzzy picture whenever the person thinks about the problem one more time, and at the end the uncertainty $(t+1)\sigma$ goes to infinity[15], meaning that the person is totally lost and every opinion looks the same for him/her (the membership value of every $x(t) \in R^n$ equals $\lim_{t\to\infty} e^{-\frac{|x(t)-x(0)|^2}{((t+1)\sigma)^2}} = 1$). This shows the importance of communicating with other people, as we will demonstrate in the next two connections.

**Connection 8 (Compromising Husband with Persistent Wife (or vice versa)):** Here, we consider the dynamic fuzzy opinion network in Fig. 14 where the Husband node dynamically updates his output by taking the weighted average of his past output and his wife's output as the new center input, while the Wife node is persistent and does not change over time. Let the initial $X(0)$ (Husband's original opinion) be a Gaussian fuzzy set with center $x(0)$ and standard deviation $\sigma_1$ (notice from Fig. 14 that the Wife's opinion is always the fuzzy set $X_2$ with membership function $\mu_{X_2}(x_2) = e^{-\frac{|x_2-c_2|^2}{\sigma_2^2}}$, where $c_2$ and $\sigma_2$ are constants) and the weights $w_1, w_2 > 0$ with $w_1 + w_2 = 1$. Our task is to determine the unconditional membership function $\mu_{X(t)}(x(t))$ of the Husband's fuzzy set $X(t)$.

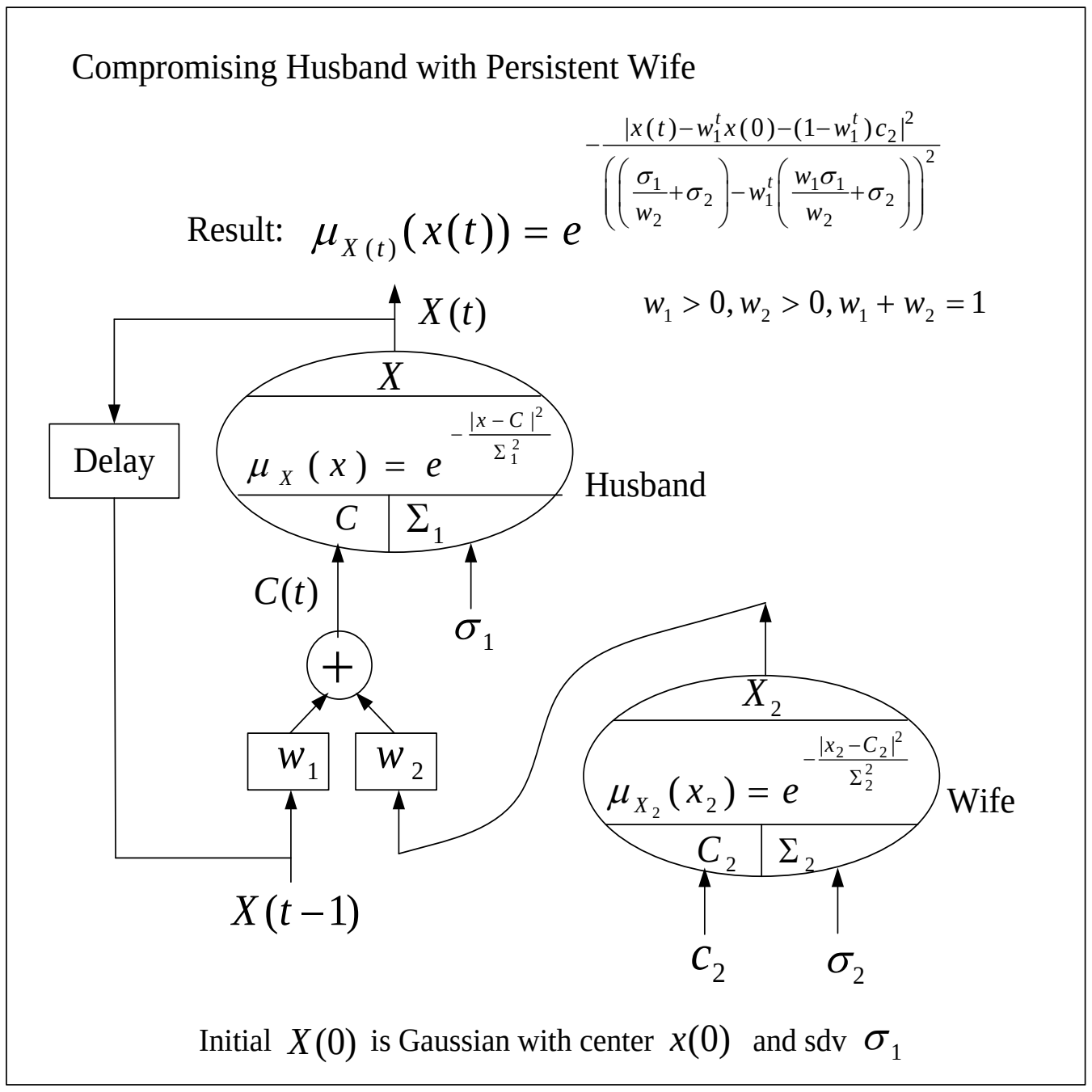

Fig. 14: Compromising Husband with Persistent Wife (Connection 8).

---

[15] An anonymous referee pointed out that self deliberation may lead to more and more uncertainty in some situations, but in other situations one could also imagine it being the other way around – after serious contemplation I become more certain about my own opinion. We agree, as physiologist William Carpenter [13] argued that merely thinking about a given behavior is sufficient to create the tendency to engage in that behavior (we call it the "action-strengthen-attitude principle"). However, the psychological research of the "illusion of control" [41] makes it clear that one's subjective experience of volition is a poor and inaccurate guide to its true causal status [8]. Overall, this problem is too hard – we, as humans ourselves, may never fully understand the ways we are thinking, and our current stage of understanding social systems is quite similar to "the blind monks touching the elephant" [69].

With this said, however, we may account both issues (action-strengthen-attitude and illusion-of-control) within our FON framework by introducing the concept of perceived uncertainty versus real uncertainty. Specifically, *perceived uncertainty* is defined as the standard deviation input to the Gaussian node, representing the uncertainty the person feels about, while *real uncertainty* is defined as the standard deviation of the output fuzzy set of the Gaussian node, characterizing the actual uncertainty of the opinion. For the self-feedback connection of Fig. 13, for example, we may choose the standard deviation input (the perceived uncertainty) to be the declining function $\frac{\sigma}{t+1}$ to account for the action-strengthen-attitude principle, and it is easy to show (following (41)-(44) with the $\sigma$ in (41) replaced by $\frac{\sigma}{t}$) that the standard deviation of the output fuzzy set (the real uncertainty) $\mathrm{Sdv}(X(t)) = \sum_{k=0}^{t}\left(\frac{\sigma}{k+1}\right)$ which is a strictly increasing function of $t$, indicating the illusion-of-control reality. This low perceived uncertainty but high real uncertainty phenomenon is very common in real life; for example, considering the ordinary Germans under the Nazi control: Most of them generally felt happy for their strong country (their perceived uncertainties were low) but actually they were in great danger (the real uncertainty was very high: The German Red Cross in 2005 put the total combined German military and civilian war dead at 7,375,800 – Wikipedia: German casualties in World War II). And the same should be true for the members of the extremist groups.

We see from Fig. 14 that

$$\mu_{X(t)|C(t)}(x(t)|C(t)) = e^{-\frac{|x(t)-C(t)|^2}{\sigma_1^2}} \quad (45)$$

$$C(t) = w_1 X(t-1) + w_2 X_2 \quad (46)$$

Since $C(t)$ is a weighted average of two fuzzy sets $X(t-1)$ and $X_2$, we need the following lemma to proceed.

**Lemma 1:** Let $X_i$ (*i=1,2, ..., n*) be fuzzy sets with Gaussian membership functions $\mu_{X_i}(x_i) = e^{-\frac{|x_i-c_i|^2}{\sigma_i^2}}$ and $w_i \geq 0$ be constant weights with $\sum_{i=1}^n w_i = 1$. Then,

$$Y_n = w_1 X_1 + w_2 X_2 + \cdots + w_n X_n \quad (47)$$

is a fuzzy set with membership function

$$\mu_{Y_n}(y_n) = e^{-\frac{|y_n - \sum_{i=1}^n w_i c_i|^2}{(\sum_{i=1}^n w_i \sigma_i)^2}} \quad (48)$$

■

Proof of this lemma is given in the Appendix.

Lemma 1 shows that the weighted average of *n* Gaussian fuzzy sets is still a Gaussian fuzzy set with center equal to the weighted average of the *n* centers and standard deviation equal to the weighted average of the *n* standard deviations. Let $\text{Center}(X)$ and $\text{Sdv}(X)$ denote the center and standard deviation of Gaussian fuzzy variable $X$, respectively, and applying Lemma 1 to (46) yields

$$\mu_{C(t)}(c(t)) = e^{-\frac{|c(t)-w_1\text{Center}(X(t-1))-w_2\text{Center}(X_2)|^2}{(w_1\text{Sdv}(X(t-1))+w_2\text{Sdv}(X_2))^2}} \quad (49)$$

Using the Compositional Rule of Inference (7) for (45) and (49), we get

$$\mu_{X(t)}(x(t)) = \max_{c(t)\in R^n} min\left[e^{-\frac{|x(t)-c(t)|^2}{\sigma_1^2}}, e^{-\frac{|c(t)-w_1\text{Center}(X(t-1))-w_2c_2|^2}{(w_1\text{Sdv}(X(t-1))+w_2\sigma_2)^2}}\right] \quad (50)$$

where the max is achieved when

$$\frac{x(t)-c(t)}{\sigma_1} = \frac{c(t) - w_1\text{Center}(X(t-1)) - w_2c_2}{w_1\text{Sdv}(X(t-1)) + w_2\sigma_2} \quad (51)$$

which gives

$$\mu_{X(t)}(x(t)) = e^{-\frac{|x(t)-w_1\text{Center}(X(t-1))-w_2c_2|^2}{(\sigma_1+w_1\text{Sdv}(X(t-1))+w_2\sigma_2)^2}} \quad (52)$$

Therefore, $\mu_{X(t)}(x(t))$ is Gaussian with center and standard deviation satisfying:

$$\text{Center}(X(t)) = w_1\text{Center}(X(t-1)) + w_2c_2 \quad (53)$$

$$\text{Sdv}(X(t)) = w_1\text{Sdv}(X(t-1)) + \sigma_1 + w_2\sigma_2 \quad (54)$$

Solving the difference equations (53) and (54) with initial condition $\text{Center}(X(0)) = x(0)$ and $\text{Sdv}(X(0)) = \sigma_1$, we obtain

$$\begin{aligned}\text{Center}(X(t)) &= w_1^t x(0) + \left(\frac{1-w_1^t}{1-w_1}\right)w_2c_2 \\ &= w_1^t x(0) + (1-w_1^t)c_2 \end{aligned} \quad (55)$$

$$\begin{aligned}\text{Sdv}(X(t)) &= w_1^t\sigma_1 + \left(\frac{1-w_1^t}{1-w_1}\right)(\sigma_1 + w_2\sigma_2) \\ &= \left(\frac{\sigma_1}{w_2} + \sigma_2\right) - w_1^t\left(\frac{w_1\sigma_1}{w_2} + \sigma_2\right)\end{aligned} \quad (56)$$

■

Since $0 < w_1 < 1$, two important observations emerge from (55) and (56): (i) as the number of feedback iterations $t$ increases, the Husband's opinion is changing from his own $x(0)$ ($\text{Center}(X(0)) = x(0)$) towards the Wife's opinion $c_2$ ($\text{Center}(X_2) = c_2$), and eventually the Husband will lose his own judgment and become totally consistent with his Wife ($\lim_{t\to\infty}\text{Center}(X(t)) = c_2$); and (ii) the Husband's uncertainty $\text{Sdv}(X(t))$ converges to the finite constant $\frac{\sigma_1}{w_2} + \sigma_2$ as $t$ goes to infinity, which is in contrast with the Self-Feedback case in Fig. 13 where the person is totally lost and the uncertainty goes to infinity. This shows the two sides of being "compromising" (in the sense of taking the weighted average feedback as the Husband in Fig. 14): on one side, the compromising person will lose his identity and becomes "brainwashed" by the persistent person; on the other side, his anxiety (measured by the uncertainty $\text{Sdv}(X(t))$) will be bounded and not increase too much as the fuzziness propagates through feedback. Since the independent-vs-security dilemma (people generally feel secure in a community but joining a community usually means losing some independence) is a common and important problem facing individuals as well as societies, our fuzzy opinion network theory provides a rigorous mathematical framework to study this issue in quantitative details.

The Compromising Husband with Persistent Wife case in Fig. 14 is special, because most loving couples are compromising with each other and learn from each other. The following connection shows what happens when the Husband and the Wife are considerate of each other.

**Connection 9 (Compromising Husband and Wife):** Here, we consider the dynamic fuzzy opinion network in Fig. 15 where both the Husband and the Wife are dynamically updating their opinions by taking the weighted averages of their outputs. Let the initial $X_i(0)$ ($i = 1$ for Husband's original opinion and $i = 2$ for Wife's original opinion) be Gaussian with center $x_i(0)$ and sdv $\sigma_i$ and the weights $w_{i1}, w_{i2} \geq 0$ with $w_{i1} + w_{i2} = 1$[16], our task is to determine the unconditional membership functions $\mu_{X_i(t)}(x_i(t))$ of the Husband's ($i = 1$) and the Wife's ($i = 2$) fuzzy sets $X_i(t)$. We see from Fig. 15 that

$$\mu_{X_i(t)|C_i(t)}(x_i(t)|C_i(t)) = e^{-\frac{|x_i(t)-C_i(t)|^2}{\sigma_i^2}} \quad (57)$$

$$C_i(t) = w_{i1}X_1(t-1) + w_{i2}X_2(t-1) \quad (58)$$

$i = 1,2$. Applying Lemma 1 to (58) yields

$$\mu_{C_i(t)}(c_i(t)) = e^{-\frac{|c_i(t)-w_{i1}\text{Center}(X_1(t-1))-w_{i2}\text{Center}(X_2(t-1))|^2}{(w_{i1}\text{Sdv}(X_1(t-1))+w_{i2}\text{Sdv}(X_2(t-1)))^2}} \quad (59)$$

Using the Compositional Rule of Inference (7) for (57) and (59), we get

$$\mu_{X_i(t)}(x_i(t)) = \max_{c_i(t)\in R^n}\min\left[e^{-\frac{|x_i(t)-c_i(t)|^2}{\sigma_i^2}}, e^{-\frac{|c_i(t)-w_{i1}\text{Center}(X_1(t-1))-w_{i2}\text{Center}(X_2(t-1))|^2}{(w_{i1}\text{Sdv}(X_1(t-1))+w_{i2}\text{Sdv}(X_2(t-1)))^2}}\right] \quad (60)$$

where the max is achieved when

[16] Notice that $w_{21} + w_{22} = 1$ excludes the $w_{21} = w_{22} = 0$ case which is Connection 8 in Fig. 14, i.e., Connection 9 cannot be reduced to Connection 8 because the Wife in Connection 9 is compromising either to herself completely ($w_{22} = 1, w_{21} = 0$), to her husband completely ($w_{21} = 1, w_{22} = 0$), or to both partially ($w_{21}, w_{22} > 0$ with $w_{21} + w_{22} = 1$), whereas the Wife in Connection 8 does not compromise to anybody, including herself.

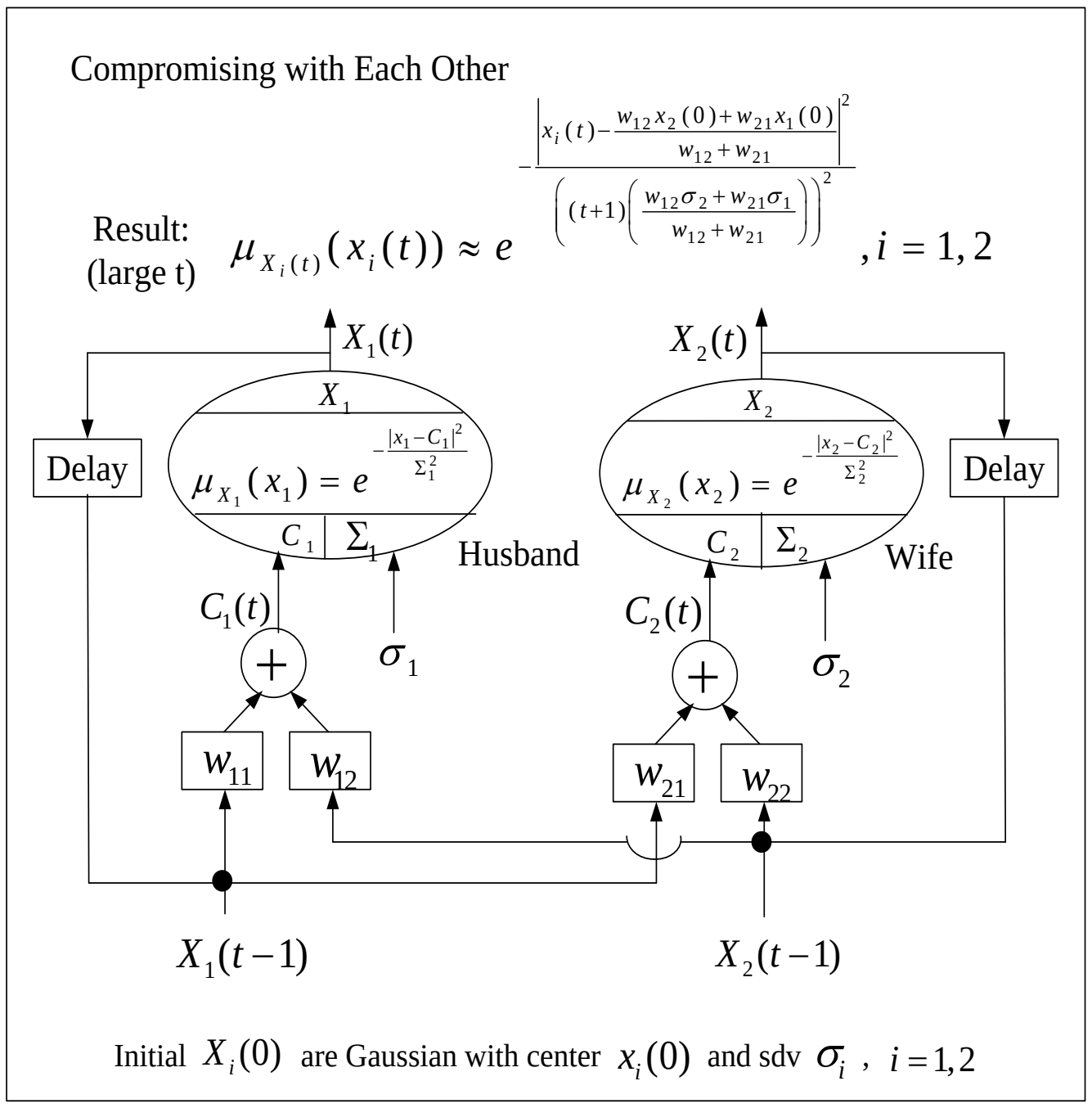

Fig. 15: Compromising with Each Other (Connection 9).

$$\frac{x_i(t)-c_i(t)}{\sigma_i} = \frac{c_i(t) - w_{i1}\mathrm{Center}(X_1(t-1)) - w_{i2}\mathrm{Center}(X_2(t-1))}{w_{i1}\mathrm{Sdv}(X_1(t-1)) + w_{i2}\mathrm{Sdv}(X_2(t-1))} \quad (61)$$

which gives

$$\mu_{X_i(t)}(x_i(t)) = e^{-\frac{|x_i(t)-w_{i1}\mathrm{Center}(X_1(t-1))-w_{i2}\mathrm{Center}(X_2(t-1))|^2}{(\sigma_i+w_{i1}\mathrm{Sdv}(X_1(t-1))+w_{i2}\mathrm{Sdv}(X_2(t-1)))^2}} \quad (62)$$

Letting $X(t) = (X_1(t), X_2(t))^T$ , $\sigma = (\sigma_1, \sigma_2)^T$ and $W = \begin{pmatrix} w_{11} & w_{12} \\ w_{21} & w_{22} \end{pmatrix}$, we have from (62) that

$$\mathrm{Center}(X(t)) = W\ \mathrm{Center}(X(t-1)) \quad (63)$$

$$\mathrm{Sdv}(X(t)) = W\ \mathrm{Sdv}(X(t-1)) + \sigma \quad (64)$$

Solving the vector difference equations (63) and (64) with initial condition $\mathrm{Center}(X(0)) = (x_1(0), x_2(0))^T$ and $\mathrm{Sdv}(X(0)) = (\sigma_1, \sigma_2)^T$, we obtain

$$\mathrm{Center}(X(t)) = W^t \begin{pmatrix} x_1(0) \\ x_2(0) \end{pmatrix} \quad (65)$$

$$\mathrm{Sdv}(X(t)) = \sum_{j=0}^{t} W^j \sigma \quad (66)$$

Noticing that $w_{11} + w_{12} = w_{21} + w_{22} = 1$, the matrix $W$ can be decomposed as

$$W = \begin{pmatrix} 1 & 1 \\ 1 & \frac{1-w_{22}}{w_{11}-1} \end{pmatrix} \begin{pmatrix} 1 & 0 \\ 0 & w_{11}+w_{22}-1 \end{pmatrix} \begin{pmatrix} 1 & 1 \\ 1 & \frac{1-w_{22}}{w_{11}-1} \end{pmatrix}^{-1} \quad (67)$$

where 1 and $w_{11} + w_{22} - 1$ are the eigenvalues of $W$, and $(1\ 1)^{\mathrm{T}}$ and $\left(1\ \frac{1-w_{22}}{w_{11}-1}\right)^{\mathrm{T}}$ are the corresponding eigenvectors. Therefore,

$$\mathrm{Center}(X(t)) = \begin{pmatrix} 1 & 1 \\ 1 & \frac{1-w_{22}}{w_{11}-1} \end{pmatrix} \begin{pmatrix} 1 & 0 \\ 0 & (w_{11}+w_{22}-1)^t \end{pmatrix} \begin{pmatrix} 1 & 1 \\ 1 & \frac{1-w_{22}}{w_{11}-1} \end{pmatrix}^{-1} \begin{pmatrix} x_1(0) \\ x_2(0) \end{pmatrix} \quad (68)$$

$$\mathrm{Sdv}(X(t)) = \begin{pmatrix} 1 & 1 \\ 1 & \frac{1-w_{22}}{w_{11}-1} \end{pmatrix} \begin{pmatrix} t+1 & 0 \\ 0 & \sum_{j=0}^{t}(w_{11}+w_{22}-1)^j \end{pmatrix} \begin{pmatrix} 1 & 1 \\ 1 & \frac{1-w_{22}}{w_{11}-1} \end{pmatrix}^{-1} \begin{pmatrix} \sigma_1 \\ \sigma_2 \end{pmatrix} \quad (69)$$

Since $|w_{11} + w_{22} - 1| < 1$ (the weights are positive and less than 1), $(w_{11} + w_{22} - 1)^t$ goes to zero as $t$ goes to infinity, so we have for very large $t$ that

$$\mathrm{Center}\begin{pmatrix} X_1(t) \\ X_2(t) \end{pmatrix} \approx \begin{pmatrix} \frac{w_{12}x_2(0)+w_{21}x_1(0)}{w_{12}+w_{21}} \\ \frac{w_{12}x_2(0)+w_{21}x_1(0)}{w_{12}+w_{21}} \end{pmatrix} \quad (70)$$

$$\mathrm{Sdv}\begin{pmatrix} X_1(t) \\ X_2(t) \end{pmatrix} \approx (t+1) \begin{pmatrix} \frac{w_{12}\sigma_2+w_{21}\sigma_1}{w_{12}+w_{21}} \\ \frac{w_{12}\sigma_2+w_{21}\sigma_1}{w_{12}+w_{21}} \end{pmatrix} \quad (71)$$

■

We see from (70) and (71) that as the number of feedback iterations $t$ increases, the couple's opinions are converging to the same value $\frac{w_{12}x_2(0)+w_{21}x_1(0)}{w_{12}+w_{21}}$ which is the weighted average of their own original central opinions $x_1(0)$ and $x_2(0)$ with weight $w_{12}$ (Husband's weight for Wife, see Fig. 14) for Wife's idea $x_2(0)$ and weight $w_{21}$ (Wife's weight for Husband) for Husband's idea $x_1(0)$; however, their uncertainties $\mathrm{Sdv}(X_i(t))$ go to infinity. This shows that if everyone is compromising and no one is firm on his/her opinion (i.e., there is no opinion leader [7]), then everyone will get more and more anxious. This is why leaders are welcome in all sorts of human societies – it is human nature to hate uncertainty [34], and strong leaders (such as the Wife in the Compromising Husband with Persistent Wife Connection of Fig. 14) give people a sense of security.

## V. Dynamic Connections with Time-Varying Confidence

Humans are social species, and we feel safe in a society. Therefore, the more we communicate with others, the less uncertain we feel about our opinions (this is the action-strengthen-attitude principle discussed in footnote 15). Consequently, the uncertainty inputs $\sigma_i$ in the dynamic Gaussian fuzzy opinion networks (such as Connections 7 to 9) should in general decrease as time moves forward when more opinion exchanges take place. For example, it is meaningful to assume that the standard deviations $\sigma_1$ of the Husband and $\sigma_2$ of the Wife in Connection 9 (Fig. 15) are decreasing functions of time; this gives the following connection.

**Connection 10 (Compromising with Each Other with Declining Uncertainties):** Here, we consider the same dynamic fuzzy opinion network in Fig. 15 except that the uncertainties $\sigma_1$ and $\sigma_2$ of the Husband and the Wife about the averaged opinions $C_1(t)$ and $C_2(t)$ are now declining with time according to[17]:

$$\sigma_1(t) = \sigma_1\, h^t, \qquad \sigma_2(t) = \sigma_2\, h^t \quad (72)$$

or

[17] According to the concept of perceived uncertainty and real uncertainty introduced in footnote 15, the $\sigma_1(t)$, $\sigma_2(t)$ of (72), (73) are perceived uncertainties which are strictly decreasing to reflect the action-strengthen-attitude principle, while the $\mathrm{Sdv}(X(t))$ of (77), (78) are real uncertainties which are strictly increasing functions of $t$, demonstrating the illusion-of-control reality.

$$\sigma_1(t) = \frac{\sigma_1}{t+1}, \qquad \sigma_2(t) = \frac{\sigma_2}{t+1} \tag{73}$$

where $t = 0,1,2,\ldots,$ and $\sigma_1$, $\sigma_2$ and $h \in (0,1)$ are constants, representing the initial uncertainties of the Husband, the Wife and the rate of declining, respectively. Our task is to determine the unconditional membership functions $\mu_{X_i(t)}(x_i(t))$ of the Husband's ($i = 1$) and the Wife's ($i = 2$) fuzzy sets $X_i(t)$ in this scenario.

Clearly, the procedure (57) to (64) in Connection 9 still applies to this time-varying sdv case, and $X_1(t)$, $X_2(t)$ are Gaussian with their centers and sdv's evolving according to the dynamical equations:

$$\text{Center}(X(t)) = W\ \text{Center}(X(t-1)) \tag{74}$$

$$\text{Sdv}(X(t)) = W\ \text{Sdv}(X(t-1)) + \sigma(t) \tag{75}$$

where $\sigma(t) = (\sigma_1(t), \sigma_2(t))^T$ are given by (72) or (73), and the other variables are the same as in Connection 9. The center equations (74) and (63) are the same and give

$$\text{Center}(X(t)) = W^t \begin{pmatrix} x_1(0) \\ x_2(0) \end{pmatrix} \tag{76}$$

and the solutions to the standard deviation equation (75) are

$$\text{Sdv}(X(t)) = \sum_{j=0}^{t} W^{t-j} \begin{pmatrix} \sigma_1 \\ \sigma_2 \end{pmatrix} h^j \tag{77}$$

for $\sigma(t) = (\sigma_1 h^t, \sigma_2 h^t)^T$ , and

$$\text{Sdv}(X(t)) = \sum_{j=0}^{t} W^{t-j} \begin{pmatrix} \frac{\sigma_1}{j+1} \\ \frac{\sigma_2}{j+1} \end{pmatrix} \tag{78}$$

for $\sigma(t) = \left(\frac{\sigma_1}{t+1}, \frac{\sigma_2}{t+1}\right)^T$. Using the similarity transformation (67) for matrix $W$ and with some straightforward computation, we get as $t$ goes to infinity that

$$\lim_{t\to\infty} \text{Center}\begin{pmatrix} X_1(t) \\ X_2(t) \end{pmatrix} = \frac{1}{w_{12}+w_{21}} \begin{pmatrix} w_{12}x_2(0) + w_{21}x_1(0) \\ w_{12}x_2(0) + w_{21}x_1(0) \end{pmatrix} \tag{79}$$

$$\lim_{t\to\infty} \text{Sdv}\begin{pmatrix} X_1(t) \\ X_2(t) \end{pmatrix} = \frac{1}{(1-h)(w_{12}+w_{21})} \begin{pmatrix} w_{12}\sigma_2 + w_{21}\sigma_1 \\ w_{12}\sigma_2 + w_{21}\sigma_1 \end{pmatrix} \tag{80}$$

for $\sigma(t) = (\sigma_1 h^t, \sigma_2 h^t)^T$ (assuming $1 - h \neq w_{12} + w_{21}$), and

$$\lim_{t\to\infty} \text{Sdv}\begin{pmatrix} X_1(t) \\ X_2(t) \end{pmatrix} = \lim_{t\to\infty} \begin{pmatrix} \dfrac{(w_{12}\sigma_2 + w_{21}\sigma_1)\sum_{j=0}^{t}\left(\frac{1}{1+j}\right) + w_{12}(\sigma_1 - \sigma_2)\sum_{j=0}^{t}\left(\frac{(w_{11}+w_{22}-1)^{t-j}}{1+j}\right)}{w_{12}+w_{21}} \\ \dfrac{(w_{12}\sigma_2 + w_{21}\sigma_1)\sum_{j=0}^{t}\left(\frac{1}{1+j}\right) + w_{21}(\sigma_2 - \sigma_1)\sum_{j=0}^{t}\left(\frac{(w_{11}+w_{22}-1)^{t-j}}{1+j}\right)}{w_{12}+w_{21}} \end{pmatrix} = \begin{pmatrix} \infty \\ \infty \end{pmatrix} \tag{81}$$

for $\sigma(t) = \left(\frac{\sigma_1}{t+1}, \frac{\sigma_2}{t+1}\right)^T$. ■

We see from (80) that if the (perceived) uncertainties of the Husband and the Wife about the averaged opinions decline according to (72), then the standard deviations of their opinions (the real uncertainties) converge to the constants in (80), rather than go to infinity as in the constant sdv case (71). However, if the declines of $\sigma_1(t)$ and $\sigma_2(t)$ are not fast enough as in (73), then the (real) uncertainties of their opinions still go to infinity as shown in (81). This result shows that speed matters; that is, if a person gains confidence very quickly after communicating with other people, then the real uncertainty will stabilize to some finite value, otherwise (if the speed of getting confidence is not fast enough) the real uncertainty will still go to infinity. In other words, it shows that quick decision and fast reaction are important, and this may provide an explanation for why people use short-cuts or heuristics rather than thorough thinking when they face complex situations [35].

The next connection studies the ring connection of $n$ Gaussian nodes with their uncertainties depending on the closeness of their opinions to the community average.

**Connection 11 (Ring Connection with State-Dependent Uncertainty):** Here, we consider the dynamic fuzzy opinion network in Fig. 16, where the center input to a Gaussian node equals the average of the delayed outputs of its two neighbors and itself, and the sdv input to the node equals the difference between its most recent opinion and the average opinion of all the nodes. Specifically, let $C_i(t)$ and $\sigma_i(t)$ be the center and sdv inputs to node $i$, respectively, then

$$C(t) = WX(t-1) \tag{82}$$

$$\sigma_i(t) = \left| \text{Center}(X_i(t-1)) - \frac{1}{n}\sum_{j=1}^{n} \text{Center}\left(X_j(t-1)\right) \right| \tag{83}$$

where $C(t) = (C_1(t), \ldots, C_n(t))^T$ , $X(t-1) = (X_1(t-1), \ldots, X_n(t-1))^T$ and

$$W = \begin{pmatrix} 1/3 & 1/3 & 0 & & \cdots & & 0 & 1/3 \\ 1/3 & 1/3 & 1/3 & 0 & & \cdots & & 0 \\ 0 & 1/3 & 1/3 & 1/3 & 0 & & \cdots & 0 \\ \vdots & & & & \vdots & & & \\ 0 & & \cdots & & 0 & 1/3 & 1/3 & 1/3 \\ 1/3 & 0 & & \cdots & & 0 & 1/3 & 1/3 \end{pmatrix} \tag{84}$$

The reason for choosing $\sigma_i(t)$ as in (83) is explained as follows. Usually, people feel safe if their opinions are close to the majority's opinion which is best represented by the average of all the opinions. Therefore, the closer the person $i$'s last opinion center $Center(X_i(t-1))$ is to the average of the opinion centers of all the people $\frac{1}{n}\sum_{j=1}^{n} Center\left(X_j(t-1)\right)$, the more confident the person $i$ is. Since $\sigma_i(t)$ represents person $i$'s confidence, (83) is a natural choice for $\sigma_i(t)$. Our task is to determine the unconditional membership functions $\mu_{X_i(t)}(x_i(t))$ of all the nodes.

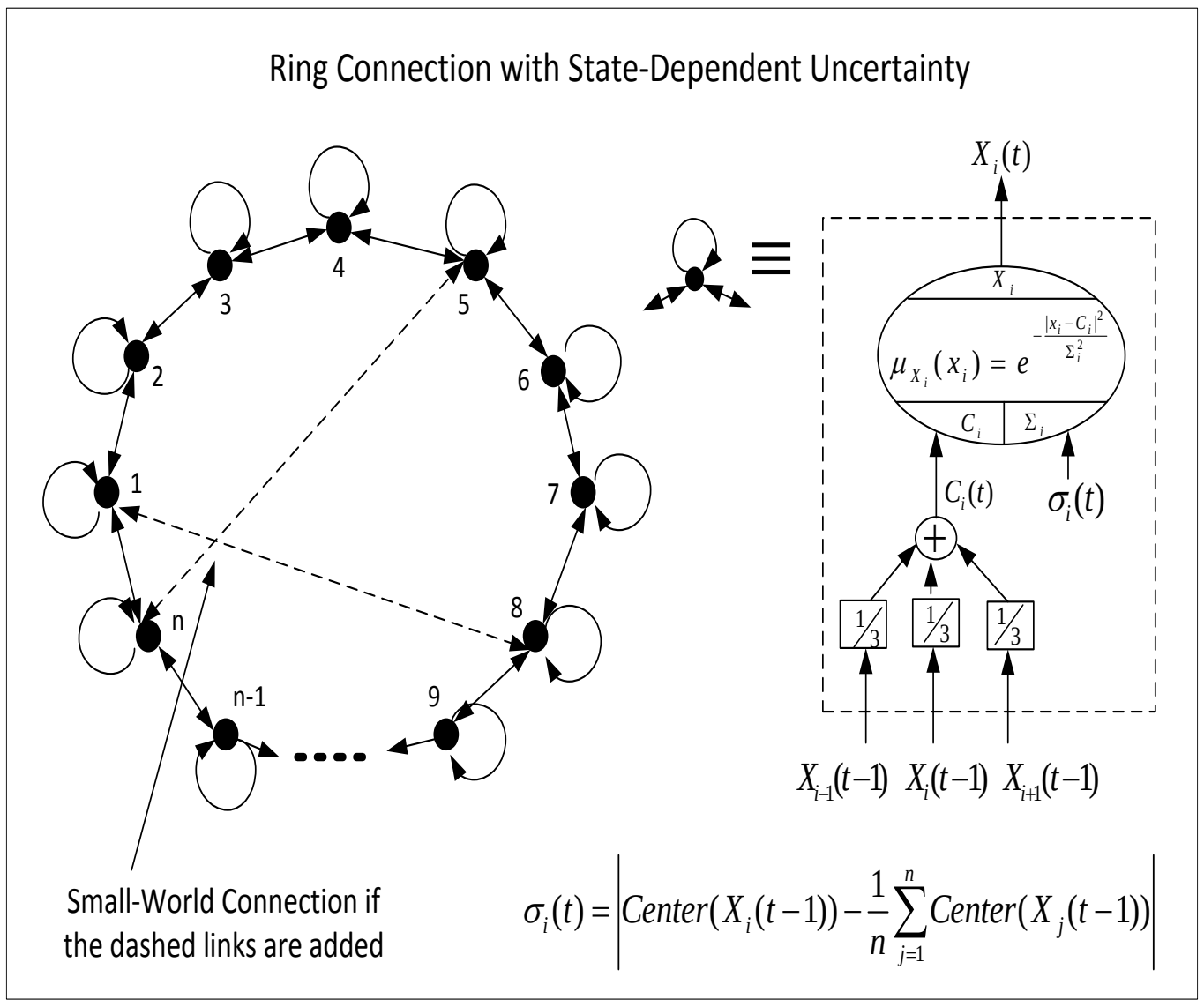

Fig. 16: Ring Connection with State-Dependent Uncertainty (Connection 11).

Using the same procedure (57)-(64) as in Connection 9 (with (58) replaced by (82)) and noticing that the $\sigma_i(t)$ of (83) is a non-fuzzy variable, we have that all $X_i(t)$ are Gaussian with their centers and sdv's evolving according to the dynamical equations:

$$\text{Center}\big(X(t)\big) = W\ \text{Center}\big(X(t-1)\big) \tag{85}$$

$$\text{Sdv}\big(X(t)\big) = W\ \text{Sdv}\big(X(t-1)\big) + \sigma(t) \tag{86}$$

where $\sigma(t) = \big(\sigma_1(t), \dots, \sigma_n(t)\big)^T$ with $\sigma_i(t)$ given by (83) and $W$ is the matrix (84). With initial condition $\text{Center}\big(X(0)\big) = x(0) = \big(x_1(0), \dots, x_n(0)\big)^T$ and $\text{Sdv}\big(X(0)\big) = \sigma(0) = (\sigma_1, \dots, \sigma_n)^T$, the solution of (85) and (86) is

$$\text{Center}\big(X(t)\big) = W^t x(0) \tag{87}$$

$$\text{Sdv}\big(X(t)\big) = W^t \sigma(0) + \sum_{j=1}^{t} W^{t-j} \sigma(j) \tag{88}$$

The following result summarizes the key properties of this Ring Connection with State-Dependent Uncertainty.

**Result 1:** For the Ring Connection with State-Dependent Uncertainty of Fig. 16, we have:

(a) the centers of all the Gaussian nodes converge to the average of the initial centers, i.e., $\lim_{t\to\infty} \text{Center}\big(X_i(t)\big) = \frac{1}{n}\sum_{j=1}^{n} x_j(0)$ for all $i = 1,2, \dots, n$;

(b) the confidence variables $\sigma_i(t)$ of all the Gaussian nodes converge to zero, i.e., $\lim_{t\to\infty} \sigma_i(t) = 0$ for all $i = 1,2, \dots, n$; and,

(c) the standard deviations $\text{Sdv}\big(X_i(t)\big)$ of all the Gaussian nodes converge to the same finite value, i.e., $\lim_{t\to\infty} \text{Sdv}\big(X_i(t)\big)$ equal the same value for all $i = 1,2, \dots, n$.

Proof of Result 1 is given in the Appendix. Fig. 17 plots the centers $\text{Center}\big(X_i(t)\big)$ and the sdv's $\text{Sdv}\big(X_i(t)\big)$ of all the nodes ($i = 1,2, \dots, n$) for a simulation run of the Ring Connection with State-Dependent Confidence with $n = 30$ nodes, where the initial $x_1(0), \dots, x_{30}(0)$ are uniformly distributed over [0,1] and $\sigma(0) = (1, \dots, 1)^T$. ■

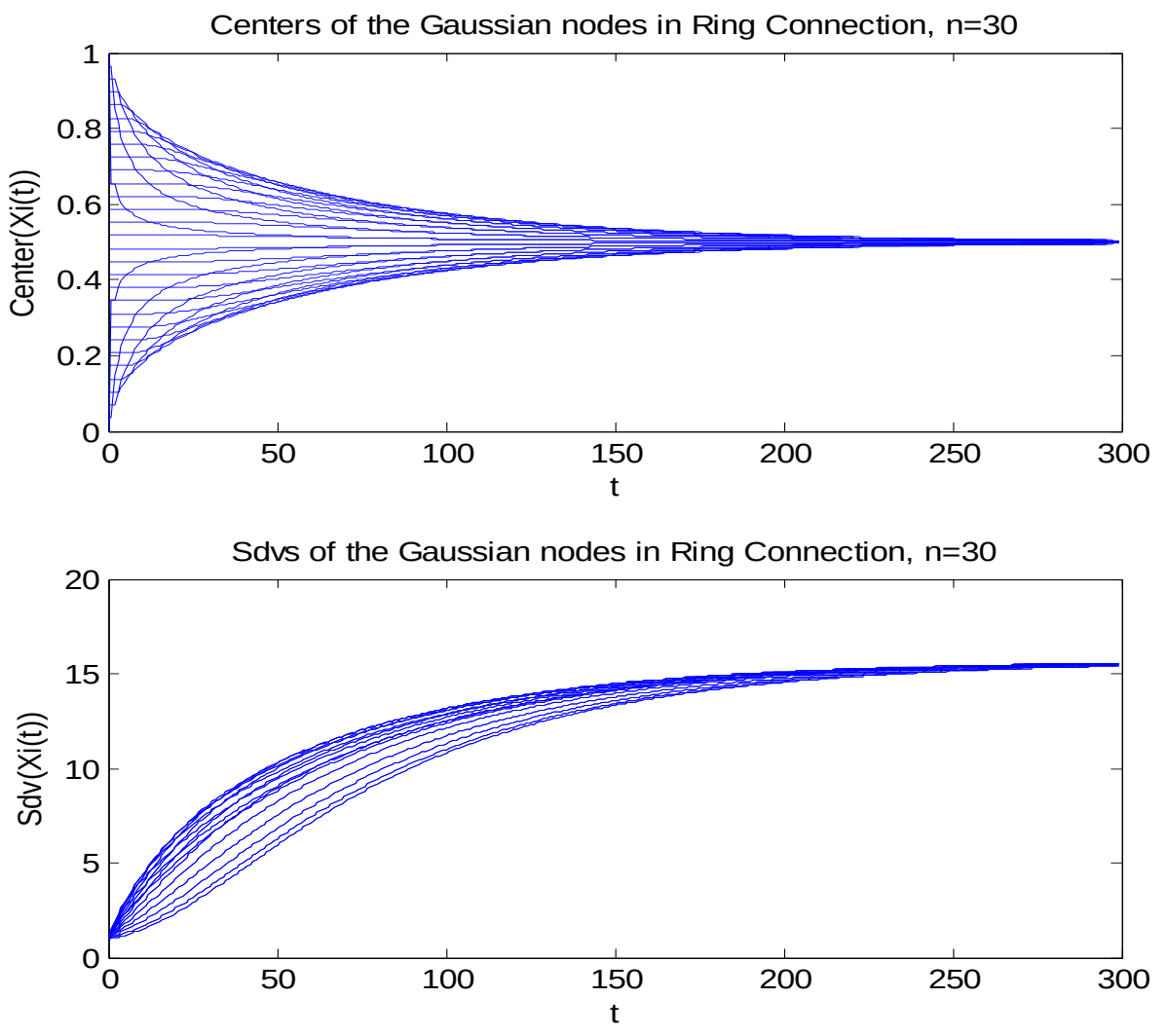

Fig. 17: A simulation run of the Ring Connection with State-Dependent Uncertainty (Connection 11) with $n = 30$ nodes, where the top and bottom sub-figures plot the centers $\text{Center}\big(X_i(t)\big)$ and the sdv's $\text{Sdv}\big(X_i(t)\big)$, respectively, $i = 1,2, \dots, 30$.

Result 1 (a), (c) show that if $n$ agents are connected in a ring as in Fig. 16, then they will eventually reach a consensus as the iteration goes to infinity, and this consensus equals the average of their original opinions. Furthermore, Result 1 (b) reveals that the $n$ agents reach this consensus very confidently, with their confidence variables $\sigma_i(t)$ converging to zero. From the proof of Result 1 in the Appendix we can see that the speed of convergence to the consensus is determined by the second largest eigenvalue of the connection matrix $W$ of (84). By adding a few random links to the Ring Connection of Fig. 16 (the dashed links), we can form a *small-world FON* whose speed of convergence will be much faster than the Ring Connection results in Fig. 17; we omit the details.

## VI. Time-Varying or State-Dependent Connections

Although feedback loops were introduced and the confidence (uncertainty inputs to the nodes) was allowed to change with time or with the status of the opinions in the G-FONs studied in the last two sections (Connections 7 to 11), the connectivity (the topology of the G-FONs) was fixed and did not change with time. In real life, however, the connectivity between people is a changing variable. For example, in a party two people are connected when they talk to each other and become disconnected when they join separate groups. Therefore, the connectivity of a fuzzy opinion network should be allowed to change with time or with the status of the opinions in order to model many interesting situations in real life. In the following we study two time-varying/state-dependent G-FONs, where Connection 12 studies the problem of two students competing for one research assistantship and Connection 13 proposes a general *bounded confidence* G-FON where two nodes are connected only when their opinions (which are fuzzy sets) are close enough to each other.

**Connection 12 ("Smart" Student versus "Stubborn" Student):** Two students are competing for one research assistantship. The professor gives them a subject and asks them to express their opinions about it (e.g., the subject might be: "What is your opinion about changing 'fuzzy sets' to 'smart sets' and 'fuzzy logic' to 'smart logic'[18] to avoid the negative connotation associated with the English word 'fuzzy' and to promote a 'new birth' of the fuzzy field when it is struggling in the mess of the middle-age (50 years old) crisis?"). The professor announces the following rule:

1) The professor tells the students his initial opinion $X_1(0)$ about the subject; then, the students tell the professor their first opinions $X_2(1)$ and $X_3(1)$ about the subject;
2) The professor updates his opinion in such a way that his central opinion at iteration $t\,(=1,2,\dots)$, $C_1(t)$, is a weighted average of his last opinion $X_1(t-1)$ and the two opinions, $X_2(t)$ and $X_3(t)$, of the two students, i.e.,

$$C_1(t) = w_1(t)X_1(t-1) + w_2(t)X_2(t) + w_3(t)X_3(t) \tag{89}$$

[18] The classical set is in-or-out, 0-or-1, which is too simple, too naïve, and exhibits no intelligence whatsoever. "Smart set" is smart, which tells you how to determine a degree of belonging, giving you flexibility and more control. "Smart logic" is smart, like a human, who is not black-and-white, who has feelings (continuous variables with inherent uncertainties), and "smart logic" gives you a collection of scientific/engineering principles and tools to reason, to control, and to make decisions in an intelligent, human-like manner.

where the weights $w_i(t)$ are positive and will change in such a way that the weight for the student whose opinion $X_i(t)$ ($i = 2$ or 3) is closer to the professor's last opinion $X_1(t-1)$ will increase, and the other student's weight will be reduced accordingly such that the three weights always sum to 1;

3) The students are totally free to update their opinions $X_i(t)$ at each iteration $t$; and;
4) The student whose first opinion center is closer to the professor's final opinion center will get the studentship, i.e., if $|\text{Center}(X_2(1)) - \text{Center}(X_1(\infty))| < |\text{Center}(X_3(1)) - \text{Center}(X1\infty)$, then Student 2 will get the research assistantship, otherwise Student 3 will get the research assistantship.

Based on the rule above, Student 2 (the "Smart" Student) chooses to always follow the professor's opinion, i.e., $X_2(t) = X_1(t-1)$, while Student 3 is "stubborn" and does not change his/her first opinion at all, i.e., $X_3(t) = X_3(1)$ for all $t = 1,2,\ldots$. Fig. 18 shows this network connection. Now the question is: Who will get the research assistantship – the "Smart" Student or the "Stubborn" Student?

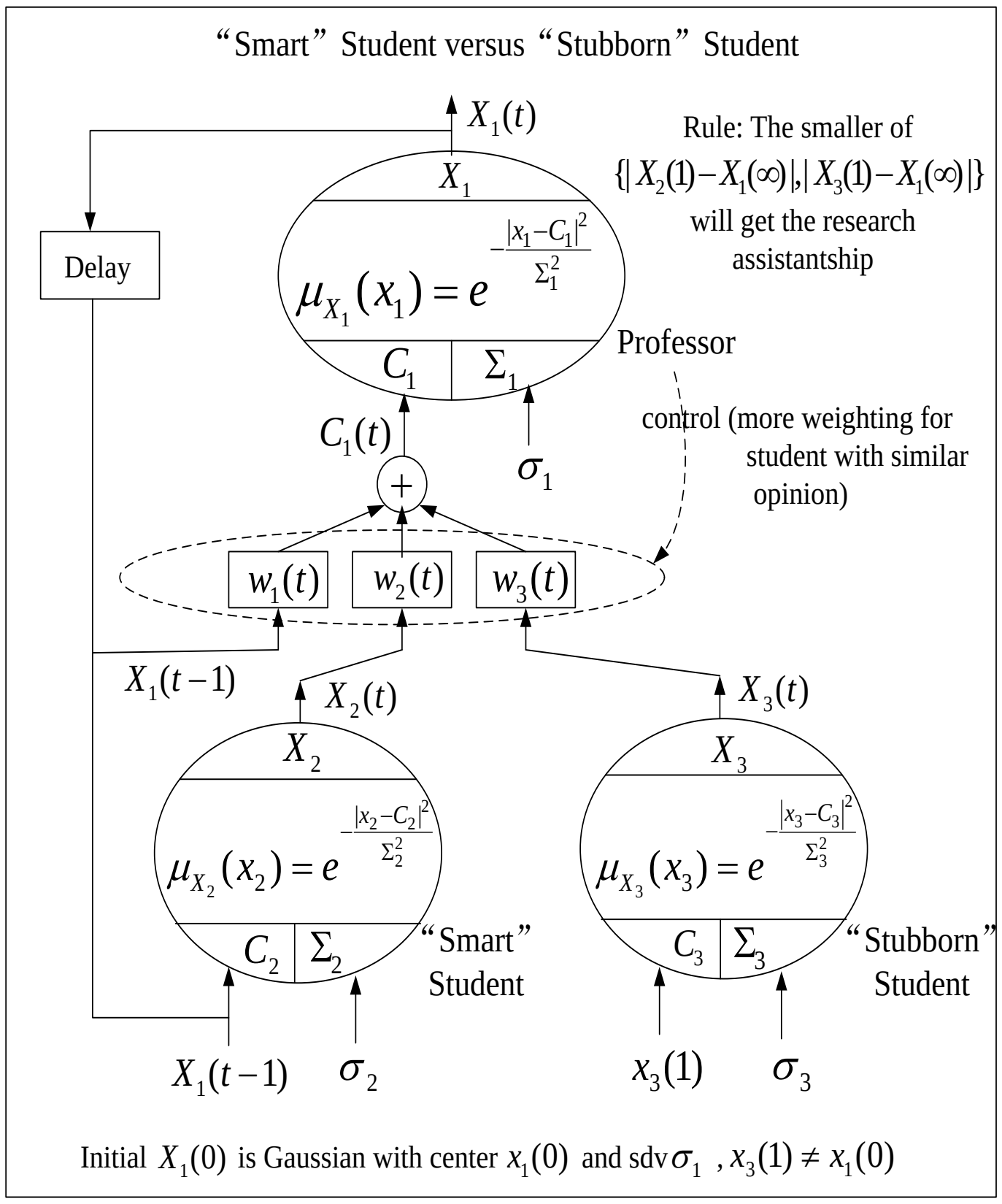

Fig. 18: "Smart" Student versus "Stubborn" Student (Connection 12).

Applying Lemma 1 to (89) gives

$$\mu_{C_1(t)}(c_1(t)) = e^{-\frac{|c_1(t)-w_1(\text{t})\text{Center}(X_1(t-1))-w_2(\text{t})\text{Center}(X_2(t))-w_3(\text{t})\text{Center}(X_3(t))|^2}{(w_1(\text{t})\text{Sdv}(X_1(t-1))+w_2(\text{t})\text{Sdv}(X_2(t))+w_3(\text{t})\text{Sdv}(X_3(t)))^2}} \quad (90)$$

Using the Compositional Rule of Inference (7) for

$$\mu_{X_1(t)|C_1(t)}(x_1(t)|C_1(t)) = e^{-\frac{|x_1(t)-C_1(t)|^2}{\sigma_1^2}} \quad (91)$$

and $\mu_{C_1(t)}(c_1(t))$ of (90), we get

$$\mu_{X_1(t)}(x_1(t)) = \max_{c_1(t)\in R^n}\min\left[e^{-\frac{|x_1(t)-c_1(t)|^2}{\sigma_1^2}}, e^{-\frac{|c_1(t)-w_1(\text{t})\text{Center}(X_1(t-1))-w_2(\text{t})\text{Center}(X_2(t))-w_3(\text{t})\text{Center}(X_3(t))|^2}{(w_1(\text{t})\text{Sdv}(X_1(t-1))+w_2(\text{t})\text{Sdv}(X_2(t))+w_3(\text{t})\text{Sdv}(X_3(t)))^2}}\right]$$

$$= e^{-\frac{|c_1(t)-w_1(\text{t})\text{Center}(X_1(t-1))-w_2(\text{t})\text{Center}(X_2(t))-w_3(\text{t})\text{Center}(X_3(t))|^2}{(\sigma_1+w_1(\text{t})\text{Sdv}(X_1(t-1))+w_2(\text{t})\text{Sdv}(X_2(t))+w_3(\text{t})\text{Sdv}(X_3(t)))^2}} \quad (92)$$

Since $X_2(t) = X_1(t-1)$ , $\text{Center}(X_3(t)) = x_3(1)$ and $\text{sdv}(X_3(t)) = \sigma_3$, we have from (92) that $X_1(t)$ is Gaussian with

$$\text{Center}(X_1(t)) = (w_1(\text{t}) + w_2(\text{t}))\text{Center}(X_1(t-1)) + w_3(\text{t})x_3(1) \quad (93)$$

$$\text{Sdv}(X_1(t)) = (w_1(\text{t}) + w_2(\text{t}))\text{Sdv}(X_1(t-1)) + \sigma_1 + w_3(\text{t})\sigma_3 \quad (94)$$

Solving (93) and (94) we obtain

$$\begin{aligned}\text{Center}(X_1(t)) &= \prod_{i=1}^{t}(1-w_3(i))\,x_1(0) + \sum_{i=1}^{t}\left(\prod_{j=i+1}^{t}(1-w_3(j))\right)w_3(i)x_3(1)\\ &= \prod_{i=1}^{t}(1-w_3(i))\,x_1(0) + \left(1-\prod_{i=1}^{t}(1-w_3(i))\right)x_3(1)\end{aligned} \quad (95)$$

$$\begin{aligned}\text{Sdv}(X_1(t)) &= \prod_{i=1}^{t}(1-w_3(i))\,\sigma_1 + \sum_{i=1}^{t}\left(\prod_{j=i+1}^{t}(1-w_3(j))\right)(\sigma_1+w_3(i)\sigma_3)\\ &= \sum_{i=0}^{t}\left(\prod_{j=i+1}^{t}(1-w_3(j))\right)\sigma_1 + \left(1-\prod_{i=1}^{t}(1-w_3(i))\right)\sigma_3 \quad (\sigma_1)\end{aligned} \quad (96)$$

To complete the computations of (95) and (96), we need the professor's weighting scheme for $w_1(t)$, $w_2(t)$ and $w_3(t)$. To reward the "Smart" Student who closely follows the professor's opinion and to punish the "Stubborn" Student who insists on his/her own opinion, the professor proposes the following weighting scheme:

$$\begin{cases} w_1(t) = 0.5 \\ w_2(t) = 0.5 - 0.25\,h^t \\ w_3(\text{t}) = 0.25\,h^t \end{cases} \quad (97)$$

where $h \in (0,1)$. That is, the professor always weights 0.5 for his own opinion and gives an equal initial weight of 0.25 for both students; then, at each iteration, the professor takes away $100(1-h)\%$ of the weight from the "Stubborn" Student and adds this deducted amount to the "Smart" Student's weight so that the three weights always sum to 1. We see from (97) that as the iteration $t$ goes to infinity, the "Stubborn" Student's weight $w_3(\text{t})$ goes to zero exponentially fast, while the "Smart" Student's weight $w_2(\text{t})$ increases towards 0.5. It seems that the "Smart" Student will get the research assistantship. Really?

Fig. 19 plots the simulation results of the professor's opinion center $\text{Center}(X_1(t))$ of (95) with the weighting scheme (97) for different values of $h$, where the "Smart" Student's first opinion center $x_2(1) = 2$ (which is also the professor's initial opinion center) and the "Stubborn" Student's first opinion center $x_3(1) = 10$. We see from Fig. 19 that when $h > 0.724$, the professor's opinion center will converge to a value above the middle point 6 $(= (x_2(1) + x_3(1))/2)$, meaning that the "Stubborn" Student will get the research assistantship; whereas when $h < 0.724$, the "Smart" Student will get the research assistantship. Since $h = 0.724$ means taking away $1 - h = 27.6\%$ of the weight from the "Stubborn" Student at every iteration, which is not a small amount based on "common sense", the results in Fig. 19 show that the "Stubborn" Student will have a good chance to get the research assistantship if the professor is misled by the "common sense" and chooses the $h$ too large (larger than 0.724). ■

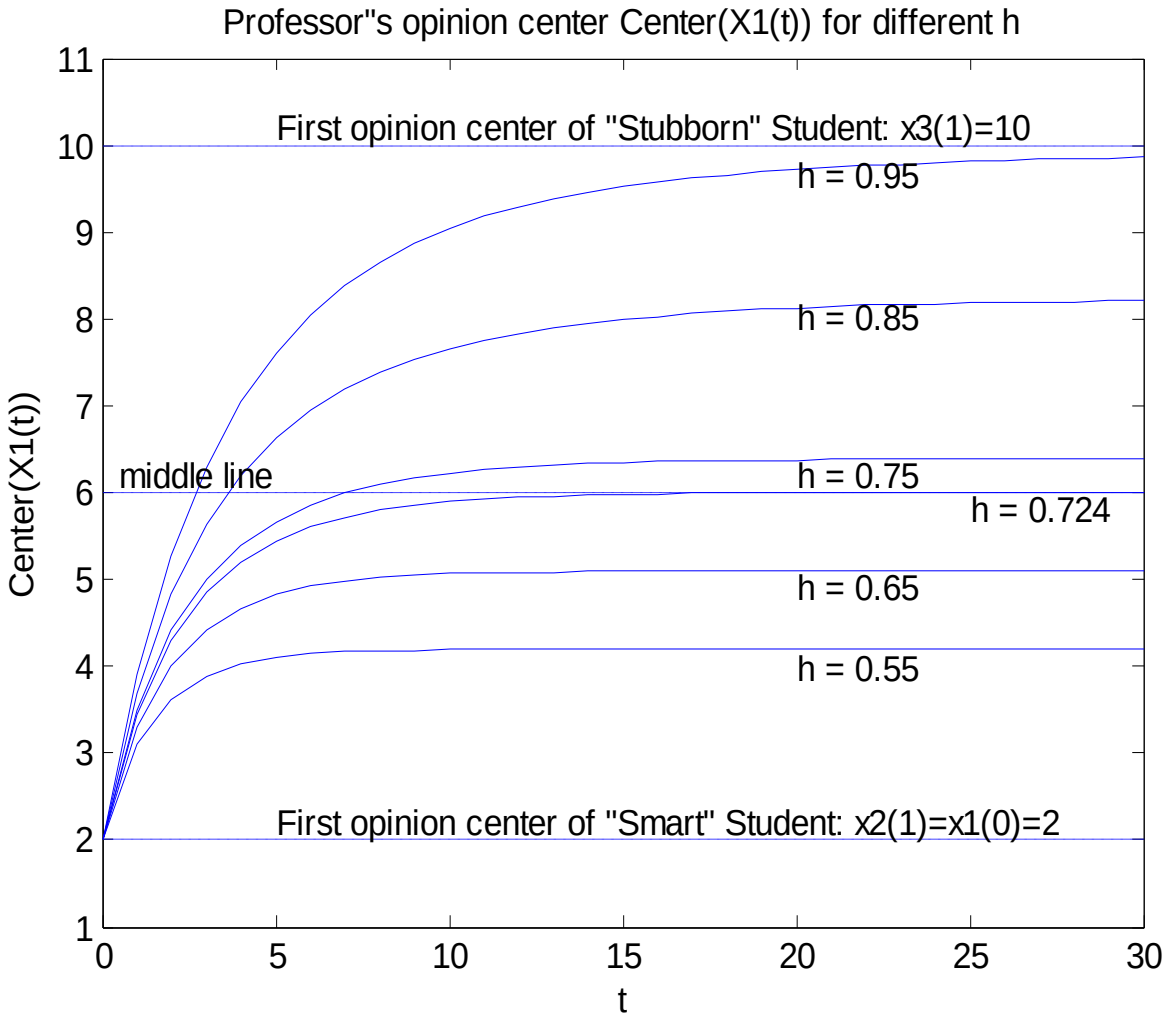

Fig. 19: Professor's opinion center $\text{Center}(X_1(t))$ with the weighting scheme (97) for different h.

This example shows again (as in Connection 10) that speed matters; that is, if the professor reacts quickly to the stubbornness and the smartness of the students (by choosing $h$ small enough), then the "Smart" Student will get the research assistantship, otherwise the stubbornness of the "Stubborn" Student will take control over everyone's opinion and secures the research assistantship. This example also shows again (as Connections 8 and 9 reveal) the important role the opinion leader (in this case the opinion leader is the "Stubborn" Student) plays in a society – stubbornness has a profound influence in opinion networks [3], [23], [74], and "repeating a lie one hundred times makes it a truth" – the tenet of propaganda machines – comes out naturally from our mathematical model.

Finally, we propose a general Bounded Confidence Connection of Gaussian nodes which generalizes the popular Hegselmann-Krause (HK) model [26] for opinion dynamics (the HK model considers only the opinions whereas our Bounded Confidence FON considers the evolution and interaction of both the opinions and their uncertainties).

**Connection 13 (Bounded Confidence Connection):** Consider the dynamic fuzzy opinion network with $n$ Gaussian fuzzy nodes in Fig. 20, where the center inputs to the $n$ Gaussian nodes $C(t) = (C_1(t), \dots, C_n(t))^T$ are weighted averages[19] of the delayed outputs of the $n$ Gaussian nodes $X(t-1) = (X_1(t-1), \dots, X_n(t-1))^T$:

$$C(t) = W(t)X(t-1) \tag{98}$$

with the time-varying weights $W(t) = [w_{ij}(t)]$ depending on the state of the network. Specifically, the weight $w_{ij}(t)$ connecting the two nodes $X_i$ and $X_j$ is non-zero only when the two fuzzy sets $X_i(t-1)$ and $X_j(t-1)$ are close enough to each other and the value of $w_{ij}(t)$ is proportional to the closeness between $X_i(t-1)$ and $X_j(t-1)$:

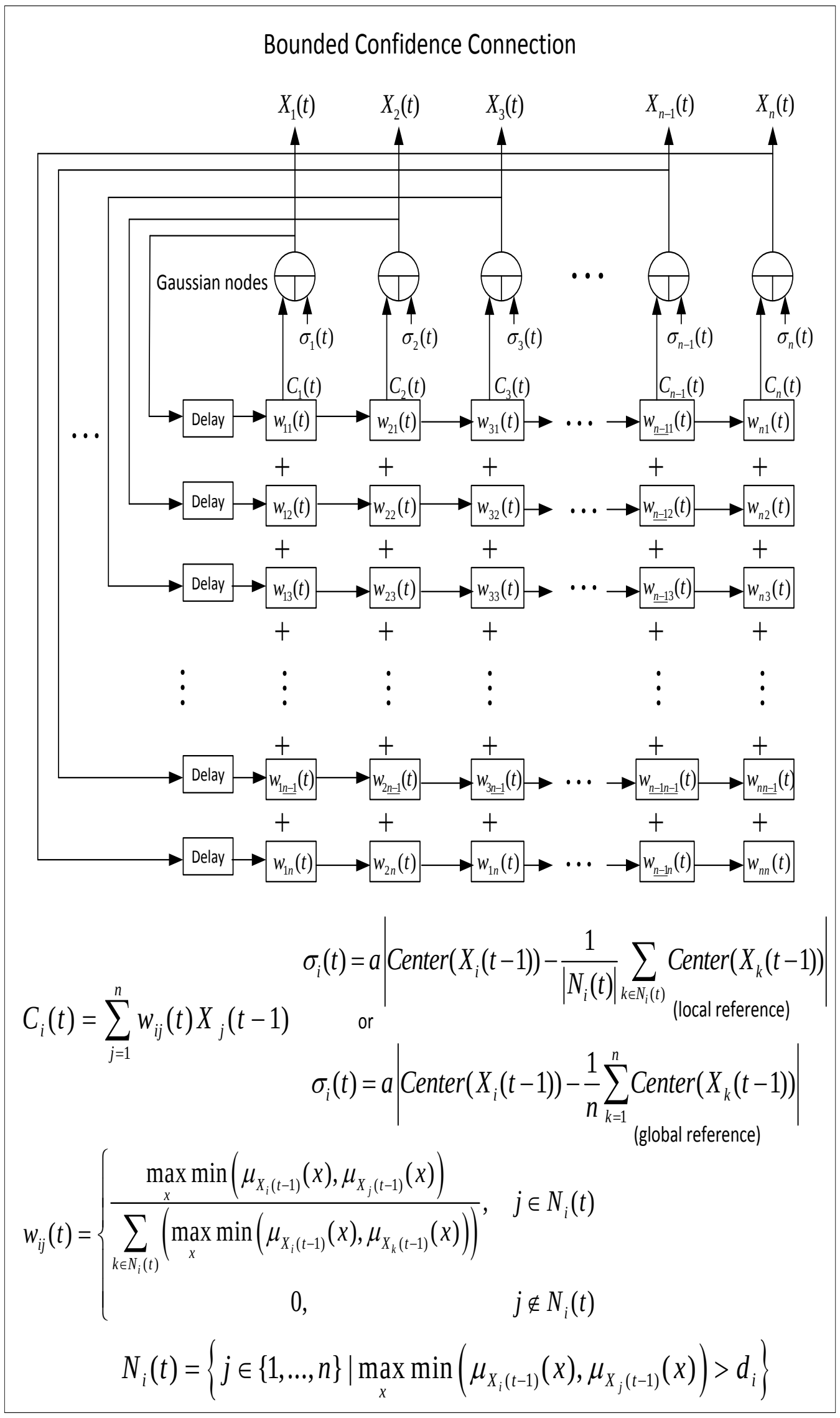

Fig. 20: Bounded Confidence Connection of Gaussian nodes (Connection 13).

[19] In DeGroot's seminal paper [18] an agent's opinion is influenced directly by the opinions of others by taking a weighted average of the other agents' probability distributions, whereas our G-FONs differ considerably as they involve influence through the centers and standard deviations of the membership functions that represent opinions and their uncertainties in our model. We thank an anonymous referee for pointing out this difference.

$$w_{ij}(t)=\begin{cases}\dfrac{\max_x min\left(\mu_{X_i(t-1)}(x),\mu_{X_j(t-1)}(x)\right)}{\sum_{j\in N_i(t)}\left(\max_x min\left(\mu_{X_i(t-1)}(x),\mu_{X_j(t-1)}(x)\right)\right)}, & j\in N_i(t)\\ 0, & j\notin N_i(t)\end{cases} \quad (99)$$

where $N_i(t)$ $(i=1,\dots,n)$ is the collection of nodes that are connected to node $i$ at time $t$, defined as:

$$N_i(t)=\left\{j\in\{1,\dots,n\}\mid \max_x min\left(\mu_{X_i(t-1)}(x),\mu_{X_j(t-1)}(x)\right)>d_i\right\} \quad (100)$$

where $\max_x min\left(\mu_{X_i(t-1)}(x),\mu_{X_j(t-1)}(x)\right)$ is the height (the largest membership value) of the intersection of the two fuzzy sets $X_i(t-1)$ and $X_j(t-1)$, representing the closeness between $X_i(t-1)$ and $X_j(t-1)$, and $d_i$ are constants. (98)-(100) mean that agent $i$ (node $i$) updates his opinion by taking a weighted average of the opinions of those agents whose opinions are close to his own, with the weights proportional to the closeness of the two opinions; this is the reason we call this fuzzy opinion network *bounded confidence* connection.

When all $d_i$'s are equal ($d_i=d$ for all $i=1,\dots,n$), we obtain a *homogeneous Bounded Confidence (BC) Fuzzy Opinion Network (FON)*, while if the $d_i$'s are different, we get a *heterogeneous BCFON*. Homogeneous BCFONs are *undirected* networks ($w_{ij}(t)=w_{ji}(t), i,j=1,\dots,n$), while heterogeneous BCFONs are *directed* networks ($w_{ij}(t)\neq w_{ji}(t)$ in general; e.g., an *open-minded agent i* (small $d_i$) may consider opinions from a *closed-minded agent j* (large $d_j$) – $w_{ij}(t)\neq 0$, but not the other way around – $w_{ji}(t)=0$). Generally speaking, homogeneous BCFONs are easier to analyze mathematically (see [70]), while heterogeneous BCFONs are closer to reality (people are different).

For the sdv inputs $\sigma_i(t)$ to the Gaussian nodes in Fig. 20, we propose two schemes:

(a) Local Reference:

$$\sigma_i(t)=a\left|\text{Center}(X_i(t-1))-\frac{1}{|N_i(t)|}\sum_{j\in N_i(t)}\left(\text{Center}\left(X_j(t-1)\right)\right)\right| \quad (101)$$

(b) Global Reference:

$$\sigma_i(t)=a\left|\text{Center}(X_i(t-1))-\frac{1}{n}\sum_{j=1}^{n}\text{Center}\left(X_j(t-1)\right)\right| \quad (102)$$

where $|N_i(t)|$ is the number of elements in $N_i(t)$ and $a$ is a scaling constant. In the Local Reference (101), agent $i$ considers only the opinions of his neighbors $N_i(t)$ and views the average of his neighbors' latest opinion centers $\frac{1}{|N_i(t)|}\sum_{j\in N_i(t)}\left(\text{Center}\left(X_j(t-1)\right)\right)$ as the correct answer, therefore the closer his last opinion center $\text{Center}(X_i(t-1))$ is to the average of the latest center opinions of his neighbors, the more confidence he has; this gives the choice of the standard deviation in (101). For the Global Reference (102), agent $i$ considers the opinions of all the agents involved and views the average of all the agents' opinion centers $\frac{1}{n}\sum_{j=1}^{n}\text{Center}\left(X_j(t-1)\right)$ as the correct answer; the closer his opinion center is to this correct answer, the more confidence he has, which gives the choice of $\sigma_i(t)$ in (102). The Global Reference case corresponds to the situation where a central agent collects the opinions from all the people involved and announces the averaged opinion back to all the people, such as the scenario in Keynes' Beauty Contest Theory (chapter 12 of [36]). The Local Reference case, on the other hand, refers to the scenario of decentralized control where a central agent does not exist and people know only the opinions of their neighbors (those they have direct contact with). Our task now is to explore how the opinions $X_i(t)$ evolve.

Using the same procedure (57)-(64) as in Connection 9 (with (58) replaced by (98)) and noticing that the $w_{ij}(t)$ of (99) and the $\sigma_i(t)$ of (101), (102) are non-fuzzy variables (real numbers for given $t$), we have that all $X_i(t)$ are Gaussian with their centers and sdv's evolving according to the dynamical equations:

$$\text{Center}(X(t))=W(t)\,\text{Center}(X(t-1)) \quad (103)$$

$$\text{Sdv}(X(t))=W(t)\,\text{Sdv}(X(t-1))+\sigma(t) \quad (104)$$

where $\sigma(t)=(\sigma_1(t),\dots,\sigma_n(t))^T$ with $\sigma_i(t)$ given by (101) or (102), and the $W(t)=[w_{ij}(t)]$, defined in (99), are computed according to

$$w_{ij}(t)=\begin{cases}\dfrac{e^{-\frac{|\text{Center}(X_i(t-1))-\text{Center}(X_j(t-1))|^2}{(\text{Sdv}(X_i(t-1))+\text{Sdv}(X_j(t-1)))^2}}}{\sum_{j\in N_i(t)}\left(e^{-\frac{|\text{Center}(X_i(t-1))-\text{Center}(X_j(t-1))|^2}{(\text{Sdv}(X_i(t-1))+\text{Sdv}(X_j(t-1)))^2}}\right)}, & j\in N_i(t)\\ 0, & j\notin N_i(t)\end{cases} \quad (105)$$

$$N_i(t)=\left\{j\in\{1,\dots,n\}\mid e^{-\frac{|\text{Center}(X_i(t-1))-\text{Center}(X_j(t-1))|^2}{(\text{Sdv}(X_i(t-1))+\text{Sdv}(X_j(t-1)))^2}}>d_i\right\} \quad (106)$$

which are obtained by substituting $\mu_X(x)=e^{-\frac{|x-\text{Center}(X)|^2}{(\text{Sdv}(X))^2}}$ ($X=X_i(t-1)$ or $X=X_j(t-1)$) into $\max_x min\left(\mu_{X_i(t-1)}(x),\mu_{X_j(t-1)}(x)\right)$ and noticing from Fig. 5 that the $\max_x$ is achieved at $\frac{x-\text{Center}(X_i(t-1))}{\text{Sdv}(X_i(t-1))}=\frac{\text{Center}(X_j(t-1))-x}{\text{Sdv}(X_j(t-1))}$ which results in

$$\max_x min\left(\mu_{X_i(t-1)}(x),\mu_{X_j(t-1)}(x)\right)=e^{-\frac{|\text{Center}(X_i(t-1))-\text{Center}(X_j(t-1))|^2}{(\text{Sdv}(X_i(t-1))+\text{Sdv}(X_j(t-1)))^2}} \quad (107)$$

We now perform some simulations of the dynamical equations (103)-(106) with Local (101) or Global (102) Reference. With the initial centers $\text{Center}(X(0))=(x_1(0),\dots,x_n(0))^T$ of the $n$ Gaussian fuzzy nodes uniformly distributed over the interval [0,1] ($x_i(0) = \frac{i-1}{n-1}, i = 1, \ldots, n$) and their initial standard deviations $\mathrm{Sdv}(X(0)) = (\sigma_1(0), \ldots, \sigma_n(0))^T$ drawn from a random uniform distribution over [0,1], Figs. 21 and 22 show a single realization of the simulation results of the Bounded Confidence Connection dynamics (103)-(106) with Local Reference (101) for $n = 12$, $a = 0.1$ and $d_i = 0.95$ ($i = 1, \ldots, n$, so we are simulating a homogeneous BCFON), where the top and bottom sub-figures in Fig. 21 plot the 12 opinion centers $\mathrm{Center}(X_i(t))$ and the 12 $\mathrm{Sdv}(X_i(t))$'s, respectively, and Fig. 22 shows the evolution of the network topology for the same simulation run. Similarly, Figs. 23 and 24 show the corresponding results with the Global Reference (102). We see from Figs. 21 and 22 that with Local Reference (101), the 12 initially uniformly distributed opinions eventually converge into two groups – nodes 1-8 form one group and nodes 9-12 form the other, and the agents in a group converge to the same opinion – the same opinion center and the same standard deviation. For the Global Reference (102), we see from Figs. 23 and 24 that the 12 agents eventually reach a consensus, i.e., they converge to the same opinion.

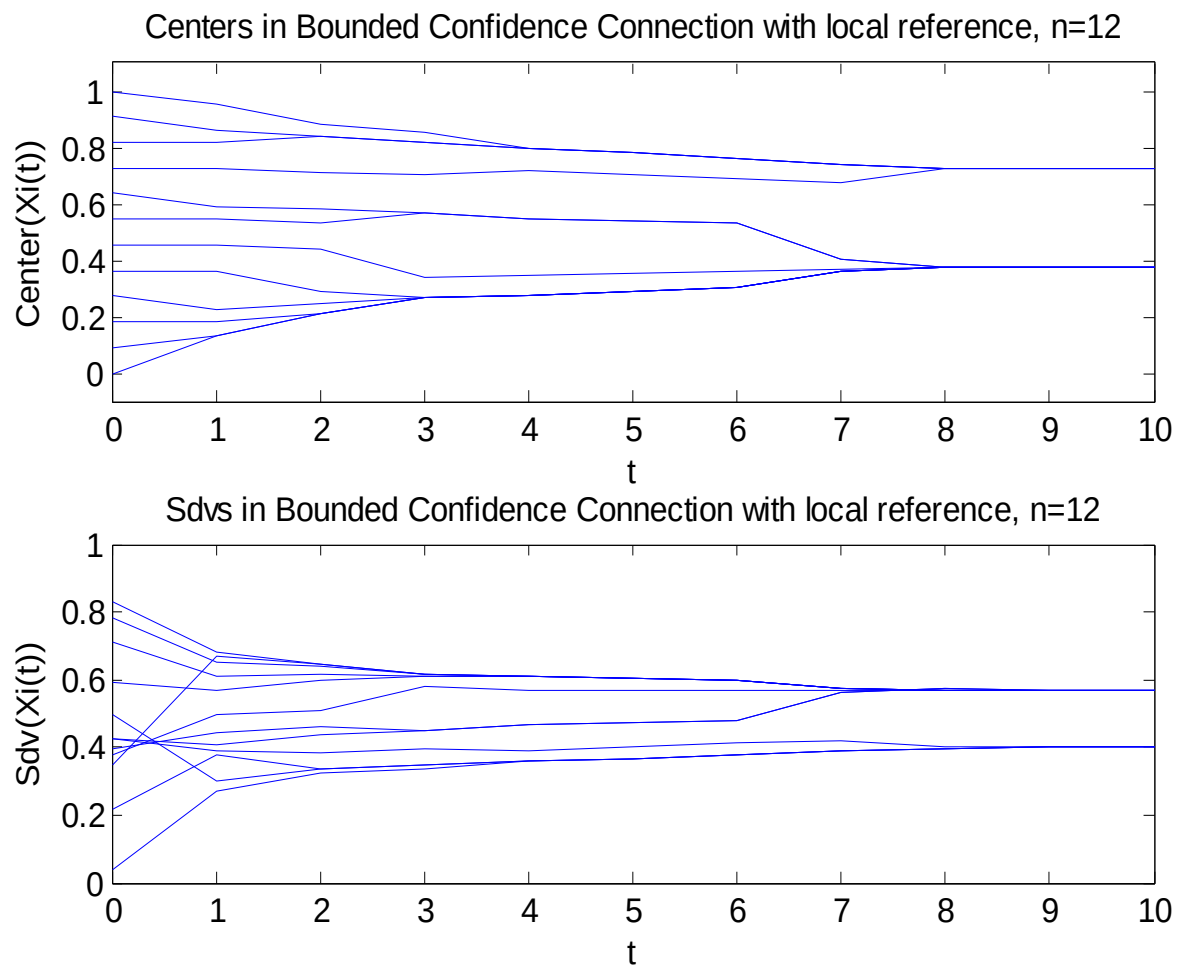

Fig. 21: A simulation run of the Bounded Confidence Connection dynamics (103)-(106) with Local Reference (101) for $n = 12$ nodes, where the top and bottom sub-figures plot $\mathrm{Center}(X_i(t))$ and $\mathrm{Sdv}(X_i(t))$, respectively.

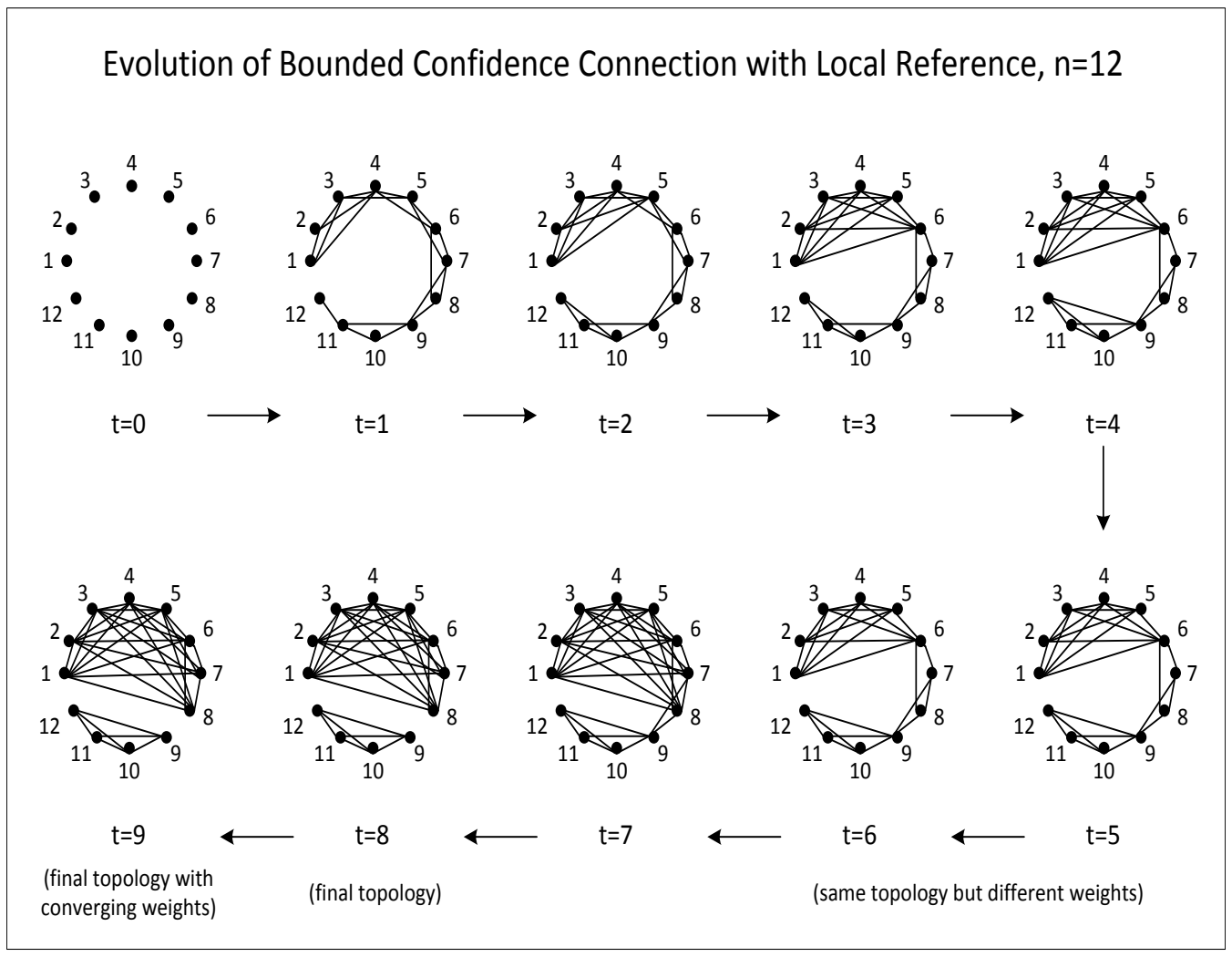

Fig. 22: Network topology evolution corresponding to the simulation in Fig. 21.

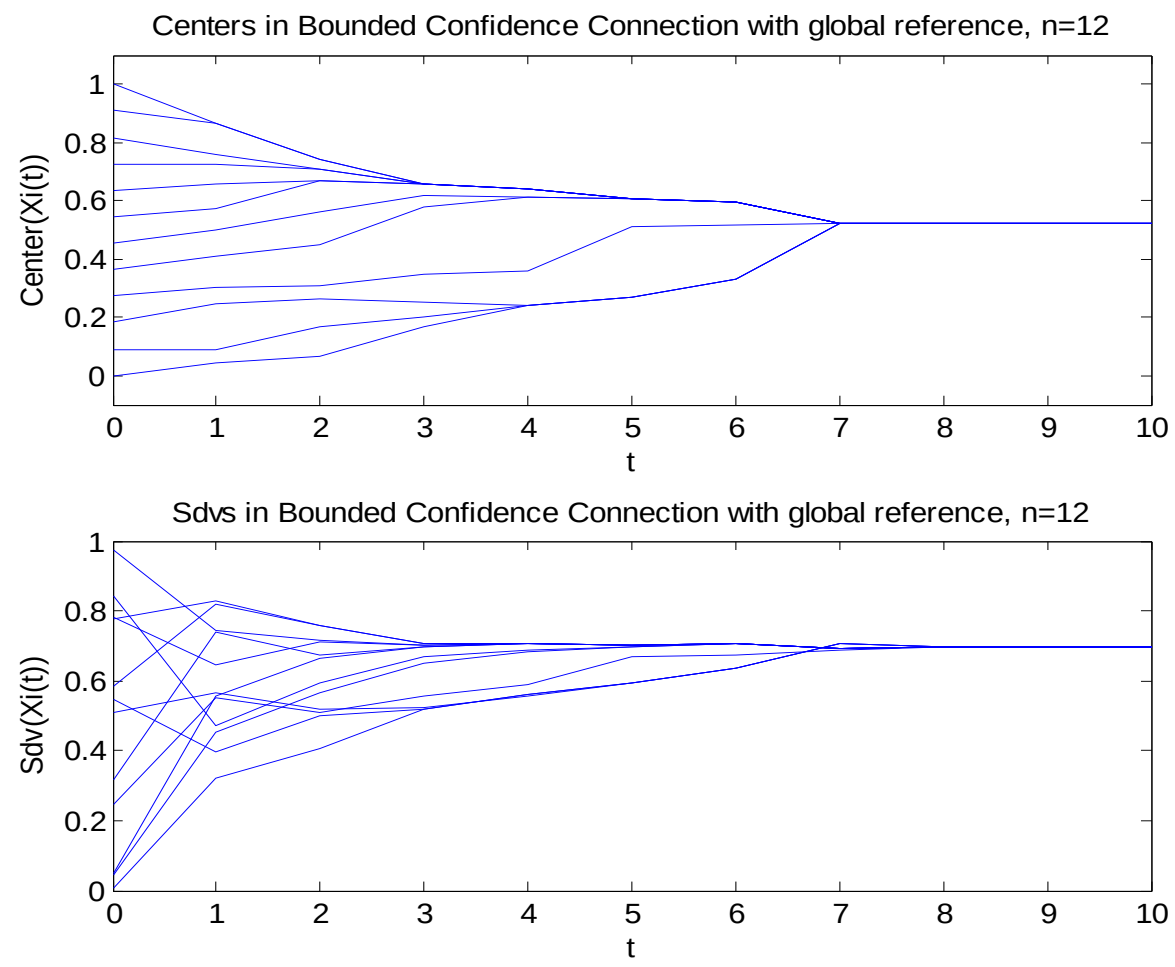

Fig. 23: A simulation run of the Bounded Confidence Connection dynamics (103)-(106) with Global Reference (102) for $n = 12$ nodes, where the top and bottom sub-figures plot $\mathrm{Center}(X_i(t))$ and $\mathrm{Sdv}(X_i(t))$, respectively.

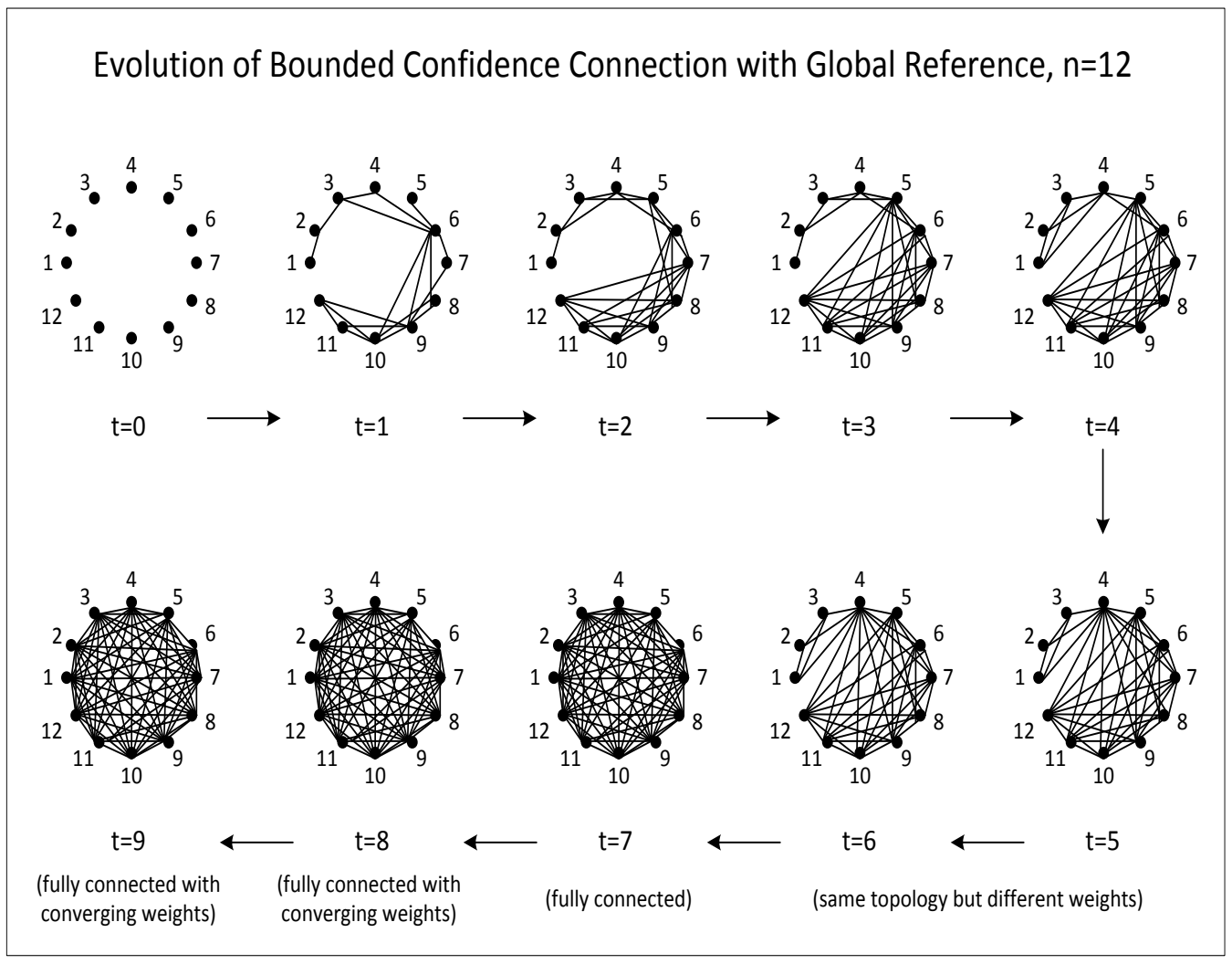

Fig. 24: Network topology evolution corresponding to the simulation in Fig. 23.

We perform one more simulation for the Global Reference case with $n = 100$ nodes, and the result is shown in Fig. 25 where the top sub-figure plots the 100 $\mathrm{Center}(X_i(t))$'s and the bottom sub-figure plots the 100 $\mathrm{Sdv}(X_i(t))$'s. We see From Fig. 25 that, again, all agents eventually converge to the same opinion. Since the main theme of this paper is to build the basics of Fuzzy Opinion Networks, theoretical analysis for the convergence properties of the Bounded Confidence GFON (BCGFON) in Fig. 20 is presented in another paper [70], where we prove that the agents in the Global Reference BCGFON eventually reach a consensus (as Figs. 23 and 25 illustrate), while the agents in the Local Reference BCGFON converge to different communities and the agents in a community reach a consensus (as Fig. 21 illustrates).

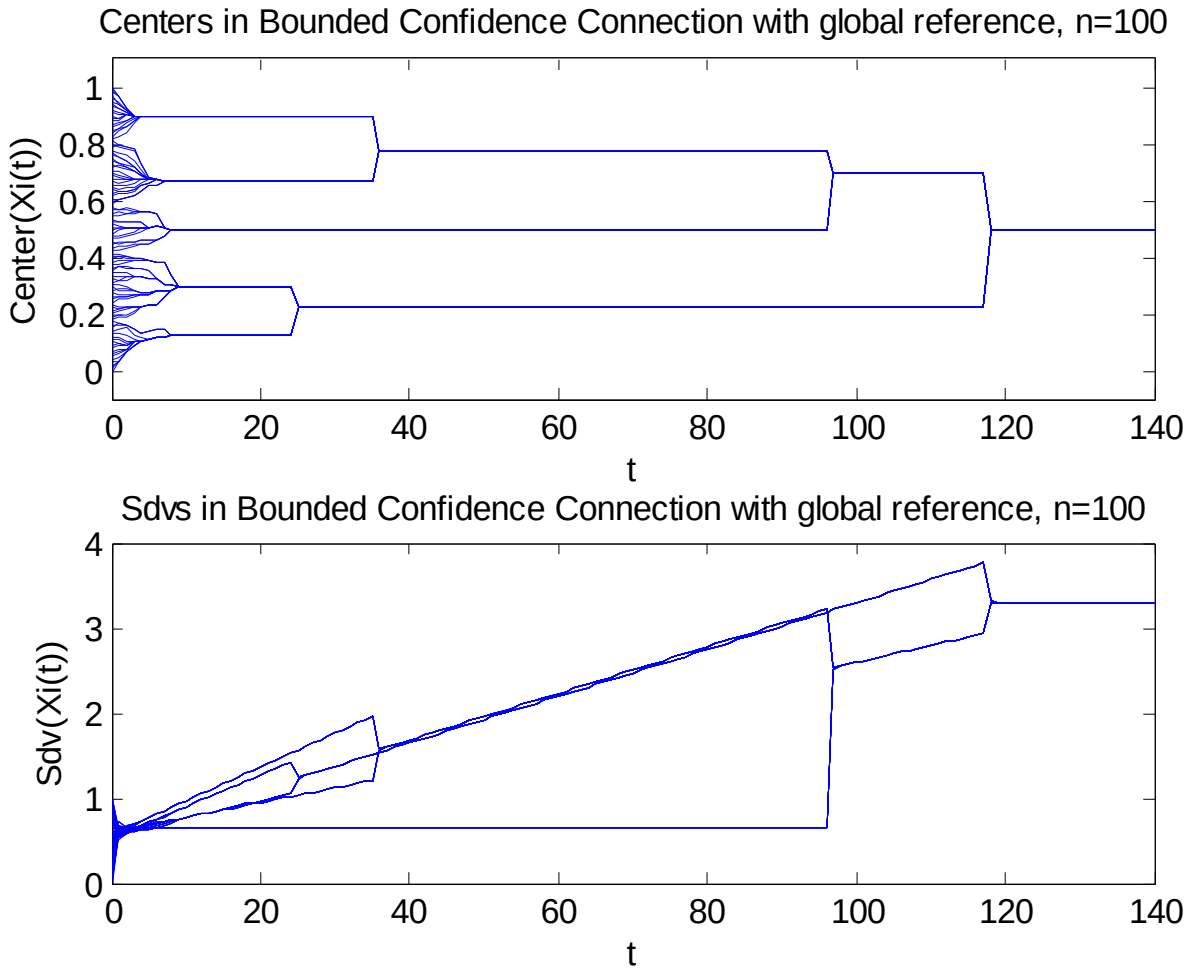

Fig. 25: A simulation run of the Bounded Confidence Connection dynamics (103)-(106) with Global Reference (102) for $n = 100$ nodes, where the top and bottom sub-figures plot Center$(X_i(t))$ and Sdv$(X_i(t))$, respectively.

## VII. Concluding Remarks

Human opinions are inherently fuzzy, because the carrier of opinions is natural language[20], thus a mathematical theory for opinion dynamics should consider the opinion and its uncertainty simultaneously. The Fuzzy Opinion Network (FON) theory proposed in this paper is such a theory, which we developed through studying in detail thirteen specific Gaussian FONs (G-FONs), starting from the basic center and sdv (standard deviation) connections and ending up with the Bounded Confidence G-FON which captures some important elements in human opinion formation and propagation. The basic properties of the G-FONs are:

(i) The center connection of Gaussian fuzzy nodes preserves the Gaussian form for the membership functions, therefore a single variable – the standard deviation (sdv) – can be used to represent the uncertainty of the opinion (which is modeled as the center of the Gaussian fuzzy set);

(ii) The center connection does not change the opinion itself but increases its uncertainty (sdv) in a linear fashion;

(iii) The sdv connection of Gaussian fuzzy nodes "stretches" the membership function from Gaussian to stretched exponential type of functions which have much heavier tails and sharper center than the Gaussian function, indicating that if the uncertainty (sdv) itself is uncertain (a fuzzy set), then the resulting membership function of the Gaussian node is no longer Gaussian so that a single variable (such as sdv) is not enough to characterize the uncertainty of the opinion (we therefore introduced the concepts of kurtosis and sharpness of a fuzzy set); and,

(iv) The weighted average of Gaussian fuzzy sets is still a Gaussian fuzzy set with its center equal to the weighted average of the centers of the individual Gaussian fuzzy sets and its sdv equal to the weighted average of the individual sdv's; this result paves the way for constructing a variety of network structures to capture many interesting situations in practice while the resulting G-FONs are still feasible for mathematical analysis.

The main insights gained from the mathematical analyses of the G-FONs in this paper include:

(a) Opinion leaders are important for a community, in the sense that the anxiety (uncertainty) of all members in the community may easily go to infinity if everyone compromises fully with all others (i.e., if there is no opinion leader in the community); when there is an opinion leader (a stubborn node in the FON), however, the anxieties of all members in the community will converge to a finite number; this may explain why strong leaders are in general uniformly welcome across culturally and politically vastly different communities [21] – the psychological origin of dictatorship[22].

(b) Speed matters for time-varying FONs; for example, depending on the speed of gaining confidence through communicating with others, the uncertainties of the members in the community may stabilize at a finite number if the speed is fast or go to infinity if the speed is slow; this may explain why the evolutionary forces make humans use short-cuts or heuristics, rather than thorough thinking, when facing complex situations [35].

(c) For state-dependent connections such as the Ring Connection or the Bounded Confidence Connection, a consensus will be reached if a central agent collects the opinions of all individuals and discloses them back to each individual (this shows that the media is a powerful machine to control people's opinion), while in the decentralized control scenario (people know only the opinions of neighbors), different communities emerge where people in the same community reach a consensus but these different consensuses remain separated forever (this explains why people in the same country tend to share similar opinions while people at different countries may have fundamentally different opinions – an origin of war).

Mathematical modeling of human opinion dynamics is a fascinating enterprise that has attracted investigators from a wide spectrum of disciplines, and the Fuzzy Opinion Network (FON) models proposed in this paper provide a new dimension for this exciting adventure (another "blind monk" touching the elephant?). While we are digging deeply into the mathematical details, it is important to keep in mind that the ultimate goal of the FON research is to provide good solutions to important real-world problems [61] such as 1) the automatic detection and allocation of extremist groups over the internet when these groups just begin to emerge (for the benefit of society) or 2) identifying the hidden operations in the investment markets and taking advantage of the knowledge (for self money making [69]). This paper is the first formal paper on FON theory and much more remains to be done. Specifically, we are considering the following future research topics:

1) Extend the Gaussian FONs of this paper to other types of fuzzy nodes and connections. For example, as discussed in footnotes 10 and 11, we may use the general triangular fuzzy nodes to construct triangular FONs to handle asymmetric uncertainties that the Gaussian FONs cannot model, and use logic operators such as max to select opinions from different people. Although these "incremental" researches are straightforward in concept, they are important because a richer content of the FON theory will provide more tools for solving real-world problems.
2) Study the Bounded Confidence FON (BCFON) of Connection 13 in detail. In addition to the convergence analysis of the opinions $\text{Center}(X(t))$ and their uncertainties $\text{Sdv}(X(t))$, we may define some key measures for the BCFON and study them in detail. For example, after redefining the connections as: $w_{ij}(t) = \begin{cases} \max_x min\left(\mu_{X_i(t-1)}(x), \mu_{X_j(t-1)}(x)\right), & j \in N_i(t) \\ 0, & j \notin N_i(t) \end{cases}$ (which can be viewed as the *threshold distance* between nodes $i$ and $j$) , we may define the *degree* of node $i$ as the summation of all the incoming weights $w_{ij}(t)$ to node $i$: $d_i(t) = \sum_{j=1}^{n} w_{ij}(t)$, and study the degree distributions in various scenarios to see, for example, under what conditions a *scale-free* [5] *BCFON* will emerge. Also, we may define the *diameter* of a BCFON as the largest distance between any two nodes and study the *small-world BCFONs*. Furthermore, we may define various concepts of centrality [72], such as *degree centrality*, *closeness centrality*, *betweenness centrality*, *prestige-*, *power-*, and *eigenvector-related centralities*, for a BCFON and study how these centrality measures change in different situations. These studies should be especially useful for understanding heterogeneous ($d_i$'s in $N_i(t)$ are different for $i = 1, \dots, n$) BCFONs (heterogeneity is a major challenge in opinion dynamics research [45], [51]).
3) Develop hierarchical FONs. Social institutions (companies, schools, military, government, etc.) are central in social control, coordination and economic growth [1], and hierarchy is the basic structure of these institutions; therefore, it is important to study FONs that connect the fuzzy nodes in a hierarchical fashion. Fig. 26 shows a hierarchical FON, where the same fuzzy node may play different roles depending on the local networks the node is connected to. Specifically, a fuzzy node in a hierarchical FON may play three different roles simultaneously: when connecting to the lower-level nodes, it is a *stubborn node* (outwards connections only); when connecting to the nodes in the same level, it is a *compromising node* (undirected connections); and when connecting to the upper-level nodes, it is an *obedience node* (inwards connections only). Based on the experiences with the stubborn nodes in Connections 8 and 12, we may easily imagine that if the three nodes in the top level of Fig. 26 reach a consensus, then all nodes in the hierarchical FON will eventually converge to this consensus; if, however, these three top nodes cannot reach a consensus, then things will get complicated. Furthermore, by adding very few connections between the nodes of different groups in the lowest level of Fig. 26 (the dashed lines in Fig. 26), we obtain a *small-world hierarchical FON* which should exhibit some very interesting properties that may have profound implications for designing hierarchical institutions for social control and coordination. Clearly, an in-depth study of these issues is a very important research direction.

[20] Perhaps a more accurate statement is the other way around: Because human opinions are inherently fuzzy, natural languages, which were developed to represent human opinions, abound with fuzzy words and vague expressions.

[21] The impact of stubbornness on opinion dynamics is an active research topic in the literature (e.g. [3], [23], [74]), and these studies generally discovered the "negative" aspects of stubbornness such as brainwashing other people (as Connections 8 and 12 of this paper demonstrate), but since the models used in these studies considered only the opinions (the uncertainties of the opinions were not considered), the "positive" aspect of stubbornness – easing the anxiety of the group members – was missed. Why a stubborn member can ease the anxieties of the other members in the group? We give the following explanation (we are all amateur social psychologists, as Myers ([52], page 17) noted): The other members of the group may think that since the stubborn member is so strong in his/her opinion, he/she must take full responsibility for the consequences of the opinion, so if the other members followed the stubborn member's opinion, they would feel less pressured, and less pressure implies less anxiety.

[22] See the masterpiece [4] for the economic origins of dictatorship and democracy.

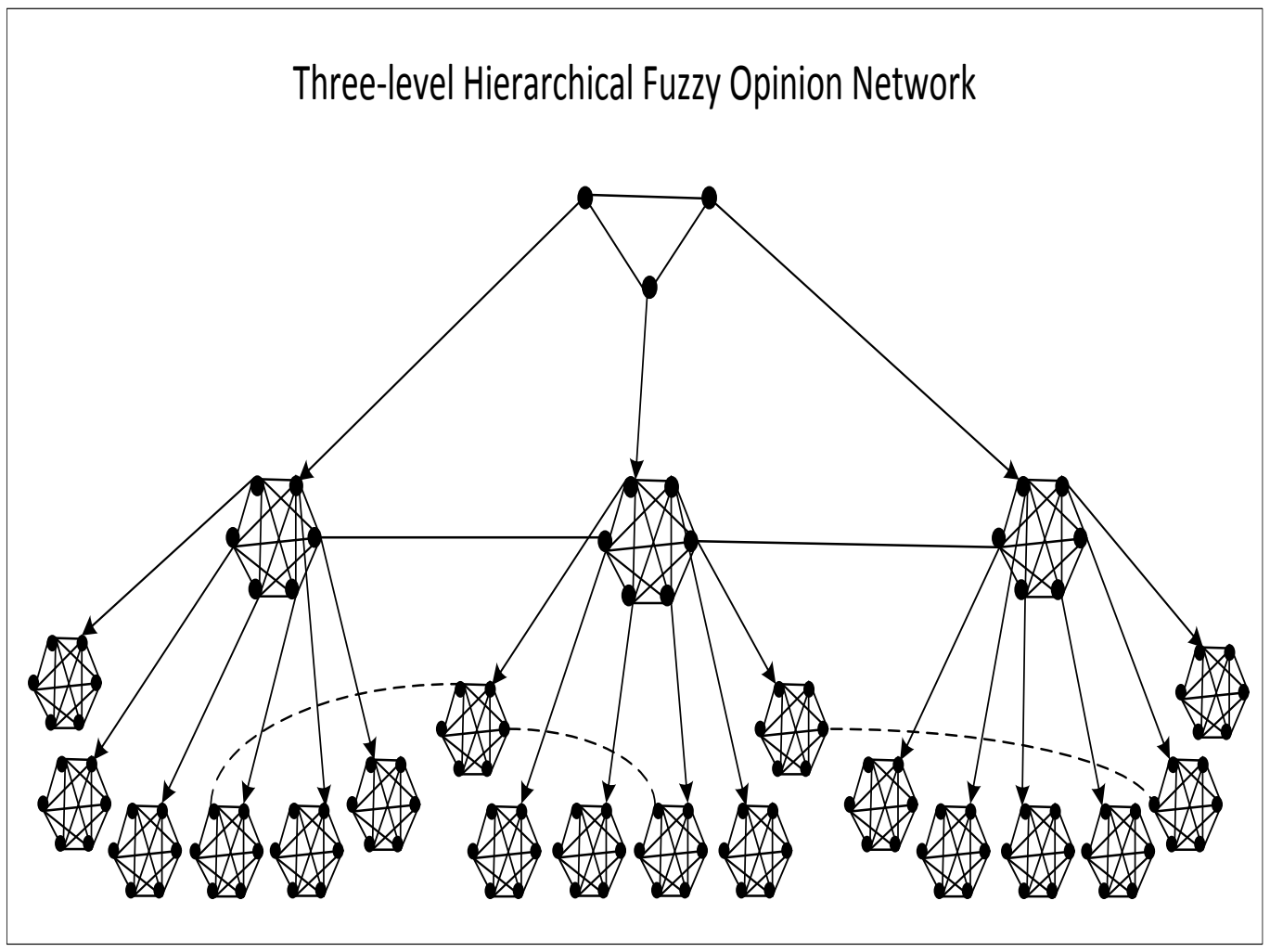

Fig. 26: A hierarchical fuzzy opinion network; when the dashed links are added, it becomes a small-world hierarchical FON.

4) How like minds go to extremes – a FON approach to group polarization [31]. The rise of the extremist movement imposes a major threat to world peace, and the internet provides a fertile land for the extremist groups to form, to grow and to prosper. To detect the formation of extremist groups over the internet, it is important to build a good mathematical model that captures the key elements of the extremist group formation process. With the huge amount of information over the internet and the fast-changing nature of the information, it is hard to imagine of developing an automatic algorithm to detect the formation of extremist groups without a good mathematical model for the group polarization dynamics. As discussed in footnote 10, there may be a variety of ways to construct FONs to model the group polarization processes.
5) Incorporate the basic principles in social psychology into the FON framework. The main difference between Natural Sciences and Social Sciences is that Natural Sciences are built on some solid foundation stones such as the Newton's Laws, whereas we lack deep underpinnings at the origin of Social Sciences [50]. What is the origin of Social Sciences? We argue that the origin of Social Sciences is human psychology because societies are nothing but human interactions. Although many psychological experiments were challenged of lacking reproducibility [11], there do exist many theories of human psychological processes that admit strong empirical supports [30]. The problem is that these psychological theories are mostly qualitative descriptions [52], and we need to convert them into mathematics in order to consolidate them as foundation stones for Social Sciences. If we construct the FONs based on these well-established psychological theories, then on one hand the constructed FONs are well founded on psychological evidences, and on the other hand it provides a framework to convert these qualitative theories into quantitative mathematics. The following social psychology theories [52] may be considered when we construct FONs (some of these theories lead to "games on FONs" to be discussed in 6)): actor-observer difference theory, adaptation-level theory, aggregation theories, altruism theory, arbitration theory, attitude inoculation theory, attribution theory, bargaining theories, behavioral confirmation theory, belief perseverance theory, bystander effect theory, cognitive dissonance theory, cohesiveness theory, collectivism theory, compliance theories, confirmation bias theory, conformity theory, correspondence bias theory, fearful attachment theory, foot-in-the-door theory, frustration-aggression theory, hindsight bias theory, hostile aggression theory, illusion of control theory, illusory correlation theory, immune neglect theory, impact bias theory, just-world theory, learned helplessness theory, locus of control theory, low-ball theory, mediation theory, mere exposure effect theory, mindguarding theory, minority slowness theory, mirror-image theory, moral exclusion theory, need to belong theory, normative influence theory, obedience theories, overconfidence theories, persuasion theories, pluralistic ignorance theory, primacy effect theory, realistic group conflict theory, recency effect theory, representativeness theory, reward attraction theory, rosy retrospection theory, self-affirmation theory, self-fulfilling prophecy, self-perception theory, self-presentation theory, sleeper effect theory, social comparison theory, social-exchange theory, social facilitation theory, social loafing theory, social trap theories, spotlight effect theory, stereotype theories, terror management theory, …[23].

6) Games on FONs. Two approaches: bottom-up or top-down. In the bottom-up approach, a specific FON is constructed to model the specific game scenario, then mathematical analysis and simulations of the model are performed to study the problem; the research assistantship competition game of Connection 12 is an example of the bottom-up approach. In the top-down approach, a payoff function [32] is assigned to each node of a general FON and the connections of the FON are determined by optimizing the payoffs of the nodes. The payoff functions may be chosen according to the social psychology theories (some of which are listed in topic 5) above), and in this way we embed the qualitative psychological theories into the quantitative mathematical models (the FONs).
7) Test and calibrate the FON models through human subject experiments. Since a key feature of FON is the simultaneous consideration of opinions and their uncertainties, the human subjects in the experiments should be asked for their opinions as well as the uncertainties about the opinions. In fact this kind of questions are very common in real life. For example, when we are asked to review a paper, two questions are usually a must: i) the overall rating of the paper, and ii) your level of expertise on the subject. Clearly, the first question asks our opinion about the paper while the second question characterizes our uncertainty about the evaluation. A human subject experiment on FON we may think of is the *course evaluation process*. As professors we are evaluated by students at the end of teaching a course, and we may redesign the evaluation process to make it a test-bed for the FON models, as follows. First, we ask the students to submit their evaluations without any discussion with the other students, and these evaluations are served as the initial opinions and the initial uncertainties in our FON model. Then, we ask the students to form different kinds of groups to discuss the evaluation and then submit their updated, after-discussion individual evaluations; these different kinds of groups may be, e.g., all-boy all-girl groups, roommate groups, state- or country-originated groups, racial groups, self-selecting groups, random groups, etc.. Finally, the data (opinions and their uncertainties) from these different grouping schemes are used to identify the parameters of the FON model, and the result is a fully specified FON model that can provide numerical-wise predictions. When we teach the same course once again for other students, we would ask the students to go through the same process, but this time we have a FON model to predict their evaluations. In this way we put the FON models under a real scientific test [21].

[23] Comments: i) so many "blind monks" touching the elephant; and ii) many theories here concern the interplay between opinions and their uncertainties, which provides support for our basic argument that an opinion and its uncertainty are two sides of the same coin and should be considered simultaneously.

## APPENDIX

**Proof of Lemma 1:** We prove it by induction. For $n = 2$, the membership function of $Y_2 = w_1 X_1 + w_2 X_2$ is obtained from the Extension Principle [76]:

$$\mu_{Y_2}(y_2) = \max_{y_2 = w_1 x_1 + w_2 x_2} min\left[\mu_{X_1}(x_1), \mu_{X_2}(x_2)\right] \qquad \text{(A1)}$$

Substituting $\mu_{X_i}(x_i) = e^{-\frac{|x_i - c_i|^2}{\sigma_i^2}}, i = 1,2$ into (A1) and replacing $x_2$ by $x_2 = \frac{1}{w_2}(y_2 - w_1 x_1)$, we have

$$\mu_{Y_2}(y_2) = \max_{x_1 \in R^n} min\left[ e^{-\frac{|x_1 - c_1|^2}{\sigma_1^2}}, e^{-\frac{\left|\frac{1}{w_2}(y_2 - w_1 x_1) - c_2\right|^2}{\sigma_2^2}} \right] \qquad \text{(A2)}$$

The max in (A2) is achieved when

$$\frac{x_1 - c_1}{\sigma_1} = \frac{\frac{1}{w_2}(y_2 - w_1 x_1) - c_2}{\sigma_2} \qquad \text{(A3)}$$

which leads to

$$\mu_{Y_2}(y_2) = e^{-\frac{|y_{n+1} - w_1 c_1 - w_2 c_2|^2}{(w_1 \sigma_1 + w_2 \sigma_2)^2}} \qquad \text{(A4)}$$

So the lemma is true for $n = 2$. Now suppose the result (48) is true for $n$, then for $n+1$ we have

$$Y_{n+1} = Y_n + w_{n+1} X_{n+1} \qquad \text{(A5)}$$

and the membership function of $Y_{n+1}$ is obtained using the Extension Principle:

$$\mu_{Y_{n+1}}(y_{n+1}) = \max_{y_{n+1} = y_n + w_{n+1} x_{n+1}} min\left[\mu_{Y_n}(y_n), \mu_{X_{n+1}}(x_{n+1})\right] \qquad \text{(A6)}$$

Substituting $\mu_{Y_n}(y_n)$ of (48) and $\mu_{X_{n+1}}(x_{n+1}) = e^{-\frac{|x_{n+1} - c_{n+1}|^2}{\sigma_{n+1}^2}}$ into (A6) and replacing $y_n$ by $y_n = y_{n+1} - w_{n+1} x_{n+1}$, we have

$$\mu_{Y_{n+1}}(y_{n+1}) = \max_{x_{n+1}\in R^n} min\left[e^{-\frac{\left|y_{n+1}-w_{n+1}x_{n+1}-\sum_{i=1}^{n} w_i c_i\right|^2}{\left(\sum_{i=1}^{n} w_i\sigma_i\right)^2}}, e^{-\frac{|x_{n+1}-c_{n+1}|^2}{\sigma_{n+1}^2}}\right] \quad (A7)$$

The max in (A7) is achieved when

$$\frac{y_{n+1} - w_{n+1}x_{n+1} - \sum_{i=1}^{n} w_i c_i}{\sum_{i=1}^{n} w_i\sigma_i} = \frac{x_{n+1} - c_{n+1}}{\sigma_{n+1}} \quad (A8)$$

which leads to

$$\mu_{Y_{n+1}}(y_{n+1}) = e^{-\frac{\left|y_{n+1}-\sum_{i=1}^{n+1} w_i c_i\right|^2}{\left(\sum_{i=1}^{n+1} w_i\sigma_i\right)^2}} \quad (A9)$$

The induction is complete and the lemma is proven. ■

**Proof of Result 1:** (a) Let $\lambda_1, \lambda_2, \ldots, \lambda_n$ be the eigenvalues of $W$ (84) and $u_1, u_2, \ldots, u_n$ be the corresponding eigenvectors. Since the $W$ is a symmetric stochastic matrix (the elements of a row of $W$ sum to 1 and all elements of the matrix are non-negative), we have from matrix theory [29] that

$$1 = \lambda_1 > |\lambda_2| \geq \ldots \geq |\lambda_n| \geq 0 \quad (A10)$$

(the largest eigenvalue $\lambda_1$ is single and equals 1 and all other eigenvalues are strictly less than 1 in absolute value) and therefore

$$W = U \operatorname{diag}(1, \lambda_2, \ldots, \lambda_n)U^T \quad (A11)$$

where $U = (u_1, u_2, \ldots, u_n)$ with $U^T U = I$ ($u_i^T u_j = 1$ if $i = j$ and $u_i^T u_j = 0$ if $i \neq j$) and $u_1 = \left(\frac{1}{\sqrt{n}}, \ldots, \frac{1}{\sqrt{n}}\right)^T$ ($Wu_1 = u_1$ since $W$ is stochastic so that $u_1$ is the eigenvector corresponding to the eigenvalue $\lambda_1 = 1$ and satisfies $u_1^T u_1 = 1$). Hence, $\text{Center}(X(t))$ in (87) becomes

$$\text{Center}(X(t)) = U \operatorname{diag}(1, \lambda_2^t, \ldots, \lambda_n^t)U^T x(0) = u_1 u_1^T x(0) + \sum_{j=2}^{n} \lambda_j^t u_j u_j^T x(0) \quad (A12)$$

Since $1 > |\lambda_2| \geq \ldots \geq |\lambda_n|$, so that the last term of (A12) approaches zero as $t$ goes to infinity, this gives

$$\lim_{t\to\infty} \text{Center}(X(t)) = u_1 u_1^T x(0) = \begin{pmatrix} \frac{1}{n}\sum_{j=1}^{n} x_j(0) \\ \vdots \\ \frac{1}{n}\sum_{j=1}^{n} x_j(0) \end{pmatrix} \quad (A13)$$

(b) Since $W = [w_{ij}]$ is symmetric and stochastic so that $\sum_{i=1}^{n} w_{ij} = 1$ for any $j = 1, \ldots, n$, we have from the dynamic equation (85) for the centers $\text{Center}(X(t))$ that

$$\sum_{i=1}^{n} \text{Center}(X_i(t)) = \sum_{i=1}^{n}\left(\sum_{j=1}^{n} w_{ij} \text{Center}\left(X_j(t-1)\right)\right) = \sum_{j=1}^{n}\left(\sum_{i=1}^{n} w_{ij}\right)\text{Center}\left(X_j(t-1)\right) = \sum_{j=1}^{n} \text{Center}\left(X_j(t-1)\right) \quad (A14)$$

Therefore,

$$\frac{1}{n}\sum_{j=1}^{n} \text{Center}\left(X_j(t-1)\right) = \frac{1}{n}\sum_{j=1}^{n} x_j(0) \quad (A15)$$

and $\sigma_i(t)$ of (83) becomes

$$\sigma_i(t) = \left|\text{Center}\left(X_i(t-1)\right) - \frac{1}{n}\sum_{j=1}^{n} x_j(0)\right| \quad (A16)$$

Since $\lim_{t\to\infty} \text{Center}\left(X_i(t)\right) = \frac{1}{n}\sum_{j=1}^{n} x_j(0)$ from (A13), we have $\lim_{t\to\infty} \sigma_i(t) = 0$.

(c) Substituting (A12) into (A16) and noticing the last equality in (A13), we have

$$\sigma_i(t) = \left|\left(\sum_{j=2}^{n} \lambda_j^{t-1} u_j u_j^T x(0)\right)_i\right| \leq \sum_{j=2}^{n} a_{ij}\left|\lambda_j\right|^{t-1} \leq a_i|\lambda_2|^{t-1} \quad (A17)$$

where $(v)_i$ denotes the $i$'th element of vector $v$, $a_{ij}$ are positive constants determined from $u_j$ and $x(0)$, $a_i = \sum_{j=2}^{n} a_{ij}$ and we use the fact $|\lambda_2| \geq |\lambda_3| \geq \ldots \geq |\lambda_n|$. With (A11) and (A10), the $\text{Sdv}(X(t))$ of (88) becomes

$$\text{Sdv}(X(t)) = u_1 u_1^T \sigma(0) + \sum_{j=2}^{n} \lambda_j^t u_j u_j^T \sigma(0) + \sum_{k=1}^{t}\left(u_1 u_1^T \sigma(k) + \sum_{j=2}^{n} \lambda_j^{t-k} u_j u_j^T \sigma(k)\right) \quad (A18)$$

Using (A17) and the fact that $1 > |\lambda_2| \geq |\lambda_3| \geq \ldots \geq |\lambda_n|$, we have

$$\sum_{k=1}^{t}\left(\sum_{j=2}^{n} \lambda_j^{t-k} u_j u_j^T \sigma(k)\right)_i \leq \sum_{k=1}^{t}\left(\sum_{j=2}^{n}\sum_{i=1}^{n} b_{ij}\left|\lambda_j\right|^{t-k} \sigma_i(k)\right) \leq \sum_{k=1}^{t} b|\lambda_2|^{t-k}|\lambda_2|^{k-1} = bt|\lambda_2|^{t-1} \quad (A19)$$

(where $b_{ij}$ and $b$ are positive constants determined from $u_j$ and the $a_i$ in (A17)) which goes to zero as $t$ goes to infinity. Hence,

$$\lim_{t\to\infty} \text{Sdv}(X(t)) = u_1 u_1^T \sigma(0) + u_1 u_1^T \lim_{t\to\infty}\sum_{k=1}^{t} \sigma(k) \quad (A20)$$

where, using (A17),

$$\lim_{t\to\infty}\sum_{k=1}^{t} \sigma_i(k) \leq \lim_{t\to\infty}\sum_{k=1}^{t}\sum_{j=2}^{n} a_{ij}\left|\lambda_j\right|^{k-1} = \sum_{j=2}^{n} a_{ij}\left(\frac{1}{1-\left|\lambda_j\right|}\right) \quad (A21)$$

Given that $\sigma_i(k) = \left|\left(\sum_{j=2}^{n} \lambda_j^{k-1} u_j u_j^T x(0)\right)_i\right|$ and the eigenvalues $\lambda_j$ are real, (A21) implies that $\sum_{k=1}^{\infty} \sigma_i(k)$ is a finite number. Since $u_1 u_1^T$ is an $n$-by-$n$ matrix with all the elements equal to $\frac{1}{n}$ (recall that $u_1 = \left(\frac{1}{\sqrt{n}}, \ldots, \frac{1}{\sqrt{n}}\right)^T$), (A20) shows that the $n$ elements of $\lim_{t\to\infty} \text{Sdv}(X(t))$ are equal to the same value. ■

## ACKNOWLEDGEMENTS

The authors wish to thank the reviewers for their very insightful and inspiring comments that helped to improve the paper.

## REFERENCES

[1] Acemoglu, D., S. Johnson and J.A. Robinson, "Institutions As A Fundamental Cause of Long-Run Growth," in P. Aghion and S.N. Durlauf, eds., *Handbook of Economic Growth* Vol. 1A, Elsevier B.V.: 385-472, 2006.

[2] Acemoglu, D. and A. Ozdaglar, "Opinion Dynamics and Learning in Social Networks," *Dynamic Games and Applications* 1.1: 3-49, 2011.

[3] Acemoglu D., A. Ozdaglar and A. ParandehGheibi, "Spread of (Mis)information in Social Networks," *Games and Economic Behavior* 70(2):194-227, 2010.

[4] Acemoglu, D. and J.A. Robinson, *Economic Origins of Dictatorship and Democracy*, Cambridge University Press, NY, 2006.

[5] Barabasi, A.L., "Scale-Free Networks: A Decade and Beyond," *Science* 325: 412-413, 2009.

[6] Bala, V. and S. Goyal, "Learning from Neighbors," *Review of Economic Studies* 65: 595-621, 1998.

[7] Ballester, C., A. Calvo-Armengol and Y. Zenou, "Who's Who in Networks. Wanted: The Key Player," *Econometrica* 74: 1403-1417, 2006.

[8] Bargh, J.A. and M.J. Ferguson, "Beyond Behaviorism: On the Automaticity of Higher Mental Processes," *Psychological Bulletin* 126(6): 925-945, 2000.

[9] Bearman, P.S. and P. Parigi, "Cloning Headless Frogs and Other Important Matters: Conversion Topics and Network Structure," *Social Forces* 83: 535-557, 2004.

[10] Blondel, V.D., J.M. Hendrickx and J.N. Tsitsiklis, "On Krause's Multi-Agent Consensus Model with State-Dependent Connectivity," *IEEE Trans. on Automatic Control* 54(11): 2586-2597, 2009.

[11] Bohannon, J., "Many Psychology Papers Fail Replication Test," *Science* 349: 910-911, 2015.

[12] Butts, C.T., "Revisiting the Foundations of Network Analysis," *Science* 325: 414-416, 2009.

[13] Carpenter, W.B., *Principles of Mental Physiology*, Appleton, NY, 1888.

[14] Charniak, E., "Bayesian Networks without Tears," *AI Magazine* 12(4): 50-63, 1991.

[15] Chamley, C., A. Scaglione, and L. Li, "Models for the Diffusion of Beliefs in Social Networks: An Overview," *Signal Processing Magazine* 30(3): 16-29, 2013.

[16] Christoff, Z. and J.U. Hansen, "A Two-tiered Formalization of Social Influence," *Logic, Rationality, and Interaction – Lecture Notes in Computer Science* 8196: 66-81, 2013.

[17] Chung, K.L., *A Course in Probability Theory (3rd Edition)*, Elsevier Inc., 2001.

[18] DeGroot, M.H., "Reaching A Consensus," *Journal of the American Statistical Association* 69: 118-121, 1974.

[19] Dubois, D. and E. Hullermeier, "Comparing Probability Measures Using Possibility Theory: A Notion of Relative Peakedness," *International Journal of Apprximate Reasoning* 45(2): 364-385, 2007.

[20] Erdos, P. and A. Renyi, "On the Evolution of Random Graphs," *Publications of the Mathematical Institute of the Hungarian Academy of Sciences* 5: 17-61, 1960.

[21] Feitelson, D.G., "From Repeatability to Reproducibility and Corroboration," *ACM SIGOPS Operating Systems Review* 49(1):3-11, 2015.

[22] Friedkin, N.E., "The Problem of Social Control and Coordination of Complex Systems in Sociology: A Look at the Community Cleavage Problem," *IEEE Control Systems Magazine* 35(3): 40-51, 2015.

[23] Ghaderi, J. and R. Srikant, "Opinion Dynamics in Social Networks with Stubborn Agents: Equilibrium and Convergence Rate," *Automatica* 50(12): 3209-3215, 2014.

[24] Golub, B. and M.O. Jackson, "Naïve Learning in Social Networks: Convergence, Influence, and the Wisdom of Crowds," http://www.standford.edu/~jacksonm /naivelearning.pdf, 2007.

[25] Gross, T. and B. Blasius, "Adaptive Coevolutionary Networks: A Review," *Journal of the Royal Society Interface* 5: 259-271, 2008.

[26] Hegselmann, R. and U. Krause, "Opinion Dynamics and Bounded Confidence: Models, Analysis, and Simulations," *Journal of Artificial Societies and Social Simulations* 5(3): http://jasss.soc.surrey.ac.uk/5/3/2.html, 2002.

[27] Hegselmann, R. and U. Krause, "Opinion Dynamics under the Influence of Radical Groups, Charismatic Leaders, and Other Constant Signals: A Simple Unifying Model," *Networks and Heterogeneous Media* 10(3): 477-509, 2015.

[28] Hommes, C.H., "Heterogeneous Agent Models in Economics and Finance," in L. Tesfatsion and K.L. Judd, eds., *Handbook of Computational Economics Vol. 2: Agent-Based Computational Economics*, Elsevier B.V.: 1109-1186, 2006.

[29] Horn, R.A. and C.R. Johnson, *Matrix Analysis (2nd Edition)*, Cambridge University Press, 2013.

[30] Hurtado-Parrado, C. and W. López-López, "Single-Case Research Methods: History and Suitability for a Psychological Science in Need of Alternatives," *Integrative Psychological and Behavioral Science* 49(3): 323-349, 2015.

[31] Isenberg, D.J., "Group Polarization: A Critical Review and Meta-Analysis," *Journal of Personality and Social Psychology* 50(6): 1141-1151, 1986.

[32] Jackson, M.O., *Social and Economic Networks*, Princeton University Press, 2008.

[33] Jensen, F.V., *Bayesian Networks and Decision Graphs (2nd Edition)*, Springer-Verlag, 2007

[34] Kahneman, D., "Maps of Bounded Rationality: Psychology for Behavioral Economics," *The American Economic Review* 93(5): 1449-1475, 2003.

[35] Kahneman,D., *Thinking: Fast and Slow*, Farrar, Straus, New York, 2011.

[36] Keynes, J.M., *The General Theory of Employment, Interest and Money*, (BN Publishing, 2009), 1935.

[37] Klir, G., "A Principle of Uncertainty and Information Invariance," *International Journal of General Systems* 17: 249-275, 1990.

[38] Krebs, V.E., "Mapping Networks of Terrorist Cells," *Connections* 24: 43-52, 2002.

[39] Kugler, T., E.E. Kausel and M.G. Kocher, "Are Groups More Rational Than Individuals? A Review of Interactive Decision Making in Groups," *CESifo working paper: Behavioural Economics* No. 3701, 2012.

[40] Laherrere, J. and D. Sornette, "Stretched Exponential Distributions in Nature and Economy: 'Fat-tails' with Characteristic Scales," *Eur. Phys. J. B* 2: 525-539, 1998.

[41] Langer, E.J., "The Illusion of Control," *Journal of Personality and Social Psychology* 32: 311-328, 1975.

[42] Liu, F, J. Seligman and P. Girard, "Logical Dynamics of Belief Change in the Community," *Synthese* 191(10): 2403-2431, 2014.

[43] Lorenz, J., "A Stabilization Theorem for Dynamics of Continuous Opinions," *Physica A* 355: 217-223, 2005.

[44] Lorenz, J., "Continuous Opinion Dynamics under Bounded Confidence: A Survey," *International Journal of Modern Physics C* 18(12): 1819–1838, 2007.

[45] Lorenz, J.,."Heterogeneous Bounds of Confidence: Meet, Discuss and Find Consensus", *Complexity* 4(15): 43–52, 2010.

[46] Martins, A.C.R., "Bayesian Updating Rules in Continuous Opinion Dynamic Models," *arXiv*:0807.4972v1, 2008.

[47] Mason, W.A., F.R. Conrey and E.R. Smith, "Situating Social Influence Processes: Dynamic, Multidirectional Flows of Influence within Social Networks," *Personality and Social Psychology Review* 11(3): 279-300, 2007.

[48] Mendel, J.M., "Type-2 Fuzzy Sets and Systems: An Overview," *IEEE Computational Intelligence Magazine* 2: 20-29, 2007.

[49] Mendel, J.M. and D. Wu, *Perceptual Computing: Aiding People in Making Subjective Judgments*, IEEE Press and Wiley, 2010.

[50] Mirowski, P., *More Heat Than Light: Economics as Social Physics, Physics as Nature's Economics*, Cambridge University Press, 1989.

[51] Mirtabatabaei, A. and F. Bullo, "On Opinion Dynamics in Heterogeneous Networks," *arXiv*:1010.2186v2, 2011.

[52] Myers, D.G., *Social Psychology (9th Edition)*, McGraw-Hill Education (Asia), 2008.

[53] Nedic, A., A. Ozdaglar and P.A. Parrilo, "Constrained Consensus and Optimization in Multi-Agent Networks," *IEEE Trans. on Automatic Control* 55(4): 922-938, 2010.

[54] Newman, M., A.L. Barabasi and D.J. Watts, Eds., *The Structure and Dynamics of Networks*, Princeton University Press, Princeton, NJ, 2006.

[55] Olfati-Saber, R., J.A. Fax and R.M. Murray, "Consensus and Cooperation in Networked Multi-Agent Systems," *Proceedings of the IEEE* 95(1): 215-233, 2007.

[56] Palla, G., A.L. Barabasi and T. Vicsek, "Quantifying Social Group Evolution," *Nature* 446: 664-667, 2007.

[57] Papoulis, A., *Probability, Random Variables, and Stochastic Processes (3rd Edition)*, McGraw-Hill Inc., 1991.

[58] Roweis, S.T. and Z. Ghahramani, "A Unifying Review of Linear Gaussian Models," *Neural Computation* 11 (2): 305-345, 1999.

[59] Russell, S.J. and P. Norvig, *Artificial Intelligence: A Modern Approach (3rd Edition)*, Pearson Education Asia Ltd., 2011.

[60] Scott, J., *Social Network Analysis (3rd Edition)*, Sage Publications Ltd., London, 2013.
[61] Sobkowicz, P., "Modelling Opinion Formation with Physics Tools: Call for Closer Link with Reality," *Journal of Artificial Societies and Social Simulation* 12(1): 11-24, 2009.
[62] Strogatz, S.H., "Exploring Complex Networks," *Nature* 410: 268-276, 2001.
[63] Surowiecki, J., *The Wisdom of Crowds: Why the Many Are Smarter Than the Few*, Abacus, London, 2005.
[64] Sunstein, C.R., "The Law of Group Polarization," *John M. Olin Program in Law and Economics Working Paper* No. 91, 1999.
[65] Sunstein, C.R., *Going to Extremes: How Like Minds Unite and Divide*, Oxford University Press, NY, 2009.
[66] Vallinder, A. and E.J. Olsson, "Trust and the Value of Overconfidence: A Bayesian Perspective on Social Network Communication," *Synthese* 191(9): 1991-2007, 2014.
[67] Vespignani, A., "Predicting the Behavior of Techno-Social Systems," *Science* 325: 425-428, 2009.
[68] Wang, L.X., *A Course in Fuzzy Systems and Control*, Prentice-Hall, NJ, 1997.
[69] Wang, L.X., "Dynamical Models of Stock Prices Based on Technical Trading Rules Part I: The Models; Part II: Analysis of the Model; Part III: Application to Hong Kong Stocks," *IEEE Trans. on Fuzzy Systems* 23(4): 787-801 (Part I); 1127-1141 (Part II); to appear (Part III), 2015.
[70] Wang, L.X., "Modeling Stock Price Dynamics with Fuzzy Opinion Networks," http://ssrn.com/abstract=2645196, 2015.
[71] Wang, L.X. and J. M. Mendel, "Fuzzy Networks: What Happens When Fuzzy People Are Connected through Social Networks," *Proc. of the 2014 IEEE Symposium Series on Computational Intelligence*, Orlando, FL, USA, Dec. 2014.
[72] Wasserman, S. and K. Faust, *Social Network Analysis: Methods and Applications*, Cambridge University Press, New York, 1994.
[73] Watts, D.J. and S. Strogatz, "Collective Dynamics of 'Small-World' Networks," *Nature* 393: 440-442, 1998.
[74] Yildiz, E., D. Acemoglu, A. Ozdaglar, A. Saberi, and A. Scaglione, "Discrete Opinion Dynamics with Stubborn Agents," http://ssrn.com/abstract=1744113, 2011.
[75] Zadeh, L.A., "Outline of a New Approach to the Analysis of Complex Systems and Decision Processes," *IEEE Trans. on Systems, Man, and Cybern.* 3: 28-44, 1973.
[76] Zadeh, L.A., "The Concept of a Linguistic Variable and Its Application to Approximate Reasoning – I," *Information Sciences*, Vol. 8, pp. 199-249, 1975.
[77] Zadeh, L.A., "Fuzzy Sets As a Basis for a Theory of Possibility," *Fuzzy Sets and Systems* 1(1): 3-28, 1978.
[78] Zadeh, L.A., "From Computing with Numbers to Computing with Words --- From Manipulation of Measurements to Manipulation of Perceptions," *IEEE Trans. on Circuits and Systems – 1, Fundamental Theory and Applications* 4: 105-119, 1999.
[79] Zimmermann, H.J., *Fuzzy Set Theory and Its Applications (4th Edition)*, Kluwer Academic Publishers, 2001.

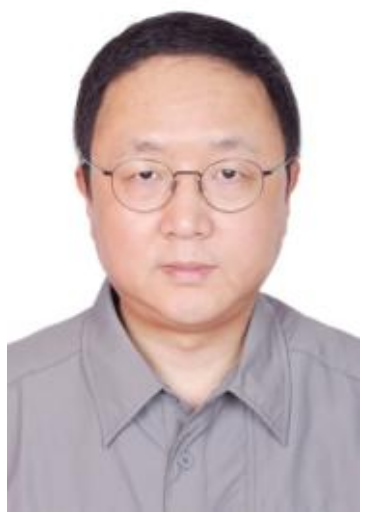

**Li-Xin Wang** received the Ph.D. degree from the Department of Electrical Engineering, University of Southern California (USC), Los Angeles, CA, USA, in 1992.

From 1992 to 1993, he was a Postdoctoral Fellow with the Department of Electrical Engineering and Computer Science, University of California at Berkeley. From 1993 to 2007, he was on the faculty of the Department of Electronic and Computer Engineering, The Hong Kong University of Science and Technology (HKUST). In 2007, he resigned from his tenured position at HKUST to become an independent researcher and investor in the stock and real estate markets in Hong Kong and China. In Fall 2013, he returned to academic and joined the faculty of the Department of Automation Science and Technology, Xian Jiaotong University, Xian, China, after a fruitful hunting journey across the wild land of investment to achieve financial freedom. His research interests are dynamical models of asset prices, market microstructure, trading strategies, fuzzy systems, and opinion dynamics.

Dr. Wang received USC's Phi Kappa Phi highest Student Recognition Award.

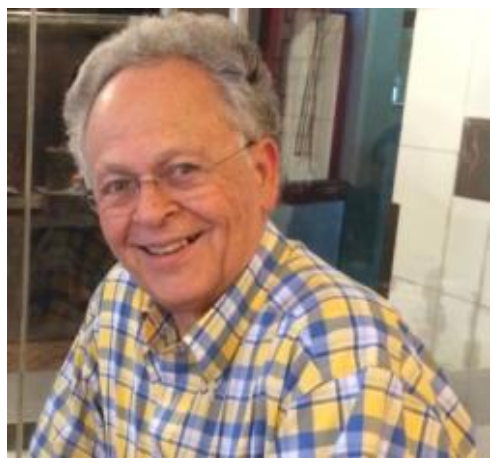

**Jerry M. Mendel** received the Ph.D. degree in electrical engineering from the Polytechnic Institute of Brooklyn, Brooklyn, NY. Currently he is Professor of Electrical Engineering at the University of Southern California in Los Angeles, where he has been since 1974. He has published over 570 technical papers and is author and/or co-author of 12 books, including *Uncertain Rule-based Fuzzy Logic Systems: Introduction and New Directions* (Prentice-Hall, 2001), *Perceptual Computing: Aiding People in Making Subjective Judgments* (Wiley & IEEE Press, 2010), and *Introduction to Type-2 Fuzzy Logic Control: Theory and Application* (Wiley & IEEE Press, 2014). His present research interests include: type-2 fuzzy logic systems and their applications to a wide range of problems, including smart oil field technology, computing with words, and fuzzy set qualitative comparative analysis. He is a Life Fellow of the IEEE, a Distinguished Member of the IEEE Control Systems Society, and a Fellow of the International Fuzzy Systems Association. He was President of the IEEE Control Systems Society in 1986, a member of the Administrative Committee of the IEEE Computational Intelligence Society for nine years, and Chairman of its Fuzzy Systems Technical Committee and the Computing With Words Task Force of that TC. Among his awards are the 1983 Best Transactions Paper Award of the IEEE Geoscience and Remote Sensing Society, the 1992 Signal Processing Society Paper Award, the 2002 and 2014 *Transactions on Fuzzy Systems* Outstanding Paper Awards, a 1984 IEEE Centennial Medal, an IEEE Third Millenium Medal, and a Fuzzy Systems Pioneer Award (2008) from the IEEE Computational Intelligence Society.